\documentclass[sigconf,nonacm]{acmart}

\usepackage{comment}
\usepackage{amsthm}
\usepackage{mathtools}
\usepackage{etoolbox}
\usepackage{multirow}

\usepackage{colortbl}
\usepackage{stmaryrd}
\usepackage[normalem]{ulem}
\usepackage{subcaption}
\usepackage{booktabs}
\usepackage[disable]{todonotes}
\usepackage{todonotes}
\usepackage{graphicx}
\usepackage{listings}
\usepackage{fancyvrb}
\usepackage{caption}
\usepackage{subcaption}
\usepackage{braket}
\usepackage[inline]{enumitem}
\usepackage{xspace}
\usepackage{xstring}
\usepackage{program}
\usepackage{tikz}
\usetikzlibrary{shapes}
\usepackage[noend,linesnumbered]{algorithm2e}
\usepackage{silence}
\usepackage{url}

\usepackage{breakurl}
\usepackage{cleveref}

\graphicspath{ {figures/} {figs/} }
\newrobustcmd{\needRewrite}[1]{\textcolor{magenta}{{#1}}}

\definecolor{revgreen}{rgb}{0,0.5,0}

\newrobustcmd{\reva}[1]{\textcolor{black}{{#1}}}
\newrobustcmd{\revb}[1]{\textcolor{black}{{#1}}}
\newrobustcmd{\revc}[1]{\textcolor{black}{{#1}}}
\newrobustcmd{\revm}[1]{\textcolor{black}{{#1}}}

\definecolor{lstpurple}{rgb}{0.5,0,0.5}
\definecolor{lstred}{rgb}{1,0,0}
\definecolor{lstreddark}{rgb}{0.7,0,0}
\definecolor{lstredl}{rgb}{0.64,0.08,0.08}
\definecolor{lstmildblue}{rgb}{0.66,0.72,0.78}
\definecolor{lstblue}{rgb}{0,0,1}
\definecolor{lstmildgreen}{rgb}{0.42,0.53,0.39}
\definecolor{lstgreen}{rgb}{0,0.5,0}
\definecolor{lstorangedark}{rgb}{0.6,0.3,0}
\definecolor{lstorange}{rgb}{0.75,0.52,0.005}
\definecolor{lstorangelight}{rgb}{0.89,0.81,0.67}
\definecolor{lstbeige}{rgb}{0.90,0.86,0.45}

\DeclareFontShape{OT1}{cmtt}{bx}{n}{<5><6><7><8><9><10><10.95><12><14.4><17.28><20.74><24.88>cmttb10}{}

\lstdefinestyle{psql}
{
tabsize=2,
basicstyle=\small\upshape\ttfamily,
language=SQL,
morekeywords={PROVENANCE,BASERELATION,INFLUENCE,COPY,ON,TRANSPROV,TRANSSQL,TRANSXML,CONTRIBUTION,COMPLETE,TRANSITIVE,NONTRANSITIVE,EXPLAIN,SQLTEXT,GRAPH,IS,ANNOT,THIS,XSLT,MAPPROV,cxpath,OF,TRANSACTION,SERIALIZABLE,COMMITTED,INSERT,INTO,WITH,SCN,UPDATED,OVER,PRECEEDING,FOLLOWING,CURRENT,ROW,ROWS,RANGE,GROUPS},
extendedchars=false,
keywordstyle=\bfseries,
mathescape=true,
escapechar=@,
sensitive=true
}

\lstdefinestyle{psqlcolor}
{
tabsize=2,
basicstyle=\small\upshape\ttfamily,
language=SQL,
morekeywords={PROVENANCE,BASERELATION,INFLUENCE,COPY,ON,TRANSPROV,TRANSSQL,TRANSXML,CONTRIBUTION,COMPLETE,TRANSITIVE,NONTRANSITIVE,EXPLAIN,SQLTEXT,GRAPH,IS,ANNOT,THIS,XSLT,MAPPROV,cxpath,OF,TRANSACTION,SERIALIZABLE,COMMITTED,INSERT,INTO,WITH,SCN,UPDATED,OVER,PRECEDING,FOLLOWING,CURRENT,ROW,ROWS,RANGE,UNBOUNDED,PARTITION,GROUPS},
extendedchars=false,
keywordstyle=\bfseries\color{lstpurple},
deletekeywords={count,min,max,avg,sum},
keywords=[2]{count,min,max,avg,sum,row_number,first,last,lag,lead,dense_rank,rank},
keywordstyle=[2]\color{lstblue},
stringstyle=\color{lstreddark},
commentstyle=\color{lstgreen},
mathescape=true,
escapechar=@,
sensitive=true
}

\lstdefinestyle{datalog}
{
basicstyle=\footnotesize\upshape\ttfamily,
language=prolog
}

\lstdefinestyle{pseudocode}
{
  tabsize=3,
  basicstyle=\small,
  language=c,
  morekeywords={if,else,foreach,case,return,in,or},
  extendedchars=true,
  mathescape=true,
  literate={:=}{{$\gets$}}1 {<=}{{$\leq$}}1 {!=}{{$\neq$}}1 {append}{{$\listconcat$}}1 {calP}{{$\cal P$}}{2},
  keywordstyle=\color{lstpurple},
  escapechar=&,
  numbers=left,
  numberstyle=\color{lstgreen}\small\bfseries,
  stepnumber=1,
  numbersep=5pt,
}

\lstdefinestyle{xmlstyle}
{
  tabsize=3,
  basicstyle=\small\upshape\ttfamily,
  language=xml,
  extendedchars=true,
  mathescape=true,
  escapechar=£,
  tagstyle=\bfseries,
  usekeywordsintag=true,
  morekeywords={alias,name,id},
  keywordstyle=\color{lstred}
}

\lstdefinestyle{xmlstyle-color}
{
  tabsize=3,
  basicstyle=\small\upshape\ttfamily,
  language=xml,
  extendedchars=true,
  mathescape=true,
  escapechar=£,
  tagstyle=\color{keywordpurple},
  usekeywordsintag=true,
  morekeywords={alias,name,id},
  keywordstyle=\color{lstred}
}

\definecolor{white}{rgb}{1,1,1}
\definecolor{black}{rgb}{0,0,0}
\definecolor{grey}{rgb}{0.7,0.7,0.7}
\definecolor{dgrey}{rgb}{0.5,0.5,0.5}
\definecolor{lightgrey}{rgb}{0.88,0.88,0.88}
\definecolor{lgrey}{rgb}{0.9,0.9,0.9}
\definecolor{llgrey}{rgb}{0.93,0.93,0.93}
\definecolor{lllgrey}{rgb}{0.96,0.96,0.96}
\definecolor{tableHeadGray}{rgb}{0.85,0.85,0.85}
\definecolor{oddRowGrey}{rgb}{0.95,0.95,0.95}
\definecolor{evenRowGrey}{rgb}{0.85,0.85,0.85}
\definecolor{yellow}{rgb}{1.0, 1.0, 0.0}
\definecolor{lightyellow}{rgb}{1.0, 1.0, 0.88}
\definecolor{selectiveyellow}{rgb}{1.0, 0.73, 0.0}
\definecolor{shadered}{rgb}{1,0.85,0.85}
\definecolor{red}{rgb}{1,0,0}
\definecolor{shadegreen}{rgb}{0.95,1,0.95}
\definecolor{green}{rgb}{0,1,0}
\definecolor{darkgreen}{rgb}{0,0.5,0}
\definecolor{shadeblue}{rgb}{0.95,0.95,1}
\definecolor{blue}{rgb}{0,0,1}
\definecolor{darkblue}{rgb}{0,0,0.5}
\definecolor{darkpurple}{rgb}{0.5,0,0.5}
\definecolor{darkdarkpurple}{rgb}{0.3,0,0.3}
\newcommand{\BG}[1]{\todo[size=\tiny]{\textbf{Boris says:$\,$} #1}}
\newcommand{\BGI}[1]{\todo[inline]{\textbf{Boris says:$\,$} #1}}

\newcommand{\OK}[1]{\todo[size=\tiny,color=grey]{\textbf{Oliver says:$\,$} #1}}

\newbool{ShowDetailedProofs}
\newcommand{\detailedproof}[2]{\ifbool{ShowDetailedProofs}{
\begin{proof}
#2
\end{proof}
}{\noindent\textit{Proof Sketch:}#1\qed}\smallskip}
\newrobustcmd{\ifnotdetailedproof}[1]{\ifbool{ShowDetailedProofs}{}{#1}}
\newrobustcmd{\ifnottechreport}[1]{\ifbool{ShowDetailedProofs}{}{#1}}
\newrobustcmd{\iftechreport}[1]{\ifbool{ShowDetailedProofs}{#1}{}}

\newcommand{\mypar}[1]{\smallskip\noindent{\bf #1.}}
\newcommand{\bfcaption}[1]{\caption{#1}}

\DeclareMathAlphabet{\mathbbold}{U}{bbold}{m}{n}

\newtheorem{Theorem}{Theorem}
\newtheorem{Definition}{Definition}

\newtheorem{Example}{Example}

\newcommand{\card}[1]{\vert\,{#1}\,\vert}
\newcommand{\defas}{\coloneqq}

\DeclareMathAlphabet{\mathbbold}{U}{bbold}{m}{n}

\newcommand{\thead}[1]{\textbf{#1}}

\newcommand{\schemasymb}{\textsc{Sch}}

\newcommand{\schema}{\schemasymb(D)}
\newcommand{\relschema}{\schemasymb(R)}
\newcommand{\schemaOf}{\mathrm{Sch}}
\newcommand{\arity}[1]{arity({#1})}
\newcommand{\db}{D}
\newcommand{\rela}{R}
\newcommand{\pRela}{\mathcal R}
\newcommand{\query}{Q}

\newcommand{\tup}{t}

\newcommand{\aDom}{\mathbb{D}}

\newcommand{\tuple}[1]{\left({#1}\right)}

\newcommand{\tconcat}{\ensuremath{\circ}}

\newcommand{\textendaggishorter}[1]{\textsf{aggres}(#1)}

\newcommand{\mathif}{\text{\textbf{if}}\,\,}
\newcommand{\mathotherwise}{\text{\textbf{otherwise}}}

\newcommand{\selection}{\sigma}
\newcommand{\projection}{\pi}

\newcommand{\aggregation}[2]{%
  \IfEqCase{#1}{%
    {}{\gamma_{{#2}}}%
    }[\gamma_{{#1},{#2}}]%
  }

\newcommand{\agga}{A}
\newcommand{\aggf}{f}
\newcommand{\gbAttrs}{G}

\newcommand{\les}[1]{<_{#1}}
\newcommand{\lest}[1]{\leq_{#1}}
\newcommand{\ales}{\les{\sortattrs}}
\newcommand{\alest}{\lest{\sortattrs}}
\newcommand{\fles}[1]{<_{#1}^{total}}

\newcommand{\afles}{\fles{\sortattrs}}

\newcommand{\sortattrs}{O}
\newcommand{\dummySortAttr}{\tau_{id}}
\newcommand{\multa}{\#}
\newcommand{\mult}{\rangeTup.\multa}
\newcommand{\ranka}{\tau}
\newcommand{\tranksymb}{\textsf{rank}}
\newcommand{\tpossymb}{\textsf{pos}}
\newcommand{\trank}[3]{\tranksymb(#1,#2,#3)}
\newcommand{\tpos}[4]{\tpossymb(#1,#2,#3,#4)}
\newcommand{\tposshort}[3]{\tpossymb(#1,#2,#3)}

\newcommand{\tcovershort}[2]{cover(#1,#2)}

\newcommand{\tboundsshort}[3]{bounds(#1,#2,#3)}
\newcommand{\rankopsymb}{\ensuremath{\text{\upshape{\textsc{sort}}}}\xspace}
\newcommand{\rank}[2]{\rankopsymb_{#1 \rightarrow #2}}
\newcommand{\winrowop}{\omega}
\newcommand{\winrangeop}{\Omega}
\newcommand{\winrow}[6]{\omega_{#1 \to #2; #3; #4}^{[#5,#6]}}
\newcommand{\winrange}[6]{\Omega_{#1 \to #2; #3; #4}^{[#5,#6]}}
\newcommand{\awinrow}{\winrow{\aggf(\agga)}{X}{\gbAttrs}{\sortattrs}{l}{u}}
\newcommand{\awinrange}{\winrange{\aggf(\agga)}{X}{\gbAttrs}{\sortattrs}{l}{u}}

\newcommand{\wrowresultshort}{\mathcal{ROW}}

\newcommand{\partition}[3]{\mathcal{P}_{#1,#2,#3}}
\newcommand{\apartition}[1]{\partition{\rela}{\gbAttrs}{#1}}
\newcommand{\partitionshort}[2]{\mathcal{P}_{#1,#2}}
\newcommand{\winsymbol}{\mathcal{W}}
\newcommand{\pwsymb}{\mathcal{PW}}
\newcommand{\windowshort}[3]{\winsymbol_{#1,#2,#3}}
\newcommand{\windowshorter}[2]{\winsymbol_{#1,#2}}
\newcommand{\window}[6]{\winsymbol^{groups}_{#1,#2,#3,#4,#5,#6}}
\newcommand{\awindow}[1]{\window{\gbAttrs}{\sortattrs}{l}{u}{\rela}{#1}}

\newcommand{\wsize}[2]{\textsf{size}([#1,#2])}
\newcommand{\awsize}{\wsize{l}{u}}

\newcommand{\flattern}[1]{\textsf{expand}(#1)}

\newcommand{\raPlus}{\ensuremath{\mathcal{RA}^{+}}}

\newcommand{\raAgg}{\ensuremath{\mathcal{RA}^{agg}}}

\newcommand{\abbrBGW}{SGW\xspace}

\newcommand{\abbrUAADB}{AU-DB\xspace}
\newcommand{\abbrUAADBs}{AU-DBs\xspace}
\newcommand{\abbrUAARel}{AU-relation\xspace}
\newcommand{\abbrAUDB}{\abbrUAADB}
\newcommand{\abbrAUDBs}{\abbrUAADBs}

\newcommand{\ptime}{\texttt{PTIME}\xspace}

\newcommand{\conpcomplete}{coNP-complete\xspace}
\newcommand{\nphard}{NP-hard\xspace}

\newcommand{\pdb}{\mathcal{D}}
\newcommand{\prel}{\mathcal{R}}

\newcommand{\ubMarker}[1]{\ensuremath{{#1}{}^{\uparrow}}}
\newcommand{\lbMarker}[1]{\ensuremath{{#1}{}^{\downarrow}}}

\newcommand{\bgMarker}[1]{\ensuremath{{#1}{}^{\bgName}}}

\newcommand{\dataDomain}{\mathbb{D}}
\newcommand{\sexpr}{e}

\newcommand{\seval}[2]{\llbracket{#1}\rrbracket_{#2}}

\newcommand{\ifte}[3]{{\bf if}\,{#1}\,{\bf then}\,{#2}\,{\bf else}\,{#3}}

\newcommand{\semN}{\ensuremath{\mathbb{N}}\xspace}

\newcommand{\semK}{\ensuremath{\mathcal{K}}\xspace}

\newcommand{\onesymbol}{\mathbbold{1}}
\newcommand{\zerosymbol}{\mathbbold{0}}
\newcommand{\multsymb}{\cdot}
\newcommand{\addsymbol}{+}

\newcommand{\bfalse}{\bot}
\newcommand{\btrue}{\top}

\newcommand{\kDom}{K}
\newcommand{\domN}{\semN}

\newcommand{\multK}{\multsymb_{\semK}}

\newcommand{\addOf}[1]{\addsymbol_{#1}}
\newcommand{\multOf}[1]{\multsymb_{#1}}

\newcommand{\oneOf}[1]{\onesymbol_{#1}}
\newcommand{\zeroOf}[1]{\zerosymbol_{#1}}
\newcommand{\zeroN}{0}

\newcommand{\liftsemiringeval}[1]{\left\llbracket #1 \right \rrbracket}

\newcommand{\dpK}[2]{{#1}^{#2}}

\newcommand{\semq}[1]{\ensuremath{\dpK{#1}{3}}}

\newcommand{\semqN}{\semq{\semN}}

\newcommand{\aggmin}{\ensuremath{\mathbf{min}}\xspace}
\newcommand{\aggmax}{\ensuremath{\mathbf{max}}\xspace}
\newcommand{\aggsum}{\ensuremath{\mathbf{sum}}\xspace}
\newcommand{\aggavg}{\ensuremath{\mathbf{avg}}\xspace}
\newcommand{\aggcount}{\ensuremath{\mathbf{count}}\xspace}

\newcommand{\mysmbNAU}[1]{\circledast_{{#1}}}

\newcommand{\smpair}{\otimes}

\newcommand{\bgName}{sg}

\newcommand{\rangeName}{I}
\newcommand{\rangeDom}{{\dataDomain_{\rangeName}}}
\newcommand{\rangeTup}{\textbf{\sffamily \tup}}
\newcommand{\rangeRel}{\mathbf{\rela}}
\newcommand{\rangeOf}[1]{\mathbf{{#1}}}
\newcommand{\rangeDB}{{\mathbf{\db}}}

\newcommand{\tmatch}{\sqsubseteq}
\newcommand{\ntmatch}{\not\sqsubseteq}

\newcommand{\dbbounds}{\sqsubset}

\newcommand{\uaaName}{AU}

\newcommand{\uaaDom}[1]{\kDom_{{\uaaName}}}

\newcommand{\uv}[3]{\ensuremath{[{#1}/{#2}/{#3}]}}
\newcommand{\ut}[3]{({#1},{#2},{#3})}

\newcommand{\certainName}{\textsc{cert}}
\newcommand{\possibleName}{\textsc{poss}}

\newcommand{\pwCertainN}{{\certainName}_{\semN}}

\newcommand{\pwPossibleN}{{\possibleName}_{\semN}}

\newcommand{\TM}{\mathcal{TM}}

\newcommand{\rewrUAA}[1]{\ensuremath{\mathrm{rewr}({#1})}}

\newcommand{\renameto}{\rightarrow}

\newcommand{\shortpart}[2]{\mathcal{P}_{#1}(#2)}

\newcommand{\awindowshort}[3]{\windowshort{#1}{#2}{#3}}
\newcommand{\awindowshorter}[2]{\windowshorter{#1}{#2}}

\newcommand{\restrwinishorter}[2]{\mathcal{RW}_{#1,#2}}

\newcommand{\winbotkname}{\textsf{min-k}\xspace}
\newcommand{\wintopkname}{\textsf{max-k}\xspace}
\newcommand{\winbotk}[2]{\text{\textsf{min-k}}(#1,#2)}
\newcommand{\wintopk}[2]{\text{\textsf{max-k}}(#1,#2)}

\newcommand{\certaincount}[1]{\text{\textsf{possn}}(#1)}

\DeclareMathSymbol{\mlq}{\mathord}{operators}{``}
\DeclareMathSymbol{\mrq}{\mathord}{operators}{`'}

\newlength{\eqheight}
\settototalheight{\eqheight}{$=$}

\booltrue{ShowDetailedProofs}

\setcopyright{acmcopyright}
\copyrightyear{2021}
\acmYear{2021}
\acmDOI{10.1145/1122445.1122456}

\acmConference[SIGMOD '21]{}{}{}
\acmBooktitle{SIGMOD '21: }
\acmPrice{15.00}
\acmISBN{978-1-4503-XXXX-X/18/06}

\title{Efficient Approximation of Certain and Possible Answers for Ranking and Window Queries over Uncertain Data \\ (Extended version)} 

\author{Su Feng}
\affiliation{\institution{Illinois Institute of Technology}}
\email{sfeng14@hawk.iit.edu}

\author{Boris Glavic}
\affiliation{\institution{Illinois Institute of Technology}}
\email{bglavic@iit.edu}

\author{Oliver Kennedy}
\affiliation{\institution{SUNY Buffalo}}
\email{okennedy@buffalo.edu}

\renewcommand{\shortauthors}{Feng, et al.}

\Crefname{Example}{Ex.}{Ex.}
\Crefname{figure}{Fig.}{Fig.}
\Crefname{section}{Sec.}{Sec.}
\Crefname{Definition}{Def.}{Def.}
\Crefname{Theorem}{Thm.}{Thm.}
\Crefname{Lemma}{Lem.}{Lem.}

\begin{document}

\lstset{style=psqlcolor}

\begin{abstract}
\label{sec:abstract}
Uncertainty arises naturally in many application domains due to, e.g., data entry errors 
and ambiguity in data cleaning. 
\revm{Prior work
in incomplete and probabilistic databases
has investigated the semantics and efficient evaluation of ranking and top-k queries over uncertain data.}
\revm{However,} most 
approaches deal with top-k and ranking in isolation 
and 
do represent uncertain input data and query results using separate, incompatible data models.
We present an efficient approach for under- and over-approximating results of ranking, top-k, and window queries over uncertain data.
Our approach 
integrates well with existing techniques for querying uncertain data, is efficient, and is  to the best of our knowledge
the first to support windowed aggregation.
We design algorithms for physical operators for uncertain sorting and windowed aggregation, and implement them in PostgreSQL.
We evaluated our approach on 
synthetic and real world datasets, demonstrating that it outperforms 
all competitors, and often produces more accurate results. 
\end{abstract}


\maketitle

\section{Introduction}\label{sec:introduction}
Many application domains need to deal with uncertainty arising from data entry/extraction errors~\cite{jeffery-06-dssdc,sarawagi2008information}, \revm{data lost because of node failures~\cite{DBLP:conf/sigmod/LangNRN14}}, ambiguous data integration~\cite{OP13,AS10,HR06a}, heuristic data wrangling~\cite{Yang:2015:LOA:2824032.2824055,F08,Beskales:2014:SRC:2581628.2581635}\revm{, and 
  bias in machine learning training data~\cite{DBLP:conf/deem/GrafbergerGS22, SP22}}.
\revm{Incomplete and probabilistic} databases~\cite{DBLP:conf/pods/ConsoleGLT20, suciu2011probabilistic} 
\revm{
  model uncertainty as a set} of so-called possible worlds.
Each world is a deterministic database representing one possible state of the real world.
\revm{The commonly used \textit{possible world semantics}~\cite{suciu2011probabilistic} returns for each world the (deterministic) query answer in this world.
Instead of this set of possible answer relations,} most systems produce either
\textit{certain answers}~\cite{DBLP:journals/jacm/ImielinskiL84} (result tuples that are returned in every world), or \textit{possible answers}~\cite{DBLP:journals/jacm/ImielinskiL84} (result tuples that are returned in at least one world).
Unfortunately, incomplete databases lack the expressiveness of deterministic databases and have high computational complexity.

Notably, uncertain versions of order-based operators like \texttt{SORT / LIMIT} (i.e., Top-K) 
have been studied extensively in the past~\cite{4221738,4812412,4221737,10.14778/1687627.1687685}. 
However,  the resulting semantics often 
lacks \emph{closure}. That is, composing such operators with other operators typically requires a complete rethinking of the entire system~\cite{DBLP:journals/tods/SolimanIC08}, \revm{because the model that the operator expects its \emph{inputs} to be encoded with differs from the model encoding the operator's \emph{outputs}.}

In \cite{FH19,FH21},
we started addressing the \revm{linked challenges of computational complexity, closure, and expressiveness in incomplete database systems,} by proposing \textbf{AU-DBs}, an approach to uncertainty management \revm{that can be competitive with deterministic query processing.
Rather than trying to encode a set of possible worlds losslessly, each \abbrAUDB tuple is defined by one range of possible values for each of its attributes and a range of (bag) multiplicities.
Each tuple of an \abbrAUDB is a hypercube that bounds a region of the attribute space, and together, the tuples bound the set of possible worlds between} an \textit{under-approximation of certain answers} and an \textit{over-approximation of possible answers}.
This model is closed under relational algebra~\cite{FH19} with aggregates~\cite{FH21} ($\raAgg$).
That is, if an \abbrAUDB $\db$ bounds a set of possible worlds, the result of any $\mathcal{RA}^{agg}$ query over $\db$ bounds the set of possible query results.
 We refer to this correctness criteria as \textbf{bound preservation}.
 In this paper, we add support for bounds-preserving order-based operators to the \abbrAUDB  model, along with a set of (nontrivial) operator implementations that make this extension efficient. \revm{The closure of the \abbrAUDB model under $\raAgg$, its efficiency, its property of bounding certain and possible answers, and its capability to compactly represent large sets of possible tuples using attribute-level uncertainty are the main factor for our choice to extend this model in this work.}

When sorting uncertain attribute values, the possible order-by attribute values of two tuples $t_1$ and $t_2$ may overlap, which leads to multiple possible sort orders.
Supporting order-based operators over \abbrAUDBs requires encoding multiple possible sort orders.
Unfortunately, a dataset can only have one physical ordering.
\revm{We address this limitation by introducing a \textbf{position} attribute, decoupling the \emph{physical} order in which the tuples are stored from the set of possible \emph{logical} orderings.}
With a tuple's position in a sort order encoded as a numerical attribute, operations that act on this order (i.e., \lstinline!LIMIT!) can be redefined in terms of standard relational operators, which, crucially already have well-defined semantics \revm{in the \abbrAUDB model}.
\revm{In short, by virtualizing sort order into a position attribute}, the existing \abbrAUDB model is sufficient to express the output of SQL's order-dependent operations \revm{in the presence of uncertainty}.

\revm{We start this paper by (i) formalizing uncertain orders within the \abbrAUDB model and present a semantics of sorting and windowed aggregation operations that can be implemented as query rewrites.
When combined with existing \abbrAUDB rewrites~\cite{FH19,FH21}, any $\mathcal{RA}^{agg}$ query with order-based operations can be executed using a deterministic DBMS.
Unfortunately, these rewrites introduce SQL constructs that necessitate computationally expensive operations, driving a central contribution of this paper: (iii) new algorithms for sort, top-k, and windowed aggregation operators for \abbrAUDBs.}

To understand the intuition behind these operators, consider the logical sort operator, which extends each input row with a new attribute storing the row's position wrt. to ordering the input relation on a list $\sortattrs$ of order-by attributes.
If the order-by attributes' values are uncertain, we have to reason about each tuple $t$'s lowest possible position (the number of tuples that certainly precede it over all possible worlds), and highest possible position (the number of tuples that possibly precede it in at least one possible world).
\revm{We can naively compute a lower (resp., upper) bound by joining every tuple $t$ with every other tuple, counting pairs where $t$ is certainly (resp., possibly) preceded by its pairing.}
\revm{We refer to this approach as} the \emph{rewrite method}, \revm{as it} can be implemented in SQL.
However, the rewrite approach has quadratic runtime. 
Inspired by techniques for aggregation over interval-temporal databases such as \cite{10.1007/978-3-319-64367-0_7}, we propose a  one-pass algorithm to compute the bounds on a tuple's position that also supports top-k queries.


\begin{Example}[Uncertain Sorting and Top-k]\label{ex:uncertain-ranking}
\revm{
  \Cref{fig:an-uncertain-brand-databa} shows a sales DB,  extracted from 3 press releases.
  Uncertainty 
  arises for a variety of reasons, including extraction errors (e.g., $D_3$ includes term 5) or missing information (e.g., only preliminary data is available for the 4th term in $D_1$).
  The task of finding the two terms with the most sales is semantically ambiguous for uncertain data.
  Several attempts to define semantics include
  (i) U-top~\cite{4221738} (\Cref{fig:u-rank-query-result}), which returns the most likely ranked order;
  (ii) U-rank~\cite{4221738} (\Cref{fig:u-rank-query-result}), which returns the most likely tuple at each position (term 4 is more likely than any other value for both the 1st and 2nd position); or
  (iii) Probabilistic threshold queries (PT-k)~\cite{10.1145/1376616.1376685,4498380}, which return tuples that appear in the top-k with a probability exceeding a threshold (PT), generalizing both possible (PT $>$ 0; \Cref{fig:pt0-query-result}) and certain (PT $\geq$ 1; \Cref{fig:pt1-query-result-tuple}) answers.
}
\end{Example}

With the exception of U-Top, none of these semantics
\revm{
return both information about certain and possible  results, making it difficult for users to gauge the (i) trustworthiness or (ii) completeness of an answer.
Risk assessment on the resulting data is difficult, preventing its use for critical applications, e.g., in the medical, engineering, or financial domains.
}
\revm{
  Furthermore, the outputs of uncertain ranking operators like U-Top are not valid as inputs to further uncertainty-aware queries, because they lose information about uncertainty in the source data.
}
\revm{
These factors motivate our choice of the \abbrAUDB data model.
First, the data model naturally encodes query result reliability. By providing each attribute value (and tuple multiplicity) as a range, users can quickly assess the precision of each answer.
Second, the data model is complete: the full set of possible answers is represented.
Finally, the model admits a closed, efficiently computable, and bounds-preserving semantics for  $\mathcal{RA}^{agg}$.
}

\iftechreport{
\begin{figure*}[t]
  \centering
\begin{minipage}{0.49\linewidth}
  \begin{subfigure}{1\linewidth}
  \begin{minipage}{\linewidth}
  \centering
  \begin{minipage}{0.3\linewidth}
  \centering
  	\scalebox{0.7}{
  	\begin{tabular}{ c|c|c}
      \footnotesize{$D_1$} & \textbf{Term}  & \textbf{Sales}  \\
      				\cline{1-3}
		& $1$ & $2$ \\
		& $2$ & $3$ \\
		\rowcolor{grey}
		& $3$ & $7$ \\
		\rowcolor{grey}
		& $4$ & $4$ \\
	\end{tabular}
	}
  \end{minipage}
  \begin{minipage}{0.3\linewidth}
  \centering
  	\scalebox{0.7}{
  	\begin{tabular}{ c|c|c}
      \footnotesize{$D_2$} & \textbf{Term}  & \textbf{Sales}  \\
      				\cline{1-3}
		& $1$ & $3$ \\
		& $2$ & $2$ \\
		\rowcolor{grey}
		& $3$ & $4$ \\
		\rowcolor{grey}
		& $4$ & $6$ \\
	\end{tabular}
	}
  \end{minipage}
	\begin{minipage}{0.3\linewidth}
	\centering
	\scalebox{0.7}{
  	\begin{tabular}{ c|c|c}
      \footnotesize{$D_3$} & \textbf{Term}  & \textbf{Sales}  \\
      				\cline{1-3}
		& $1$ & $2$ \\
		& $2$ & $2$ \\
		\rowcolor{grey}
		& $5$ & $4$ \\
		\rowcolor{grey}
		& $4$ & $7$ \\
	\end{tabular}
	}
  \end{minipage}
  \end{minipage}
  \begin{minipage}{\linewidth}
  	\centering
  	\begin{minipage}{0.3\linewidth}
  	\centering
  	\scalebox{0.7}{
  	\begin{tabular}{ c|c|c}
      \textbf{Term}  & Sales & \textbf{Sum}  \\
      				\cline{1-3}

		$1$  &	$2$ &  $5$ \\
		$2$  &	$3$ &  $10$ \\
		$3$  &	$7$ &  $11$ \\
		$4$  &	$4$ &  $4$ \\
	\end{tabular}
	}
  \end{minipage}
  \begin{minipage}{0.3\linewidth}
  \centering
  \scalebox{0.7}{
  	\begin{tabular}{ c|c|c}
      \textbf{Term} & Sales & \textbf{Sum}  \\
      				\cline{1-3}

		$1$  &	$3$ &  $5$ \\
		$2$  &	$2$ &  $6$ \\
		$3$  &	$4$ &  $10$ \\
		$4$  &	$6$ &  $6$ \\
	\end{tabular}
 	}
	\end{minipage}
	\begin{minipage}{0.3\linewidth}
	\centering
	\scalebox{0.7}{
  	\begin{tabular}{ c|c|c}
      \textbf{Term} & Sales & \textbf{Sum}  \\
      				\cline{1-3}

		$1$ &	$2$ & $4$ \\
		$2$ &	$2$ & $9$ \\
		$5$ &	$4$ & $4$ \\
		$4$ &	$7$ & $11$ \\
	\end{tabular}
	}
 \end{minipage}
 \end{minipage}
 \\[-3mm]
 \caption{An uncertain sales database with three possible worlds (with probability .4, .3  and .3 respectively) with top-2 highest selling  term in each world high-lighted and the result of the rolling sum of sales for the current and next term.}\label{fig:an-uncertain-brand-databa}
 \end{subfigure} %
 \end{minipage}
 \begin{minipage}{0.49\linewidth}
 \centering
 \begin{minipage}{0.49\linewidth}
 \centering
 \begin{subfigure}{\linewidth}
 \centering
 \begin{tabular}{   c}
 \textbf{Term}      \\
 \cline{1-1}
 $4$                \\
 $3$                \\
 \end{tabular}
 \\[-3mm]
 \caption{U-Top Top-k result}\label{fig:u-top-query-result}
 \end{subfigure}
 \end{minipage}
 \begin{minipage}{0.49\linewidth}
 \centering
 \begin{subfigure}{\linewidth}
 \centering
 \begin{tabular}{   c}
 \textbf{Term}      \\
 \cline{1-1}
 $4$                \\
 $4$                \\
 \end{tabular}
 \\[-3mm]
 \caption{U-Rank query result}\label{fig:u-rank-query-result}
 \end{subfigure}
 \end{minipage}
 \begin{minipage}{0.49\linewidth}
 \centering
 \begin{subfigure}{\linewidth}
 \captionsetup{width=.7\linewidth}
 \centering
 \begin{tabular}{   c}
 \textbf{Term}      \\
 \cline{1-1}
 $3$                \\
 $4$                \\
 $5$                \\
 \end{tabular}
 \\[-3mm]
 \caption{PT(0) query result (possible answers)}\label{fig:pt0-query-result}
 \end{subfigure}
 \end{minipage}
 \begin{minipage}{0.49\linewidth}
 \centering
 \begin{subfigure}{\linewidth}
 \captionsetup{width=.7\linewidth}
 \centering
 \begin{tabular}{   c}
 \textbf{Term}      \\
 \cline{1-1}
 $4$                \\
 $ $                \\
 $ $                \\
 \end{tabular}
 \\[-3mm]
 \caption{PT(1) Top-k result (certain answers)}\label{fig:pt1-query-result-tuple}
 \end{subfigure}
 \end{minipage}
 \end{minipage}
 \begin{minipage}{0.49\linewidth}
 \begin{subfigure}{1\linewidth}
 \centering
 \begin{minipage}{0.4\linewidth}
   \centering
   {\small
 \begin{tabular}{c|c|c}
 \textbf{Term}      & \textbf{Sales}  & $\semN^3$          \\
 \cline{1-3}
 $1$                & $\uv{2}{2}{3}$ & \ut{1}{1}{1} \\
 $2$                & $\uv{2}{3}{3}$ & \ut{1}{1}{1}        \\
 \rowcolor{grey}
 $\uv{3}{3}{5}$     & $\uv{4}{7}{7}$ & \ut{1}{1}{1}        \\
 \rowcolor{grey}
 $4$                & $\uv{4}{4}{7}$ & \ut{1}{1}{1}        \\
 \end{tabular}
 }
 \end{minipage}
 \begin{minipage}{0.59\linewidth}
   \centering
   {\small
 \begin{tabular}{c|c|cc}
 \textbf{Term}      & \textbf{Sales} & \textbf{Position} & $\semN^3$          \\
 \cline{1-3}
 $1$                & $\uv{2}{2}{3}$ & \uv{2}{3}{3}     & \ut{0}{0}{0}        \\
 $2$                & $\uv{2}{3}{3}$ & \uv{2}{2}{3}     & \ut{0}{0}{0}        \\
 \rowcolor{grey}
 $\uv{3}{3}{5}$     & $\uv{4}{7}{7}$ & \uv{0}{0}{1}     & \ut{1}{1}{1}        \\
 \rowcolor{grey}
 $4$                & $\uv{4}{4}{7}$ & \uv{0}{1}{1}     & \ut{1}{1}{1}        \\
 \end{tabular}
 }
 \end{minipage}
 \\[-2mm]
 \caption{\abbrAUDB bounding the worlds and top-2 result produced by our approach}\label{fig:au-db-result-topk}
 \end{subfigure}
 \end{minipage}
 \begin{minipage}{0.49\linewidth}
 \begin{subfigure}{1\linewidth}
   \centering
   \captionsetup{width=.8\linewidth}
   {\small
 \begin{tabular}{   c|c|c|c}
 \textbf{Term}      & \textbf{Sales} & \textbf{Sum}     & $\semN^3$ \\
 \cline{1-4}
 $1$                & $\uv{2}{2}{3}$ & $\uv{4}{5}{6}$   & \ut{1}{1}{1} \\
 $2$                & $\uv{2}{3}{3}$ & $\uv{6}{10}{10}$ & \ut{1}{1}{1} \\
 $\uv{3}{3}{5}$     & $\uv{4}{7}{7}$ & $\uv{4}{11}{14}$ & \ut{1}{1}{1} \\
 $4$                & $\uv{4}{4}{7}$ & $\uv{4}{4}{14}$  & \ut{1}{1}{1} \\
 \end{tabular}
 }
 \\[-2mm]
 \caption{\abbrAUDB windowed aggregation result produced by our approach}\label{fig:au-db-result-window}
 \end{subfigure}
 \end{minipage}
 \\[-3mm]
 \setlength{\belowcaptionskip}{-15pt}
 \caption{Ranking, Top-k, and Window Queries over an Incomplete (Probabilistic) Database. We get different results for the various semantics proposed in related work. Our approach stands out in that it bounds both certain and possible answers and is closed not just under these specific query types, but also \raAgg.}\label{fig:ranking-top-k-and-window-}
\end{figure*}
}
\ifnottechreport{
	\begin{figure}[t]
  \centering
\begin{minipage}{0.99\linewidth}
  \begin{subfigure}{1\linewidth}
  \begin{minipage}{\linewidth}
  \centering
  \begin{minipage}{0.3\linewidth}
  \centering
  	\scalebox{0.7}{
  	\begin{tabular}{ c|c|c}
      \footnotesize{$D_1$} & \textbf{Term}  & \textbf{Sales}  \\
      				\cline{1-3}
		& $1$ & $2$ \\
		& $2$ & $3$ \\
		\rowcolor{grey}
		& $3$ & $7$ \\
		\rowcolor{grey}
		& $4$ & $4$ \\
	\end{tabular}
	}
  \end{minipage}
  \begin{minipage}{0.3\linewidth}
  \centering
  	\scalebox{0.7}{
  	\begin{tabular}{ c|c|c}
      \footnotesize{$D_2$} & \textbf{Term}  & \textbf{Sales}  \\
      				\cline{1-3}
		& $1$ & $3$ \\
		& $2$ & $2$ \\
		\rowcolor{grey}
		& $3$ & $4$ \\
		\rowcolor{grey}
		& $4$ & $6$ \\
	\end{tabular}
	}
  \end{minipage}
	\begin{minipage}{0.3\linewidth}
	\centering
	\scalebox{0.7}{
  	\begin{tabular}{ c|c|c}
      \footnotesize{$D_3$} & \textbf{Term}  & \textbf{Sales}  \\
      				\cline{1-3}
		& $1$ & $2$ \\
		& $2$ & $2$ \\
		\rowcolor{grey}
		& $5$ & $4$ \\
		\rowcolor{grey}
		& $4$ & $7$ \\
	\end{tabular}
	}
  \end{minipage}
  \end{minipage}
  \begin{minipage}{\linewidth}
  	\centering
  	\begin{minipage}{0.3\linewidth}
  	\centering
  	\scalebox{0.7}{
  	\begin{tabular}{ c|c|c}
      \textbf{Term}  & Sales & \textbf{Sum}  \\
      				\cline{1-3}

		$1$  &	$2$ &  $5$ \\
		$2$  &	$3$ &  $10$ \\
		$3$  &	$7$ &  $11$ \\
		$4$  &	$4$ &  $4$ \\
	\end{tabular}
	}
  \end{minipage}
  \begin{minipage}{0.3\linewidth}
  \centering
  \scalebox{0.7}{
  	\begin{tabular}{ c|c|c}
      \textbf{Term} & Sales & \textbf{Sum}  \\
      				\cline{1-3}

		$1$  &	$3$ &  $5$ \\
		$2$  &	$2$ &  $6$ \\
		$3$  &	$4$ &  $10$ \\
		$4$  &	$6$ &  $6$ \\
	\end{tabular}
 	}
	\end{minipage}
	\begin{minipage}{0.3\linewidth}
	\centering
	\scalebox{0.7}{
  	\begin{tabular}{ c|c|c}
      \textbf{Term} & Sales & \textbf{Sum}  \\
      				\cline{1-3}

		$1$ &	$2$ & $4$ \\
		$2$ &	$2$ & $9$ \\
		$5$ &	$4$ & $4$ \\
		$4$ &	$7$ & $11$ \\
	\end{tabular}
	}
 \end{minipage}
 \end{minipage}
 \\[-3mm]
 \caption{An uncertain sales database with three possible worlds (with probability .4, .3  and .3 respectively) with top-2 highest selling terms high-lighted and the result of the rolling sum of sales for the current and next term.}\label{fig:an-uncertain-brand-databa}
 \end{subfigure} %
 \end{minipage}
 \begin{minipage}{0.99\linewidth}
 \centering
 \begin{minipage}{0.24\linewidth}
 \centering
 \begin{subfigure}{\linewidth}
 \centering
 \begin{tabular}{   c}
 \textbf{Term}      \\
 \cline{1-1}
 $4$                \\
 $3$                \\
 \end{tabular}
 \\[-3mm]
 \caption{U-Top}\label{fig:u-top-query-result}
 \end{subfigure}
 \end{minipage}
 \begin{minipage}{0.24\linewidth}
 \centering
 \begin{subfigure}{\linewidth}
 \centering
 \begin{tabular}{   c}
 \textbf{Term}      \\
 \cline{1-1}
 $4$                \\
 $4$                \\
 \end{tabular}
 \\[-3mm]
 \caption{U-Rank}\label{fig:u-rank-query-result}
 \end{subfigure}
 \end{minipage}
 \begin{minipage}{0.24\linewidth}
 \centering
 \begin{subfigure}{\linewidth}
 \captionsetup{width=.7\linewidth}
 \centering
 \scalebox{0.8}{
 \begin{tabular}{   c}
 \textbf{Term}      \\
 \cline{1-1}
 $3$                \\
 $4$                \\
 $5$                \\
 \end{tabular}
 }
 \\[-3mm]
 \caption{PT(0)}\label{fig:pt0-query-result}
 \end{subfigure}
 \end{minipage}
 \begin{minipage}{0.24\linewidth}
 \centering
 \begin{subfigure}{\linewidth}
 \captionsetup{width=.7\linewidth}
 \centering
 \begin{tabular}{   c}
 \textbf{Term}      \\
 \cline{1-1}
 $4$                \\
 $ $                \\
 \end{tabular}
 \\[-3mm]
 \caption{PT(1)}\label{fig:pt1-query-result-tuple}
 \end{subfigure}
 \end{minipage}
 \end{minipage}
 \begin{minipage}{0.99\linewidth}
 \begin{subfigure}{1\linewidth}
 \centering
 \begin{minipage}{0.4\linewidth}
   \centering
   {\small
  \scalebox{0.9}{
 \begin{tabular}{c|c|c}
 \textbf{Term}      & \textbf{Sales}  & $\semN^3$          \\
 \cline{1-3}
 $1$                & $\uv{2}{2}{3}$ & \ut{1}{1}{1} \\
 $2$                & $\uv{2}{3}{3}$ & \ut{1}{1}{1}        \\
 \rowcolor{grey}
 $\uv{3}{3}{5}$     & $\uv{4}{7}{7}$ & \ut{1}{1}{1}        \\
 \rowcolor{grey}
 $4$                & $\uv{4}{4}{7}$ & \ut{1}{1}{1}        \\
 \end{tabular}
 }}
 \end{minipage}
 \begin{minipage}{0.59\linewidth}
   \centering
   {\small
   \scalebox{0.9}{
 \begin{tabular}{c|c|cc}
 \textbf{Term}      & \textbf{Sales} & \texttt{Position} & $\semN^3$          \\
 \cline{1-4}
 $1$                & $\uv{2}{2}{3}$ & \uv{2}{3}{3}     & \ut{0}{0}{0}        \\
 $2$                & $\uv{2}{3}{3}$ & \uv{2}{2}{3}     & \ut{0}{0}{0}        \\
 \rowcolor{grey}
 $\uv{3}{3}{5}$     & $\uv{4}{7}{7}$ & \uv{0}{0}{1}     & \ut{1}{1}{1}        \\
 \rowcolor{grey}
 $4$                & $\uv{4}{4}{7}$ & \uv{0}{1}{1}     & \ut{1}{1}{1}        \\
 \end{tabular}
 }
 }
 \end{minipage}
 \\[-2mm]
 \caption{\abbrAUDB bounding the worlds and top-2 result produced by our approach}\label{fig:au-db-result-topk}
 \end{subfigure}
 \end{minipage}
 \begin{minipage}{0.99\linewidth}
 \begin{subfigure}{1\linewidth}
   \centering
   \captionsetup{width=.9\linewidth}
   {\small
 \begin{tabular}{   c|c|c|c}
 \textbf{Term}      & \textbf{Sales} & \textbf{Sum}     & $\semN^3$ \\
 \cline{1-4}
 $1$                & $\uv{2}{2}{3}$ & $\uv{4}{5}{6}$   & \ut{1}{1}{1} \\
 $2$                & $\uv{2}{3}{3}$ & $\uv{6}{10}{10}$ & \ut{1}{1}{1} \\
 $\uv{3}{3}{5}$     & $\uv{4}{7}{7}$ & $\uv{4}{11}{14}$ & \ut{1}{1}{1} \\
 $4$                & $\uv{4}{4}{7}$ & $\uv{4}{4}{14}$  & \ut{1}{1}{1} \\
 \end{tabular}
 }
 \\[-2mm]
 \caption{\abbrAUDB windowed aggregation result produced by our approach}\label{fig:au-db-result-window}
 \end{subfigure}
 \end{minipage}
 \\[-3mm]
 \setlength{\belowcaptionskip}{-15pt}
 \caption{Ranking, top-k, and windowed aggregation queries over an incomplete (probabilistic) database , including \abbrAUDBs.
 }
\end{figure}
}

\begin{Example}[\abbrAUDB top-2 query]\label{ex:au-db-top-2-query}
  \Cref{fig:au-db-result-topk} (left) shows an \abbrAUDB, \revm{which uses triples, consisting of a lower bound, a selected-guess value (defined shortly), and an upper bound to bound the value range of an attribute (\textbf{Term}, \textbf{Sales}) and the multiplicity of a tuple ($\mathbb N^3$)}.
  The \abbrAUDB \emph{bounds} all of the possible worlds of our running example. Intuitively, each world's tuples fit into the ranges defined by the \abbrAUDB.
  The selected-guess values encode one distinguished world (here, $D_1$)
\revm{--- supplementing the bounds with an educated guess about which possible world correctly reflects the real world~\footnote{\revm{
  The process of obtaining a selected-guess world is domain-specific, but \cite{FH19,FH21} suggest the most likely world, if it can be feasibly obtained.
}}, providing backwards compatibility with existing systems, and a convenient reference point for users~\cite{BS20,kumari:2016:qdb:communicating}.}
  \Cref{fig:au-db-result-topk} (right) shows the result of computing the top-2 answers sorted on \emph{term}.
  The rows marked in grey encode all tuples that could exist in the top-2 result in \revm{some possible world. For example, the tuples $\tuple{3, 4}$ ($D_1$), $\tuple{3, 7}$ ($D_2$), and $\tuple{5, 7}$ ($D_3$) are all encoded by the \abbrAUDB tuple $\tuple{\uv{3}{3}{5}, \uv{4}{7}{7}} \rightarrow \ut{1}{1}{1}$}.
  Results with a row multiplicity range of \ut{0}{0}{0} are certainly not in the result.
  The \abbrAUDB compactly represents an \emph{under-approximation} of \emph{certain answers} and an \emph{over-approximation} of all the \emph{possible answers}, e.g., for our example, the \abbrAUDB admit additional worlds
  with 5 sales in term 4.
\end{Example}

Implementing windowed aggregation requires determining the (uncertain) membership of each window, \revm{which may be affected both by uncertainty in sort position, and in group-by attributes. Furthermore, we have to reason about which of the tuples possibly belonging to a window minimize / maximize the aggregation function result.
}
It is possible implemented this reasoning in SQL, albeit at the cost of range self-joins on the relation (%
this \emph{rewrite method} is discussed in detail in~\cite{techreport} and
evaluated in \Cref{sec:experiment}).
\revm{We propose a one-pass algorithm for windowed aggregation over \abbrAUDBs, which we will refer to as the \emph{native method}.}

The intuition behind our algorithm is to share state between multiple windows.
For example, consider the window \lstinline!ROWS BETWEEN 3! \lstinline!PRECEDING AND CURRENT ROW!.
In the deterministic case, with each new window one row enters the window and one row leaves.
Sum-based aggregates (\lstinline{sum}, \lstinline{count}, \lstinline{average}) can  leverage commutativity and associativity of addition, i.e.,  updating the window requires only constant time.
\revm{Similar techniques~\cite{DBLP:conf/sigmod/AlbrightDGGLKSSW08} can maintain} of \lstinline{min}/\lstinline{max} aggregates in time logarithmic in the window size.

Non-determinism in the row position makes such resource sharing problematic.
First, tuples with non-deterministic positions \revm{do not necessarily leave the window in FIFO order};
We need iteration over tuples sorted on both the upper- and lower-bounds of their position.
Second, \revm{the number of tuples that} could \emph{possibly} belong to the window may be significantly larger than the window size.
\revm{Considering all possible rows for a $k$-row window (using the naive \abbrAUDB aggregation operator~\cite{FH21}) results in a looser bound than if only} subsets of size $k$ \revm{are considered}.
For that, we need access to rows possibly in a window sorted on the bounds of the aggregation attribute values (e.g., to find the $k$-subset with the minimal/maximal sum) in both decreasing order of their upper bound and increasing order of their lower bound. Furthermore, we have to  \revm{separate maintain} tuples that certainly belong to a window (\revm{which must contribute} to both bounds).
To efficiently maintain sets of tuples such that they can be accessed in several sort orders efficiently, we develop a new data structure which we refer to as a \emph{connected heap}.
A connected heap is a set of heaps 
where an element popped from one heap can be efficiently \revm{($O(\log n)$)} removed from the other heaps even if their sort orders differ from the heap we popped the element from.
This data structure allows us to efficiently maintain sufficient state for computing \abbrAUDB results for windowed aggregation. 
\revm{In preliminary experiments, we demonstrated that, connected heaps significantly outperform a solution based on classical heaps.}
\begin{Example}[Windowed Aggregation]\label{ex:windowed-aggregation}
  Consider the following windowed aggregation query:
\begin{lstlisting}
SELECT *, sum(Sales) OVER (ORDER BY term ASC
BETWEEN CURRENT ROW AND 1 FOLLOWING) as sum FROM R;
\end{lstlisting}
\Cref{fig:au-db-result-window} shows the result of this query over our running example \abbrAUDB. The column \texttt{Sum} bounds all possible windowed aggregation results for each \abbrAUDB tuple and the entire \abbrAUDB relation bounds the windowed aggregation result for all possible worlds. Notice that \abbrAUDBs ignore correlations which causes an over-approximation of ranges in the result. For example, term 1 has a maximum aggregation result value of 6 according to the \abbrAUDB representation but the maximum possible aggregation value across all possible world is 5.
\end{Example}

\section{Related Work}
\label{sec:related-work}
\iftechreport{
We build on prior research in 
incomplete and probabilistic databases, uncertain aggregation, uncertain top-k, uncertain sorting, and temporal databases.}
\ifnottechreport{
We build on prior research in 
incomplete and probabilistic databases, uncertain aggregation, uncertain top-k and uncertain sorting.
}
\mypar{Probabilistic/Incomplete databases}
Certain answer semantics~\cite{DBLP:journals/jacm/ImielinskiL84,AK91,L16a,GL16,GL17,L79a} only returns answers that are guaranteed to be correct.
Computing certain answers is \conpcomplete in data-complexity~\cite{AK91,DBLP:journals/jacm/ImielinskiL84}. However, under-approximations~\cite{R86,GL17,L16a,GL16,CG19,FH19} can be computed in \ptime.
\abbrUAADBs~\cite{FH21} build on the selected-guess and lower bounds-based approach of~\cite{FH19}, adding an upper bound on possible answers and attribute-level uncertainty with ranges 
to support aggregation. 
MCDB~\cite{jampani2008mcdb} and Pip~\cite{5447879} sample from the set of possible worlds to generate expectations of possible outcomes, but can not generally obtain bounds on their estimates.
Queries over symbolic models for incomplete data like C-tables~\cite{DBLP:journals/jacm/ImielinskiL84} and 
m-tables~\cite{sundarmurthy_et_al:LIPIcs:2017:7061} often have \ptime data complexity, but obtaining 
certain answers from query results
is 
intractable. 

\mypar{Aggregation in Incomplete/Probabilistic Databases}
General solutions for non-windowed aggregation over uncertain data remain an open problem~\cite{DBLP:conf/pods/ConsoleGLT20}.
Due to the complexity of uncertain aggregation, most approaches focus on identifying tractable cases and producing lossy representations~\cite{5447879,DBLP:journals/tkde/MurthyIW11,DBLP:conf/icdt/AbiteboulCKNS10,DBLP:journals/tods/SolimanIC08,CC96,DBLP:conf/soda/JayramKV07,DBLP:journals/vldb/BurdickDJRV07,DBLP:conf/sigmod/YangWCK11, liang-20-frmdcanp}.
These result encodings are not closed (i.e., not useful for subsequent queries), and are also expensive to compute (often \nphard).
Symbolic  models~\cite{amsterdamer2011provenance,DBLP:journals/pvldb/FinkHO12,DBLP:journals/jiis/LechtenborgerSV02} that are closed under aggregation permit 
\ptime data complexity, but extracting certain / possible answers is still intractable.  
We proposed \abbrAUDBs~\cite{FH21} which are closed under $\raAgg$ and achieve efficiency through approximation. 

\mypar{Uncertain Top-k}
A key challenge in uncertain top-k ranking is defining a meaningful semantics. 
The set of tuples certainly (resp., possibly) in the top-k may have fewer (more) than k tuples.
U-Top~\cite{4221738} picks the top-k set with the highest probability.
U-Rank~\cite{4221738} assigns to each rank the tuple which is most-likely to have this rank.
Global-Topk~\cite{4498380} first ranks tuples by their probability of being in the top-k and returns the k most likely tuples.
Probabilistic threshold top-k (PT-k)~\cite{10.1145/1376616.1376685} returns all tuples that have a probability of being in the top-k that exceeds a pre-defined threshold.
Expected rank~\cite{4812412} calculates the expected rank for each tuple across all possible worlds and picks the k tuples with the highest expected rank.
R\'{e} et al.~\cite{4221737} proposed a multi-simulation algorithm that stops when a guaranteed top-k probability can be guaranteed. Soliman et al.~\cite{10.1145/1386118.1386119} proposed a framework that integrates tuple retrieval, grouping, aggregation, uncertainty management, and ranking in a pipelined fashion.
\iftechreport{Li et al.~\cite{10.14778/1687627.1687685} proposed a unified ranking approach for top-k based on generating functions which use and/xor trees to reason about complex correlations.}
Each of these generalizations necessarily breaks some intuitions about top-k, producing more (or fewer) than k tuples, or producing results that are not the top-k in any world.

\mypar{Uncertain Order}
Amarilli et. al. extends the relational model with a partial order to encode uncertainty in the sort order of a relation~\cite{AB17, AB19}.
For more general use cases where posets can not represent all possible worlds, Amarilli et. al. also develop a symbolic model of provenance~\cite{AB14} whose expressions encode possible orders. 
Both approaches are limited to set semantics.
\iftechreport{
\mypar{Temporal Aggregation}
Temporal databases must reason about tuples associated with partially overlapping intervals.
For example, a window aggregate 
may be recast as an interval self-join, where one table defines the set of windows and each of its tuples is joined with the tuples in the window.
We take inspiration from temporal databases for our own operator implementations.
\BG{If we need to save space, then we can limit the discussion of temporal aggregation.}
The first temporal aggregation algorithm was given in~\cite{tuma1993implementing}. 
Moon et al.~\cite{839401} proposed a balanced tree algorithm for \aggcount, \aggsum and \aggavg aggregates, and a divide-and-conquer algorithm for \aggmin and \aggmax.
Kline and Snodgrass proposed the \textit{aggregation tree}~\cite{380389}, an in-memory data structure that supports incremental computation of temporal aggregates.
Yang et al.~\cite{914813} proposed a materialized version called the \textit{SB-tree} that can be used as an index for incremental temporal aggregation computations.
 The \textit{MVSB-tree}~\cite{10.1145/375551.375600} is and extension of the SB-tree that supports predicates in the aggregation query.
Piatov and Helmer~\cite{10.1007/978-3-319-64367-0_7} proposed a sweep-line based approach the reduces the space needed to compute \aggmin and \aggmax aggregates over temporal data. 
}

\section{Notation and Background}\label{sec:background}

A database schema $\schema = \{\schemaOf(\rela_1), \ldots, \schemaOf(\rela_n)\}$ is a set of relation schemas  $\schemaOf(\rela_i) = \tuple{A_1,\; \ldots,\; A_n}$. 
Use $\arity{\relschema}$ to denote the number of attributes in $\relschema$. An instance $\db$ for schema $\schema$ is a set of relation instances with one relation per schema in $\schema$: $\db=\{\rela_1, \dots, \rela_n\}$.
Assuming a universal value domain $\aDom$, a tuple with schema $\relschema$ is an element from $\aDom^{\arity{\relschema}}$.

A 
$\semK$-relation~\cite{Green:2007:PS:1265530.1265535} annotates each tuple with an element of a (commutative) semiring.
In this paper, we focus on 
$\semN$-relations. An $\semN$-relation of arity $n$ is a function that maps each tuple ($\aDom^n$) in the relation to an annotation in $\semN$ representing the tuple's multiplicity. Tuples not in the relation are mapped to multiplicity $0$.
$\semN$-relations have finite support (tuple not mapped to $0$).
Since $\semK$-relations are functions from tuples to annotations, it is customary to denote the annotation of a tuple $\tup$ in relation $\rela$ as $\rela(\tup)$.
A $\semK$-database is a set of $\semK$-relations.
Green et al.~\cite{Green:2007:PS:1265530.1265535} did use the semiring operations to express positive relational algebra ($\raPlus$) operations over $\semK$-relations as shown in \Cref{fig:semKLift}.
Notably for us, for the natural numbers semiring $\semN = \tuple{\domN, +, \times, 0, 1}$, 
 this semantics are equivalent to those of positive bag-relational algebra.

\begin{figure}
\centering
\noindent
\begin{minipage}{0.45\columnwidth}
  $$\liftsemiringeval{\pi_A(\rela)}(\tup) = \sum_{\tup':\; \tup = \pi_A \tup'} \rela(\tup')$$
  $$\liftsemiringeval{\rela \cup S}(\tup) = \rela(\tup) + S(\tup)$$
\end{minipage}
\begin{minipage}{0.45\columnwidth}
  $$\liftsemiringeval{\sigma_\theta(\rela)}(\tup) = \begin{cases}
    \rela(\tup)&\textbf{ if }\theta(\tup)\\
    0 &\textbf{ otherwise}
  \end{cases}$$
  $$\liftsemiringeval{\rela \times S}(\tup) = \rela(\tup) \times S(\tup)$$
\end{minipage}
\\[-2mm]
\setlength{\belowcaptionskip}{-15pt}
\caption{Evaluation semantics $\liftsemiringeval{\cdot}$ that lift the operations of a semiring $\semK$ to $\raPlus$ operations over $\semK$-relations.}
\label{fig:semKLift}
\end{figure}


\subsection{Incomplete N-Relations}
\label{sec:incompl-prob-k}

An incomplete $\semN$-database  $\pdb = \{\db_1, \ldots, \db_n\}$ (resp., incomplete $\semN$-relation $\pRela = \{\rela_1, \ldots, \rela_n\}$) is a set of $\semN$-databases $\db_i$ (resp., $\semN$-relations $\rela_i$) called possible worlds.
Queries over incomplete $\semN$-databases use possible world semantics:
The result of a  query $\query$ over an incomplete $\semN$-database $\pdb$ is the set of relations $\prel$ (possible worlds) derived by evaluating $\query$ over every world in $\pdb$ using the semantics of \Cref{fig:semKLift}.
In addition to enumerating all possible query results, past work has introduced the concept of \textit{certain} and \textit{possible answers} for set semantics, which are respectively the set of tuples present in all worlds or in at least one world.
Certain and possible answers have been generalized~\cite{GL19a,FH19} to bag semantics as the extrema of the tuple's annotations across all possible worlds.
Formally, the certain and possible annotations of a tuple $\tup$ in $\pRela$ are:
\\[-10mm]
\begin{center}
  \begin{align*}
    \pwCertainN(\pRela,\tup)  &\defas \min(\{\rela(\tup) \mid \rela \in \pRela \})\\
    \pwPossibleN(\pRela,\tup) &\defas \max(\{\rela(\tup) \mid \rela \in \pRela \})
  \end{align*}
\end{center}

\subsection{AU-Databases (\abbrAUDBs)} \label{sec:AU-model}

Using $\semK$-relations 
,
we 
 introduced \textit{\abbrAUDBs}~\cite{FH19} (\textit{attribute-annotated uncertain databases}), a special type of $\semK$-relation that summarizes an incomplete $\semK$-relation by bounding its set of possible worlds.
An \abbrAUDB differs from the classical relational model in two key ways:
First, tuples are not defined as individual points $\aDom^n$, but rather as a bounding hypercube specified as upper and lower bounds (and selected-guess) for each attribute value.
Every such hypercube can represent zero or more tuples contained inside it.
Second, the annotation of each hypercube tuple is also a range of possible annotations (e.g., multiplicities for range-annotated $\semN$-relations).
Intuitively, an \abbrAUDB \textit{bounds} a possible world if the hypercubes of its tuples contain all of the possible world's tuples, and the total multiplicity of tuples in the possible world fall into the range annotating the hypercube.
An \abbrAUDB \textit{bounds} an incomplete $\semK$-database $\pdb$ if it bounds all of $\pdb$'s possible worlds.
\revm{To be able to model, e.g., the choice of repair made by a heuristic data repair algorithm, }
the value and annotation domains of an \abbrAUDB also contain a third component: a \textit{selected-guess} (\abbrBGW) that encodes one world. 

Formally, in an \abbrAUDB, attribute values are \textit{range-annotated values} $c = \uv{\lbMarker{c}}{\bgMarker{c}}{\ubMarker{c}}$ from a \textit{range-annotated domain} $\rangeDom$ that encodes the selected-guess value $\bgMarker{c} \in \dataDomain$ and two values ($\lbMarker{c}$, $\ubMarker{c} \in \aDom$) that bound $\bgMarker{c}$ from below and above.
For any $c \in \rangeDom$ we have $\lbMarker{c} \leq \bgMarker{c} \leq \ubMarker{c}$.
We call a value $c \in \rangeDom$ certain if $\lbMarker{c}=\bgMarker{c}=\ubMarker{c}$.
 \abbrAUDBs encode bounds on the multiplicities of tuples by using $\semqN = \tuple{\semqN, \addOf{\semqN}, \multOf{\semqN}, \zeroOf{\semqN}, \oneOf{\semqN}}$ annotations on tuples in $\rangeDom^n$.
The annotation $\ut{\lbMarker{k}}{\bgMarker{k}}{\ubMarker{k}}$ encodes a lower bound on the certain multiplicity of the tuple, the multiplicity of the tuple in the \abbrBGW, and an over-approximation of the tuple's possible multiplicity.
\iftechreport{We note that an \abbrAUDB can be encoded in a relational database by encoding each annotated value as three columns encoding the lower bound, selected-guess and upper bound value.}
\revm{Consider the AU-DB relation $\rangeRel(A,B)$ with a tuple $(\uv{1}{3}{5}, \uv{a}{a}{a})$ annotated with $\ut{1}{1}{2}$. This tuple represents the fact that each world consists of either 1 and 2 tuples with $B=a$ and $A$ between $1$ and $5$. The \abbrBGW contains a tuple $(3,a)$ with multiplicity 1.}

\paragraph{Bounding Databases}

As noted above, an \abbrAUDB summarizes an incomplete $\semN$-relation by defining bounds over the possible worlds that comprise it.
To formalize bounds over $\semN$-relations, we first define what it means for a range-annotated tuple to bound a set of deterministic tuples.
Let  $\rangeTup$ be a  range-annotated tuple with schema $\tuple{a_1, \ldots, a_n}$ and $\tup$ be a tuple  with the same schema as $\rangeTup$.
$\rangeTup$ bounds $\tup$ (denoted $\tup \tmatch \rangeTup$) iff
$\forall{i \in \{1, \ldots ,n\}}:
 \lbMarker{\rangeTup.a_i} \leq \tup.a_i \leq \ubMarker{\rangeTup.a_i}
$

Note that a single \abbrAUDB tuple may bound multiple deterministic tuples, and conversely that a single deterministic tuple may be bound by multiple \abbrAUDB tuples.  
Informally, an \abbrUAARel bounds a possible world if we can distribute the multiplicity of each tuple in the possible world over the \abbrUAARel's tuples.
This idea is formalized through \textit{tuple matchings}.
A tuple matching $\TM$ from an $n$-ary \abbrUAARel $\rangeRel$ to an $n$-ary relation $\rela$ is a function $(\rangeDom)^{n} \times \dataDomain^n \to \semN$ that fully allocates the multiplicity of every tuple of $\rela$:
\begin{align*}
  &\forall \rangeTup \in \rangeDom^n: \forall \tup \ntmatch \rangeTup: \TM(\rangeTup, \tup) = \zeroN
  &\forall \tup \in \dataDomain^n: \hspace*{-1mm}\sum_{\rangeTup \in \rangeDom^n} \hspace*{-1mm}\TM(\rangeTup, \tup)=\rela(\tup)
\end{align*}
$\rangeRel$ bounds $\rela$ (denoted $\rela \dbbounds \rangeRel$) iff there exists a tuple matching $\TM$ where the total multiplicity allocated to each $\rangeTup \in \rangeRel$ falls within the bounds annotating $\rangeTup$:
\begin{align*} 
  &\forall \rangeTup \in \rangeDom^n:\sum_{\tup \in \dataDomain^n} \TM(\rangeTup,\tup) \geq \lbMarker{\rangeRel(\rangeTup)}
  \;\text{\textbf{and}}\; 
     \sum_{\tup \in \dataDomain^n} \TM(\rangeTup,\tup) \leq \ubMarker{\rangeRel(\rangeTup)}
\end{align*}

An \abbrAUDB relation $\rangeRel$ bounds an incomplete $\semN$-relation $\prel$ (denoted $\prel \dbbounds \rangeRel$) iff it bounds every possible world (i.e., $\forall \rela \in \prel: \rela \dbbounds \rangeRel$), and if projecting down to the selected guess attribute of $\rangeRel$ results in a possible world of $\prel$.
As 
 shown in~\cite{FH19,FH21}, (i) \abbrAUDB query semantics is closed under $\raPlus$, set difference and aggregations, and (ii) queries preserve bounds.
That is, if every relation $\rangeRel_i \in \rangeDB$ bounds the corresponding relation of an incomplete database $\prel_i \in \pdb$ (i.e., $\forall i: \prel_i \dbbounds \rangeRel_i$), then for any query $\query$, the results over $\rangeDB$ bound the results over $\pdb$ (i.e., $\query(\pdb) \dbbounds \query(\rangeDB)$).


\paragraph{Expression Evaluation}
In~\cite{FH21},
we defined a semantics $\seval{\sexpr}{\rangeTup}$ for evaluating primitive-valued expressions $\sexpr$ over the attributes of a range tuple $\rangeTup$.  These semantics 
preserves bounds:
given any expression $\sexpr$ and any deterministic tuple $\tup$ bounded by $\rangeTup$ (i.e., $\tup \tmatch \rangeTup$), the result of deterministically evaluating the expression ($\seval{\sexpr}{\tup}$) is guaranteed to be bounded by the ranged evaluation $\seval{\sexpr}{\rangeTup}$.
$$\forall \tup \tmatch \rangeTup : c = \seval{\sexpr}{\tup}, \ut{\lbMarker{c}}{\bgMarker{c}}{\ubMarker{c}} = \seval{\sexpr}{\rangeTup} \;\; \rightarrow \;\; \lbMarker{c} \leq c \leq \ubMarker{c}$$
\cite{FH21} proved this property for any $e$ composed of attributes, constants, arithmetic and boolean operators, and comparisons.
\revm{For example, 
  $\uv{\lbMarker{a}}{\bgMarker{a}}{ \ubMarker{a}} + \uv{\lbMarker{b}}{\bgMarker{b}}{ \ubMarker{b}} = \uv{\lbMarker{a} + \lbMarker{b}}{\bgMarker{a} + \bgMarker{b}}{\ubMarker{a} + \ubMarker{b}}$}


\section{Deterministic Semantics}
\label{sec:det-materialized-sort}

Before introducing the AU-DB semantics for ranking and windowed aggregation, we first formalize the corresponding deterministic  algebra operators that
materialize sort positions of rows  \emph{as data}.

\mypar{Sort order}
Assume a total order $<$ for the domains of all attributes.
For simplicity, we only consider sorting in ascending order.
The extension for supporting both ascending and descending order is straightforward.
For any two tuples $t$ and $t'$ with schema $(A_1, \ldots, A_n)$ and sort attributes $\sortattrs = (A_{i_1}, \ldots, A_{i_m})$ we define:\\[-5mm]
\begin{align*}
  \tup \ales \tup' \Leftrightarrow  \exists &j \in \{ 1, \ldots, m \} : \\
  &\forall  k \in \{1, \ldots, j-1 \}: \tup.A_{i_k} = \tup'.A_{i_k} \land \tup'.A_{i_j} < \tup.A_{i_j}
\end{align*}\\[-4mm]
The less-than or equals comparison operator $\alest$ generalizes this definition in the usual way.  Note that SQL sorting (\lstinline!ORDER BY!) and some window bounds (\lstinline!ROW BETWEEN ...!) may be non-deterministic.
For instance, consider a relation $\rela$ with schema $(A,B)$ with two rows $t_1 = (1,1)$ and $t_2 = (1,2)$ each with multiplicity 1;
Sorting this relation on attribute $A$ (the tuples are indistinguishable on this attribute), can return the tuples in either order.
Without loss of generality, we ensure a fully deterministic semantics (up to tuple equivalence) by extending the ordering on attributes $\sortattrs$, using the remaining attributes of the relation as a tiebreaker:
The total order $\tup \afles \tup'$ for tuples from a relation $\rela$ is defined as $\tup \les{\sortattrs,\schemaOf(\rela)-\sortattrs} \tup'$ (assuming some arbitrary order of the attributes in $\schemaOf(\rela)$).
\iftechreport{
\begin{Example}[Sorting]\label{ex:ranking}
Consider the relation $\rela$ shown on the left below. The multiplicity from \semN assigned to each tuple is shown on the right. The result of sorting the relation on attribute $A$ using our deterministic semantics and storing the sort positions in column \texttt{pos} is shown below on the right. Note the order of tuples $t_1 = (3,15)$ and $t_2 = (1,1)$ is made deterministic by $t_2 \fles{A} t_1$, because $t_2.B < t_1.B$. Note also that the two copies of $t_2$ are each assigned a different position.

  \begin{minipage}{0.35\linewidth}
    \centering
    \begin{tabular}{c|cc}
      \thead{A} & \thead{B} & \semN \\
      \hline
      3 & 15 & 1 \\
      1 & 1 & 2\\
    \end{tabular}
  \end{minipage}
  \begin{minipage}{0.64\linewidth}
    \centering
    \begin{tabular}{c|c|cc}
        \thead{A} & \thead{B} & \thead{pos} & \semN\\
        \hline
      1 & 1 & 0 & 1\\
      1 & 1 & 1 & 1\\
      3 & 15 & 4 & 1\\
      \end{tabular}
  \end{minipage}
\end{Example}
}
We first introduce operators for windowed aggregation, because  sorting 
can be defined as a special case of windowed aggregation.

\subsection{Windowed Aggregation}
\label{sec:windowed-aggregation}

A windowed aggregate is defined by an aggregate function, a sort order (\lstinline!ORDER BY!), and a window bound specification.
A window boundary is relative to the defining tuple, by the order-by attribute value (\lstinline!RANGE BETWEEN...!), or by position (\lstinline!ROWS BETWEEN!).
In the interest of space, we will limit our discussion to row-based windows, as range-based windows are strictly simpler.
A window includes every tuple within a specified interval of the defining tuple.
%
%
Windowed aggregation extends each input tuple with the aggregate value computed over the tuple's window.
If a \lstinline!PARTITION BY! clause is present, then window boundaries are evaluated within a tuple's partition.
In SQL, a single query may define a separate window for each aggregate function (SQL's \lstinline!OVER! clause).
This can be modeled by applying multiple window operators in sequence. 

\begin{Example}[Row-Based Windows]\label{ex:row-based-windows}
  Consider the bag relation below and consider the windowed aggregation $sum(B)$ sorting on $A$ with bounds $[-2,0]$ (including the two preceding tuples and the tuple itself).
  The window for the first duplicate of $t_1 = (a,5,3)$ contains tuple $t_1$ with multiplicity $1$, the window for the second duplicate of $t_1$ contains $t_1$ with multiplicity $2$ and so on.
  Because each duplicate of $t_1$ ends up in a different window, there are three result tuples produced for $t_1$, each with a different $sum(B)$ value. Furthermore, tuples $t_2 = (b,3,1)$ and $t_3 = (b,3,4)$ have the same position in the sort order, demonstrating the need to use $\afles$ to avoid non-determinism in what their windows are. We have $t_2 \afles t_3$ and, thus, the window for $t_2$ contains $t_2$ with multiplicity $1$ and $t_1$ with multiplicity $2$ while the window for $t_3$ contains $t_1$, $t_2$ and $t_3$ each with multiplicity $1$.

\begin{minipage}{0.35 \linewidth}
  \centering
  \scalebox{0.8}{
  \begin{tabular}{c|c|cc}
    \thead{A} & \thead{B} & \thead{C} & \thead{$\semN$}\\ \hline
    a & 5 & 3 & 3 \\
    b & 3 & 1 & 1 \\
    b & 3 & 4 & 1\\
  \end{tabular}
  }
\end{minipage}
\begin{minipage}{0.64 \linewidth}
  \centering
  \scalebox{0.8}{
  \begin{tabular}{c|c|c|cc}
    \thead{A} & \thead{B} & \thead{C} & \thead{sum(B)} & \thead{$\semN$}\\ \hline
    a & 5 & 3 & 5 & 1 \\
    a & 5 & 3 & 10 & 1 \\
    a & 5 & 3 & 15 & 1 \\
    b & 3 & 1 & 13 & 1\\
    b & 3 & 4 & 11 & 1\\
  \end{tabular}
  }
\end{minipage}

\end{Example}
  The semantics of the row-based window aggregate operator $\winrowop$ is shown in \Cref{fig:windowed-aggregation-oper}.
The parameters of $\winrowop$ are partition-by attributes $\gbAttrs$, order-by attributes $\sortattrs$, an aggregate function $f(A)$ with $A \subseteq \schemaOf(R)$, and an interval $[l,u]$. For simplicity, we hide some arguments ($\gbAttrs$,$\sortattrs$,$l$,$u$) in the  definitions and assume they passed to intermediate definitions where needed.
The operator outputs a relation with schema $\schemaOf(R) \circ X$.

The heavy lifting occurs in the definition of relation
$\wrowresultshort(\rela)$, which ``explodes'' relation $\rela$, adding an attribute $i$ to replace each tuple of multiplicity $n$ with $n$ distinct tuples.
$\wrowresultshort(\rela)$ computes the windowed aggregate over the window defined for the pair $(\tup, i)$, denoted as
$\windowshort{\rela}{t}{i}(\tup')$.
To construct this window, we define the range of the sort positions the tuple $\tup$ covers (
$\tcovershort{\rela}{\tup}$), and the range of positions in its window ($\tboundsshort{R}{t}{i}$).
The multiplicity of tuple $\tup'$ in the partition of $\tup$ (denoted $\partitionshort{R}{t}(\tup')$) is the size of the overlap between the bounds of $\tup$, and the cover of $\tup'$.

\iftechreport{
  $\winrangeop$ computes the dense-rank windowed aggregate.
  $\trank{\rela}{\sortattrs}{\tup}$ computes the dense rank of tuple $\tup \in \rela$ using $\ales$: the number of tuple groups (tuples with matching values of $O$) preceding $\tup$.
  We define the window for tuple $\tup$ (denoted $\awindow{\tup}$) point-wise for each tuple by considering the tuple's dense rank within its partition (denoted $\apartition{t}$).

  Because dense-rank windows are computed over tuple groups, the multiplicity of teach tuple in the window is taken directly from the relation.
  For sparse rank windows, we need to ensure that the window contains exactly the desired number of rows, e.g., a window with bounds $[-2,0]$ should contain exactly 3 rows, e.g., 1 row with multiplicity 2 and one row with multiplicity 1. 
  Since we need to ensure that windows contain a fixed number of tuples, it may be the case that a tuple will be included with a multiplicity in the window that is less than the tuple's multiplicity in the input relation. Furthermore, the window for one duplicate of a tuple may differ from the window of another duplicate of the same tuple.
} 

\begin{figure}[t]
  \centering

\iftechreport{
  \begin{align*}
    \awinrange(\rela)(t) &=
                           \begin{cases}
                             \rela(t') &\mathif t = t' \tconcat \aggf(\projection_{\agga}(\awindow{t'})) \\
                             0 &\mathotherwise\\
                           \end{cases} \\
    \awindow{t}(t') &=
                      \begin{cases}
                        R(t') &\mathif (\trank{\apartition{t}}{\sortattrs}{t} \\
                        & \hspace{6mm} - \trank{\apartition{t}}{\sortattrs}{t'}) \in [l,u]\\
                        0 &\mathotherwise\\
                      \end{cases}\\
    \trank{\rela}{\sortattrs}{t} &= \card{ \{\; t'.\sortattrs \mid R(t') > 0 \land t' \ales t\; \}}
  \end{align*}
} 

  \begin{align*}
    \awinrow(\rela)(t) &= \projection_{\schemaOf(R),X}(\wrowresultshort(\rela))\\
    \wrowresultshort(\rela)(t) &=
                           \begin{cases}
                             1 &\mathif t = t' \tconcat \aggf(\projection_{\agga}(\windowshort{R}{t'}{i})) \tconcat i\\ &\hspace{6mm} \land i \in [0,R(t')-1]\\
                             0 &\mathotherwise\\
                           \end{cases} \\
    \partitionshort{R}{t}(t') &=
                         \begin{cases}
                           R(t') &\mathif t'.\gbAttrs = t.\gbAttrs\\
                           0 &\mathotherwise\\
                         \end{cases}
    \\
    \windowshort{R}{t}{i}(t') &= \vert\, \tcovershort{\partitionshort{R}{t}}{t'} \cap \tboundsshort{\partitionshort{R}{t}}{t}{i} \,\vert\\
\tposshort{\rela}{t}{i} &= i + \sum_{t' \afles t} \rela(t')\\
    \tcovershort{\rela}{t} &= [ \tposshort{\rela}{t}{0}, \tposshort{\rela}{t}{\rela(t) - 1} ]\\
    \tboundsshort{\rela}{t}{i} &= [ \tposshort{\rela}{t}{i} +l, \tposshort{\rela}{t}{i} + u ]
  \end{align*}
    \\[-4.5mm]
	\setlength{\belowcaptionskip}{-10pt}
  \caption{Windowed Aggregation}\label{fig:windowed-aggregation-oper}
\end{figure}

\subsection{Sort Operator}\label{sec:ranking-operato}







We now define a sort operator $\rank{\sortattrs}{\ranka}(\rela)$ which extends each row of $\rela$ with an attribute $\ranka$ that stores the position of this row in $\rela$ according to $\afles$. This operator is just ``syntactic sugar'' as it can be expressed using  windowed aggregation.

\begin{Definition}[Sort Operator]\label{def:rank-operator}
  Consider a relation $\rela$ with schema $(A_1, \ldots, A_n)$, list of attributes $\sortattrs = (B_1, \ldots, B_m)$ where each $B_i$ is in $\schemaOf(\rela)$. The \emph{sort operator} $\rank{\sortattrs}{\ranka}(\rela)$ returns a relation with schema $(A_1, \ldots, A_n, \ranka)$ as defined below.
\begin{align*}
\rank{\sortattrs}{\ranka}(\rela) &= \projection_{\schemaOf(\rela), \ranka - 1 \to \ranka}( \winrow{count(1)}{\ranka}{\emptyset}{\sortattrs}{-\infty}{0}(\rela))
\end{align*}
\end{Definition}


Top-k queries can be expressed using the sort operator followed by a selection. For instance, 
the SQL query shown below can be written as $\projection_{A,B}(\selection_{r \leq 3}(\rank{A}{r}(\rela)))$.

\begin{lstlisting}
SELECT A,B FROM R ORDER BY A LIMIT 3;
\end{lstlisting}


\section{AU-DB Sorting and Top-k Semantics}\label{sec:ranking}

We now develop a bound-preserving semantics for sorting and top-k queries over AU-DBs.
Recall that each tuple in an \abbrAUDB is annotated with a triple of multiplicities and that each (range-annotated) value is likewise a triple. Elements of a range-annotated value $\mathbf{c} = \uv{c_1}{c_2}{c_3}$ or multiplicity triple $\ut{n_1}{n_2}{n_3}$ are accessed as:  $\lbMarker{\mathbf{c}} = c_1$, $\bgMarker{\mathbf{c}} = c_2$, and $\ubMarker{\mathbf{c}} = c_3$.
\revm{We use bold face to denote range-annotated tuples, relations, values, and databases.}
Both the uncertainty of a tuple's multiplicity and the uncertainty of the values of order-by attributes create uncertainty in a tuple's position in the sort order. \revm{The former, because it determines how many duplicates of a tuple appear in the sort order which affects the position of tuples which may be larger wrt. the sort order and the latter because it affects which tuples are smaller than a tuple wrt. the sort order.}
As mentioned before, a top-k query is a selection over the result of a sort operator \revm{which checks that the sort position of a tuple is less than or equal to $k$}.
A bound-preserving semantics for selection was already presented in~\cite{FH21}.
Thus, we focus on sorting and use the existing selection semantics 
for top-k queries.

\mypar{Comparison of Uncertain Values}
Before introducing sorting over \abbrAUDBs, we first discuss \revm{the evaluation of} $\ales$ over tuples with uncertain values (recall that $\afles$ is defined in terms of $\ales$).
Per \cite{FH21}, a Boolean expression over range-annotated values evaluates  to a bounding triple (using the order $\bfalse < \btrue$ \revm{where $\bfalse$ denotes false and $\btrue$ denotes true}). \revm{The result of an evaluation  of an expression $e$ is denoted as  $\seval{e}{}$.}
For instance, $\seval{\uv{1}{1}{3} < \uv{2}{2}{2}} = \uv{\bfalse}{\btrue}{\btrue}$, because the expression may evaluate to false (e.g., if the first value is $3$ and the second values is $2$), evaluates to true in the selected-guess world, and may evaluate to true (if the $1^{st}$ value is $1$ and the $2^{nd}$ value is $2$). The extension of $<$ to comparison of tuples on attributes $\sortattrs$ using $\ales$ is shown below.
\revm{For example, consider tuples $\rangeOf{t_1} = (\uv{1}{1}{3}, \uv{a}{a}{a})$ and $\rangeOf{t_2} = (\uv{2}{2}{2}, \uv{b}{b}{b})$ over schema $R(A,B)$. We have $\rangeOf{t_1} \les{A,B} \rangeOf{t_2} = \uv{\bfalse}{\btrue}{\btrue}$, because $\rangeOf{t_1}$ could be ordered before $\rangeOf{t_2}$ (if $\rangeOf{t_1}.A$ is 1), is ordered before $\rangeOf{t_2}$ in the selected-guess world ($1 < 2$), and may be ordered after $\rangeOf{t_2}$ (if $\rangeOf{t_1}.A$ is 3).}

\vspace{-3mm}
\begin{equation*}
\begin{aligned}
	\lbMarker{ \seval{\rangeTup \ales \rangeOf{t'}}{}  } &=  \exists i \in \{ 1, \ldots, n \}:  \forall j \in \{1, \ldots, i-1 \}:  \\
                                     &\hspace{10mm}\lbMarker{\seval{\rangeTup.A_j = \rangeOf{t'}.A_j}{}} \land \lbMarker{\seval{\rangeTup.A_i < \rangeOf{t'}.A_i}{}}\\
  \bgMarker{ \seval{\rangeTup \ales \rangeOf{t'}}{}  } &= \exists i \in \{ 1, \ldots, n \}:  \forall j \in \{1, \ldots, i-1 \}:  \\
	 &\hspace{10mm}\bgMarker{\seval{\rangeTup.A_j = \rangeOf{t'}.A_j}{}} \land \bgMarker{\seval{\rangeTup.A_i < \rangeOf{t'}.A_i}{}}\\
	\ubMarker{ \seval{\rangeTup \ales \rangeOf{t'}}{}  } &=  \exists i \in \{ 1, \ldots, n \}:  \forall j \in \{1, \ldots, i-1 \}:  \\
                                     &\hspace{10mm}\ubMarker{\seval{\rangeTup.A_j = \rangeOf{t'}.A_j}{}} \land \ubMarker{\seval{\rangeTup.A_i < \rangeOf{t'}.A_i}{}}\\
\end{aligned}
\end{equation*}
\vspace{-1mm}
\noindent
To simplify notation, we will use $\rangeTup \ales \rangeOf{t'}$ instead of $\seval{\rangeTup \ales \rangeOf{t'}}{}$.

\BG{DELETED (see whether the second part needs to be merged into the following sections)
each tuple attribute values
Recall that for ranking on \abbrAUDBs, each tuple may have a range of possible rankings. In order to get an over-estimation of the top-K result, computing top-K for \abbrAUDB may return more than K tuples. In this section we define top-K semantics for \abbrAUDBs (where sorting is a special case of top-K where $K=\mid \rangeRel \mid $) and propose a one pass algorithm that implement ranking and top-K on \abbrAUDB relational encodings.}

\mypar{Tuple Rank and Position}
To define windowed aggregation and sorting over \abbrAUDBs, we generalize 
$\tpossymb$ using the uncertain version of $\ales$.
The lowest possible position of the first duplicate of a tuple $\rangeTup$ in an \abbrAUDB relation $\rangeRel$ is the total multiplicity of tuples $\rangeOf{t'}$ that certainly exist ($\lbMarker{\rangeRel(\rangeOf{t'})} > 0$) and are certainly smaller than $\rangeTup$ (i.e., $\lbMarker{\seval{\rangeOf{t'} \ales \rangeTup}{}} = \btrue$).
The selected-guess position of a tuple is the position of the tuple in the selected-guess world, and the greatest possible position of $\rangeTup$ is the total multiplicity of tuples that possibly exist ($\ubMarker{\rangeRel(\rangeOf{t'})} > 0$) and possibly precede $\rangeTup$ (i.e., $\ubMarker{\seval{\rangeOf{t'} \ales \rangeTup}{}} = \btrue$). \revm{The sort position of the $i^{th}$ duplicate (with the first duplicate being $0$) is computed by adding $i$ to the position bounds of the first duplicate.}
\vspace{-1mm}
\begin{align}
     \lbMarker{\tpos{\rangeRel}{\sortattrs}{\rangeTup}{i}}  &= i + \textstyle\sum_{\lbMarker{ ( \revb{ \rangeOf{t'} \ales \rangeTup } ) }} \lbMarker{ \rangeRel(\rangeOf{t'}) } \label{eq:upos-lb}\\
    \bgMarker{\tpos{\rangeRel}{\sortattrs}{\rangeTup}{i}}  &= i + \textstyle\sum_{ \bgMarker{ ( \revb{ \rangeOf{t'} \ales \rangeTup } ) }} \bgMarker{ \rangeRel(\rangeOf{t'}) } \label{eq:upos-sg}\\
  \ubMarker{\tpos{\rangeRel}{\sortattrs}{\rangeTup}{i}}  &=  i + \textstyle\sum_{ \ubMarker{ ( \revb{ \rangeOf{t'} \ales \rangeTup } ) }} \ubMarker{ \rangeRel(\rangeOf{t'}) }\label{eq:upos-ub}
\end{align}
\vspace{-6mm}


\subsection{\abbrAUDB Sorting Semantics}

To define \abbrAUDB sorting, we split the \revm{possible duplicates} of a tuple and extend the resulting tuples with a range-annotated value denoting the tuple's (possible) positions in the sort order.
The certain multiplicity of the $i^{th}$ duplicate of a tuple $\rangeTup$ in the result is either $1$ for duplicates that are guaranteed to exist ($i < \lbMarker{\rangeRel(\rangeTup)}$) and $0$ otherwise. \revm{The selected-guess multiplicity is $1$ for duplicates that do not certainly exist (in some possible world there may be less than $i$ duplicates of the tuple), but are in the selected-guess world (the selected-guess world has $i$ or more duplicates of the tuple). Finally, the possible multiplicity is always $1$.}

\begin{Definition}[\abbrAUDB Sorting Operator]\label{def:au-db-sorting-operat}
Let $\rangeRel$ be an \abbrAUDB relation and $\sortattrs \subseteq \schemaOf(\rangeRel)$. The result of applying the sort operator $\rank{\sortattrs}{\ranka}$ to $\rangeRel$ is defined in \Cref{fig:uadbSortSemantics}
\end{Definition}

\begin{figure}
  \begin{align*}
    \rank{\sortattrs}{\ranka}&(\rangeRel)(\rangeTup) =  \\ & \hspace{-15pt}
    \begin{cases}
      (1,1,1) &  \mathif \rangeTup = \parbox[t]{.5\textwidth}{$
      \rangeTup' \circ  \tpos{\rangeRel}{\sortattrs}{\rangeTup'}{i}
       \land i \in \left[0, \lbMarker{\rangeRel(\rangeOf{t'})} \right)$}\\
      (0,1,1) & \mathif \rangeTup =\parbox[t]{.5\textwidth}{$
      \rangeTup' \circ  \tpos{\rangeRel}{\sortattrs}{\rangeTup'}{i}
       \land i \in \left[\lbMarker{\rangeRel(\rangeOf{t'})}, \bgMarker{\rangeRel(\rangeOf{t'})} \right) $} \\
      (0,0,1) & \mathif \rangeTup = \parbox[t]{.5\textwidth}{$
      \rangeTup' \circ  \tpos{\rangeRel}{\sortattrs}{\rangeTup'}{i}
       \land \revb{ i \in \left[\bgMarker{\rangeRel(\rangeOf{t'})}, \ubMarker{\rangeRel(\rangeOf{t'})} \right) }
$}\\
      (0,0,0) & \mathotherwise\\
    \end{cases}
  \end{align*}
  \vspace*{-5mm}
  \setlength{\belowcaptionskip}{-15pt}
  \caption{Range-annotated sort operator semantics.}
  \label{fig:uadbSortSemantics}
\end{figure}

\revm{Every tuple in the result of sorting is constructed by extending an input tuple $\rangeTup'$ with the range of positions $\tpos{\rangeRel}{\sortattrs}{\rangeTup'}{i}$ it may occupy wrt. the sort order.}
The definition decomposes $\rangeTup$ into a base tuple $\rangeTup'$, and a position triple for each 
duplicate of $\rangeTup$ in $\rangeRel$.
We annotate all certain duplicates as certain $\ut{1}{1}{1}$, remaining selected-guess (but uncertain) duplicates as uncertain $\ut{0}{1}{1}$ and non-selected guess duplicates as possible $\ut{0}{0}{1}$.
\BG{The example below illustrates the semantics of $\rankopsymb$ over an \abbrAUDB relation.}

\begin{Example}[\abbrAUDB Sorting]\label{ex:audb-ranking}
  Consider the \abbrAUDB relation $\rangeRel$ shown on the left below with certain, selected guess and possible multiplicities from $\semN^3$ assigned to each tuple. For values or multiplicities that are certain, we write only the certain value instead of the triple.
  The result of sorting the relation on attributes $A,B$ using \abbrAUDB sorting semantics and storing the sort positions in column \texttt{pos} (\revm{$\rank{A,B}{pos}(\rangeRel)$}) is shown below on the right. \revm{Observe how the $1^{th}$ input tuple $\rangeOf{t_1} = (1, \uv{1}{1}{3})$ was split into two result tuples occupying adjacent sort positions. The $3^{rd}$ input tuple $\rangeOf{t_3} = (\uv{1}{1}{2},2)$ could be the $1^{th}$ in sort order (if its $A$ value is $1$ and the $B$ values of the duplicates of $\rangeOf{t_1}$ are equal to $3$) or be at the $3^{rd}$ position if two duplicates of $\rangeOf{t_1}$ exist and either $A$ is $2$ or the $B$ values of $\rangeOf{t_1}$ are all $<3$.}

\noindent
\resizebox{1\linewidth}{!}{
  \begin{minipage}[t]{1.09\linewidth}
  \begin{minipage}{0.4\linewidth}
      \begin{tabular}{c|cc}
        \thead{A} & \thead{B} & $\semN^3$ \\
        \hline
        1 & \uv{1}{1}{3} & (1,1,2)\\
        \uv{2}{3}{3} & 15 & (0,1,1) \\
        \uv{1}{1}{2} & 2 & (1,1,1)\\
      \end{tabular}
    \end{minipage}
    \begin{minipage}{0.63\linewidth}
      \centering
      \begin{tabular}{c|c|cc}
        \thead{A} & \thead{B} & \thead{pos} & $\semN^3$ \\
        \hline
        1 & \uv{1}{1}{3} & \uv{0}{0}{1} & (1,1,1)\\
        1 & \uv{1}{1}{3} & \uv{1}{1}{2} & (0,0,1)\\
        \uv{1}{1}{2} & 2 & \uv{0}{1}{2} & (1,1,1)\\
        \uv{2}{3}{3} & 15 & \uv{2}{2}{3} & (0,1,1)\\
      \end{tabular}
    \end{minipage}
  \end{minipage}
   }
\end{Example}
\vspace{-3mm}
\subsection{Bound Preservation}
We now prove that our semantics for the sorting operator on AU-DB relations is bound preserving, i.e., given an AU-DB $\rangeRel$ that bounds an incomplete bag database $\prel$, the result of a sort operator $\rank{\sortattrs}{\ranka}$ applied to $\rangeRel$ bounds the result of $\rank{\sortattrs}{\ranka}$ evaluated over $\prel$.
\begin{Theorem}[Bound Preservation of Sorting]\label{theo:bound-preservation-o}
Given an \abbrAUDB relation $\rangeRel$ and incomplete bag relation $\prel$ 
such that $\prel \dbbounds \rangeRel$, and $\sortattrs \subseteq \schemaOf(\prel)$. We have:
\vspace{-2mm}
  \[\rank{\sortattrs}{\ranka}(\prel) \dbbounds \rank{\sortattrs}{\ranka}(\rangeRel)\] 
\end{Theorem}
\vspace{-2mm}
\detailedproof{
We prove the theorem by taking a tuple matching $\TM$ for each possible world $\rela$ in the input (that is guaranteed to exist, because $\prel \dbbounds \rangeRel$) and construct a tuple matching $\TM'$ for the output of sorting based on which $\rank{\sortattrs}{\ranka}(\prel) \dbbounds \rank{\sortattrs}{\ranka}(\rangeRel)$ holds. In the proof we make use of the fact that the sort operator distributes the multiplicity of an input tuple $t$ to multiple output tuples which each are extensions of $t$ with a sort position, keeping all other attributes the same as in the input.}{
Since $\prel \dbbounds \rangeRel$, for every possible world $R \in \prel$, there has to exist a tuple matching $\TM_{\rangeRel}$ based on which this property holds. We will show that based on $\TM_{\rangeRel}$ we can generate a tuple matching for $\rangeRel_{res} = \rank{\sortattrs}{\ranka}(\rangeRel)$ and $\rank{\sortattrs}{\ranka}(\prel)$. The existence of such a tuple matching for every $\rank{\sortattrs}{\ranka}(\rela) \in \rank{\sortattrs}{\ranka}(\prel)$ implies that $\rank{\sortattrs}{\ranka}(\prel) \dbbounds \rank{\sortattrs}{\ranka}(\rangeRel)$. WLOG consider tuple $\rangeTup \in \rangeRel$ and output tuples $\rangeTup_{res} = \rangeTup \circ \tpos{\rangeRel}{\sortattrs}{\rangeTup}{i}$.

We first show that split the tuple $\rangeTup$ preserves tuple multiplicities. Applying the definitions from \Cref{fig:uadbSortSemantics}, 
\begin{align*}
	\lbMarker{\rangeRel(\rangeTup)} = & \sum_{i} \lbMarker{\rangeRel_{res}(\rangeTup \circ \tpos{\rangeRel}{\sortattrs}{\rangeTup}{i})} \\
	\bgMarker{\rangeRel(\rangeTup)} = & \sum_{i} \bgMarker{\rangeRel_{res}(\rangeTup \circ \tpos{\rangeRel}{\sortattrs}{\rangeTup}{i})} \\
	\ubMarker{\rangeRel(\rangeTup)} = & \sum_{i} \ubMarker{\rangeRel_{res}(\rangeTup \circ \tpos{\rangeRel}{\sortattrs}{\rangeTup}{i})}
\end{align*}
Thus, the split preserves upper and lower multiplicity bounds.  \\
Then we propose a tuple matching $\TM_{res}$ that maps each split tuple to a specific deterministic tuple.
Let \(S_{\rangeTup}=\{t_1, \ldots, t_n\} \in \rela\) be the only tuples such that \(\mathcal{\TM_{\rangeRel}}(\mathbf{t},t_i) > 0\) and let us assume that \(t_i \leq_{O} t_j\) if \(i < j\) which we can ensure by sorting these tuples based on \(O\). Recall that both AU-DB sorting and deterministic sorting splits each tuple \(\rangeTup\) (\(t\)) into individual tuples \(\rangeOf{t_i}\) (\(t_{ik}\)).\\
Let us denote \(n_i\) the total multiplicity of tuples orders before \(t_i\), i.e., \(n_i = \sum_{j < i} \allowbreak S_{\mathbf{t}}(t_j)\). Note that \(n_i\) is the relative sort position of the first duplicate of \(t_j\) wrt. the first duplicate of \(t_1 (t_{11})\). Then in the result tuple matching we will define $$\TM_{res}(\rangeTup \circ \tpos{\rangeRel}{\sortattrs}{\rangeTup}{n_i+k}, \tup_i \circ \tpos{\rela}{\sortattrs}{\tup_i}{k}) = 1$$ for all \(k < \rela(t_i)\). \\
Because of the definitaiton of ranged comparison that utilizes range-based scalar expression semantics which is bound preserving we know that if \(t \sqsubseteq \mathbf{t}\) and \(\rangeOf{t'} \sqsubseteq \rangeOf{t'}\) then \((t' <_O t) \leftarrow \lbMarker{(\rangeOf{t'} <_O \rangeOf{t})}\). First we use the definition of tuple matching. Because only part of \(R(t')\) may be matched against \(\mathbf{t}\) we have \(\TM_{\rangeRel}(\mathbf{t}',t') \leq R(t')\)
\begin{align*}
	\lbMarker{\tpos{\rangeRel}{\sortattrs}{\rangeTup}{0}} & =\sum_{\lbMarker{(\rangeOf{t'} <_O \rangeOf{t})}} \lbMarker{\rangeRel(\rangeOf{t'})} \\
	& \leq \sum_{\lbMarker{(\rangeOf{t'} <_O \rangeOf{t})}}\TM_{\rangeRel}(\rangeOf{t'},t') \\
	& \leq \sum_{t': t' <_{O} t_1} R(t') = \tpos{\rela}{\sortattrs}{\tup_1}{0}
\end{align*}
this is only for first tuple matched against \(\mathbf{t}\). For \(t_i\) we know that $$\tpos{\rela}{\sortattrs}{\tup_i}{k} = n_i+k+\tpos{\rela}{\sortattrs}{\tup_1}{0}+\sum_{t'' \geq_O t_1 \land t'' <_O t_i \land t'' \notin S_{\rangeTup}} R(t'')$$
The position of k-th duplicate of $t_i$ is constructed by adding number of matched preceding tuples to the position of the first matched tuple $t_i$, and we also need to add all tuples that are not matched to $\rangeTup$ but is in preceding of $t_i$. \\
For $\rangeTup$ we know that 
\begin{align*}
	\lbMarker{\tpos{\rangeRel}{\sortattrs}{\rangeTup}{n_i+k}} = n_i + k + & \lbMarker{\tpos{\rangeRel}{\sortattrs}{\rangeTup}{0}} \\
	& \leq n_i+k+\tpos{\rela}{\sortattrs}{\tup_1}{0} \\
	& \leq n_i+k+\tpos{\rela}{\sortattrs}{\tup_1}{0} + \\
	& \hspace{10mm} \sum_{t'' \geq_O t_1 \land t'' <_O t_i \land t'' \notin S_{\rangeTup}} R(t'') \\
	& = \tpos{\rela}{\sortattrs}{\tup_i}{k}
\end{align*}
Thus, $\tpos{\rangeRel}{\sortattrs}{\rangeTup}{n_i+k}$ lower bounds $\tpos{\rela}{\sortattrs}{\tup_i}{k}$. \\
In an analog way, we have 
\begin{align*}
	\ubMarker{\tpos{\rangeRel}{\sortattrs}{\rangeTup}{0}} & =\sum_{\ubMarker{(\rangeOf{t'} <_O \rangeOf{t})}} \ubMarker{\rangeRel(\rangeOf{t'})} \\
	& \geq \sum_{\ubMarker{(\rangeOf{t'} <_O \rangeOf{t})}}\TM_{\rangeRel}(\rangeOf{t'},t') \\
	& \geq \sum_{t': t' <_{O} t_n} R(t') = \tpos{\rela}{\sortattrs}{\tup_n}{0} + \\
	& \hspace{10mm} \sum_{t'' \geq_O t_1 \land t'' <_O t_n \land t'' \notin S_{\rangeTup}} R(t'')
\end{align*}
Following with
\begin{align*}
	& \ubMarker{\tpos{\rangeRel}{\sortattrs}{\rangeTup}{n_i+k}} = n_i + k + \ubMarker{\tpos{\rangeRel}{\sortattrs}{\rangeTup}{0}} \\
	& \geq n_i+k+\tpos{\rela}{\sortattrs}{\tup_n}{0}+\sum_{t'' \geq_O t_1 \land t'' <_O t_n \land t'' \notin S_{\rangeTup}} R(t'') \\
	& \geq n_i+k+\tpos{\rela}{\sortattrs}{\tup_1}{0}+\sum_{t'' \geq_O t_1 \land t'' <_O t_i \land t'' \notin S_{\rangeTup}} R(t'') \\
	& = \tpos{\rela}{\sortattrs}{\tup_i}{k}
\end{align*}
Thus, $\tpos{\rela}{\sortattrs}{\tup_i}{k} \sqsubseteq \tpos{\rangeRel}{\sortattrs}{\rangeTup}{n_i+k}$. Also since $\tup_i \sqsubseteq \rangeTup$, $\TM_{res}$ is valid.
}
\section{AU-DB Windowed Aggregation}\label{sec:audb-windowed-agg-sem}

\BG{For windowed aggregation, the a tuple's membership in a partition may be uncertain, and the membership or multiplicity of a tuple in a window may be uncertain.
Furthermore, both types of uncertainty and the uncertainty of attribute values we are aggregating over may cause uncertainty of an aggregation function result.}

We now introduce a bound preserving semantics for windowed aggregation over \abbrAUDBs.
We have to account for three types of uncertainty:
(i) uncertain partition membership if a tuple may not exist ($\lbMarker{\rangeRel(\rangeTup)} = 0$) or has uncertain partition attributes;
(ii) uncertain window membership if a tuple's partition membership, position, or multiplicity are uncertain; and
(iii) uncertain aggregation results from either preceding type of uncertainty, or if we are aggregating over uncertain values.
We compute the windowed aggregation result for each input tuple in multiple steps: (i) we first use \abbrAUDB sorting to split each input tuple into tuples whose multiplicities are at most one. This is necessary, because the aggregation function result may differ among the duplicates of a tuple (as is already the case for deterministic windowed aggregation); (ii) we then compute for each tuple $\rangeTup$ an \abbrAUDB relation $\shortpart{\rangeTup}{\rangeRel}$ storing the tuples that certainly and possibly belong to the partition for that tuple; (iii) we then compute an \abbrAUDB relation $\awindowshorter{\rangeRel}{\rangeTup}$ encoding which  tuples certainly and possibly belong to the tuple's window; (iv) since row-based windows contain a fixed number of tuples, we then determine from the tuples that possibly belong to the window, the subset that together with the tuples that certainly belong to the window (these tuples will be in the window in every possible world) minimizes / maximizes the aggregation function result. This then  enables us to bound the aggregation result for each input tuple from below and above. For instance, for a row-based window $[-2,0]$, we know that the window for a tuple $\rangeTup$ will never contain more than 3 tuples. If we know that two tuples certainly belong to the window, then at most one additional possible tuple can belong to the window.



\subsection{Windowed Aggregation Semantics}\label{sec:audb-win-agg-semantics}

As before, we omit windowed aggregation parameters ($\gbAttrs$,$\sortattrs$,$l$,$u$,$f$,$\agga$) from the arguments of  intermediate constructs  and assume they are passed along where needed. 

\textbf{Partitions}
\revm{We start by defining \abbrAUDB relation $\shortpart{\rangeTup}{\rangeRel}$ which  encodes the multiplicity of tuple $\rangeOf{t'}$ in the partition for $\rangeTup$ based on partition-by attributes $\gbAttrs$. This is achieved using selection, comparing a tuple's values in $\gbAttrs$ with the values of $\rangeTup.\gbAttrs$ on equality. \abbrAUDB selection sets the certain (selected-guess, or possible multiplicity) of a tuple to $0$ if the tuple possibly (in the selected-guess world, or certainly) does not fulfill the selection condition.}

%
\vspace{-1mm}
$$
\revm{\shortpart{\rangeTup}{\rangeRel} = \seval{\selection_{\gbAttrs = \rangeTup.\gbAttrs}(\rangeRel)}{}}
$$
\vspace{-1mm}
\mypar{Certain and Possible Windows}
We need to be able to reason about which tuples (and with which multiplicity) belong certainly to the window for a tuple and which tuples (with which multiplicity) could possibly belong to a window.
For a tuple $\rangeTup$, \revm{we model the window's tuples as an \abbrAUDB relation $\awindowshorter{\rangeRel}{\rangeTup}$ where a tuple's lower bound multiplicity encodes the number of duplicates of the tuple that are certainty in the window, the selected-guess multiplicity encodes the multiplicity of the tuple in the selected-guess world, and the upper bound encodes the largest possible multiplicity with which the tuple may occur in the window minus the certain multiplicity. In the remainder of this paper we omit the definition of the select-guess, because it can be computed using the deterministic semantics for windowed aggregation. For completeness, we include it in the extended version of this paper~\cite{techreport}.
  We formally define $\awindowshorter{\rangeRel}{\rangeTup}$ in \Cref{fig:certain-and-poss-row-based-audb-win}.} \revm{Recall that in the first step we used sort to split the duplicates of each tuple into tuples with multiplicity upper bound of $1$. Thus, the windows we are constructing here are for tuples instead of for individual duplicates of a tuple.} A tuple $\rangeOf{t'}$ is guaranteed to belong to the window for of a tuple $\rangeTup$ with a multiplicity of $n = \lbMarker{\rangeRel(\rangeOf{\t'})}$ (the number of duplicates of the tuple that certainly exist)
if the tuple certainly belongs to the partition for $\rangeTup$ and all possible positions that these $n$ duplicates of the tuple occupy in the sort order are guaranteed to be contained in the smallest possible interval of sort positions contained in the bounds of the window for $\rangeTup$. Tuple $\rangeOf{t'}$ possibly belongs to the window 
of $\rangeTup$ if any of its possible positions falls within the interval of all possible positions 
of $\rangeTup$. \revm{As an example consider \Cref{fig:possible-and-certain-wind} which shows the sort positions that certainly (red) and possibly (green) belong to tuple $\rangeTup$'s window (window bounds [-1,4]). For any window $[l,u]$, sort positions certainly covered by the window start from latest possible starting position for $\rangeTup$'s window which is $\ubMarker{\rangeTup.\ranka} + l$ ($6 + (-1) = 5$ in our example) and end at the earliest possible upper bound for the window which is
  $\lbMarker{\rangeTup.\ranka} +u$ ($4 + 4 = 8$ in our example).
  Furthermore, \Cref{fig:possible-and-certain-wind} shows the membership of three tuples in the window. Tuple $\rangeOf{t_1}$ does certainly not belong to the window, because none of its possible sort positions are in the window's set of possible sort positions, $\rangeOf{t_{2}}$ does certainly belong to the window, because all of its possible sort positions are in the set of positions certainly in the window. Finally, $\rangeOf{t_{3}}$ possibly belongs to the window, because some of its sort positions are in the set of positions possibly covered by the window.}

\begin{figure}[t]
  \centering
  \scalebox{0.9}{
  \begin{tikzpicture}
    [
    edg/.style={->,line width=0.4mm},
    ]
    \def\xscaler{0.5}
    \def\yscaler{0.4}
    \def\axispos{-5.5 * \yscaler}

    \draw[|->, thick] (-0,\axispos) -- (13 * \xscaler,\axispos);

    \foreach \x in {0,...,12}
        \draw[thick] (\x * \xscaler,0.3 * \yscaler + \axispos) -- (\x * \xscaler,-0.3 * \yscaler + \axispos) node[below] {\x};

		\draw[thick,blue,|-|]   (4 * \xscaler,-1 * \yscaler * 1)  node[left,black]{} -- node[above,black]{\rangeTup} (6 * \xscaler,-1 * \yscaler * 1)
    ;

        \draw[edg,darkgreen,|-|] (3  * \xscaler, -1 * \yscaler * 2) -- node[below,black]{\textcolor{darkgreen}{possible window}} (10  * \xscaler, -1 * \yscaler * 2);

    \draw[edg,red,|-|] (5  * \xscaler, -1 * \yscaler * 2) -- node[above,black]{\textcolor{red}{certain window}} (8  * \xscaler, -1 * \yscaler * 2);

    \foreach \b/\s/\pos/\expl/\tup in {0/2/3/not in window/$\rangeOf{t_1}$,5/7/4/certainly in/$\rangeOf{t_2}$,9/12/4/possibly in/$\rangeOf{t_3}$}
    \draw[thick,blue,|-|]   (\b * \xscaler,-1 * \yscaler * \pos)  node[left,black]{} -- node[below,black]{\expl} node[above,black]{\tup} (\s * \xscaler,-1 * \yscaler * \pos) node[right,black]{}
    ;
  \end{tikzpicture}
  }
  \vspace{-10pt}
  \setlength{\belowcaptionskip}{-15pt}
  \caption{Possible and certain window membership of tuples in the window [-1,4] for $\rangeTup$ based on their possible sort positions.}\label{fig:possible-and-certain-wind}
\end{figure}


\begin{figure*}[t]
  \centering
  \begin{align*}
    \lbMarker{\awindowshorter{\rangeRel}{\rangeTup}(\rangeOf{t'})} & =
              \begin{cases}
		\revb{\lbMarker{\shortpart{\rangeTup}{\rangeRel}(\rangeOf{t'})}} & \mathif \revb{[\lbMarker{\tpos{\shortpart{\rangeTup}{\rangeRel}}{\sortattrs}{\rangeOf{t'}}{0}}, \ubMarker{\tpos{\shortpart{\rangeTup}{\rangeRel}}{\sortattrs}{\rangeOf{t'}}{0}}] \subseteq [\ubMarker{\tpos{\shortpart{\rangeTup}{\rangeRel}}{\sortattrs}{\rangeTup}{0}} +l, \lbMarker{\tpos{\shortpart{\rangeTup}{\rangeRel}}{\sortattrs}{\rangeTup}{0}} + u] } \\
                      0 &\mathotherwise \\
              \end{cases}\\
    \ubMarker{\awindowshorter{\rangeRel}{\rangeTup}(\rangeOf{t'})} & =
                             \begin{cases}
          \revb{\ubMarker{\shortpart{\rangeTup}{\rangeRel}(\rangeOf{t'})} -  \lbMarker{\awindowshort{\rangeRel}{\rangeTup}{i}(\rangeOf{t'})}}
           &\mathif \revb{([\lbMarker{\tpos{\shortpart{\rangeTup}{\rangeRel}}{\sortattrs}{\rangeOf{t'}}{0}}, \ubMarker{\tpos{\shortpart{\rangeTup}{\rangeRel}}{\sortattrs}{\rangeOf{t'}}{0}}]
             \cap [\lbMarker{\tpos{\shortpart{\rangeTup}{\rangeRel}}{\sortattrs}{\rangeTup}{0}} +l, \ubMarker{\tpos{\shortpart{\rangeTup}{\rangeRel}}{\sortattrs}{\rangeTup}{0}} + u]) \neq \emptyset } \\
                                           0 &\mathotherwise\\
                                         \end{cases}
   \iftechreport{
   \\
   \bgMarker{\awindowshorter{\rangeRel}{\rangeTup}(\rangeOf{t'})} & =
                             \begin{cases}
          \bgMarker{\shortpart{\rangeTup}{\rangeRel}(\rangeOf{t'})}           &\mathif \bgMarker{\tpos{\shortpart{\rangeTup}{\rangeRel}}{\sortattrs}{\rangeOf{t'}}{0}}
             \subseteq [\bgMarker{\tpos{\shortpart{\rangeTup}{\rangeRel}}{\sortattrs}{\rangeTup}{0}} +l, \bgMarker{\tpos{\shortpart{\rangeTup}{\rangeRel}}{\sortattrs}{\rangeTup}{0}} + u] \\
                                           0 &\mathotherwise\\
                                         \end{cases}
   }
  \end{align*}
  \vspace*{-5mm}
  \setlength{\belowcaptionskip}{-10pt}
  \caption{Certain and possible window membership for row-based windowed aggregation over \abbrAUDBs}
  \label{fig:certain-and-poss-row-based-audb-win}
\end{figure*}
\mypar{Combining and Filtering Certain and Possible Windows}
As mentioned above, row-based windows contain a fixed maximal number of tuples based on their bounds. We use $\awsize$ to denote the size of a window with bounds $[l,u]$, i.e., $\awsize = (u - l) + 1$. This limit on the number of tuples in a window should be taken into account when computing bounds on the result of an aggregation function.
For that, we combine the tuples certainly in the window
(say there are $m$ such tuples) with a selected bag of up to $\awsize - m$ rows possibly in the window that  minimizes (for the lower aggregation result bound) or maximizes (for the upper aggregation result bound) the aggregation function result for an input tuple. Let us use $\certaincount{\rangeRel, \rangeTup}$ to denote $\awsize - m$:

\[
  \certaincount{\rangeRel,\rangeTup} = \awsize - \sum_{\rangeOf{t'}}{\lbMarker{\awindowshorter{\rangeRel}{\rangeTup}(\rangeOf{t'})}}
\]

Which bag of up to $\certaincount{\rangeRel, \rangeTup}$ tuples minimizes / maximizes the aggregation result depends on what aggregation function is applied.
For \aggsum, the up to $\certaincount{\rangeRel, \rangeTup}$ rows with the smallest negative values are included in the lower bound and the up to $\certaincount{\rangeRel,\rangeTup}$ rows with the greatest positive values for the upper bound.
For \aggcount no additional row are included for the lower bound and up to \revm{$\certaincount{\rangeRel}$} rows for the upper bound.

\revm{ For each tuple $\rangeTup$, we define \abbrAUDB relation $\restrwinishorter{\rangeRel}{\rangeTup}$ where each tuple's lower/upper bound multiplicities encode the multiplicity of this tuple contributing to the lower and upper bound aggregation result, respectively.
  We only show the definition for \aggsum, the definitions for other aggregation functions are similar. In the definition, we make use $\lbMarker{\rangeRel}$ and $\ubMarker{\rangeRel}$:}
\revm{
\begin{align*}
  \rangeRel^{\downarrow}(\rangeTup) &= \rangeRel(\rangeTup)^{\downarrow} &
                                                                           \rangeRel^{\uparrow}(\rangeTup) &= \rangeRel(\rangeTup)^{\uparrow}
\end{align*}
}

Note that $\rangeRel^{\downarrow}$ and $\rangeRel^{\uparrow}$ are bags ($\semN$-relations) over range-annotated tuples.
Furthermore, we define $\winbotk{\rangeRel, \rangeTup}{\agga}$ (and $\wintopk{\rangeRel, \rangeTup}{\agga}$)  that are computed by restricting $\awindowshorter{\rangeRel}{\rangeTup}$ to the tuples with the smallest negative values (largest positive values) as lower (upper) bounds  on attribute $\agga$ that could contribute to the aggregation, keeping tuples with a total multiplicity of up to $\certaincount{\rangeRel,\rangeTup}$.
Note that
the deterministic conditions / expressions in the definition of $\winbotk{\rangeRel, \rangeTup}{\agga}$ (and $\wintopk{\rangeRel, \rangeTup}{\agga}$) are well-defined, because single values are extracted from all range-annotated values.
For \aggmax (resp., \aggmin) and similar idempotent aggregates, it suffices to know the greatest (resp., least) value possibly in the window.
\vspace{-10pt}

\begin{align*}
  \lbMarker{\restrwinishorter{\rangeRel}{\rangeTup}(\rangeOf{t'})} &=                                                   \lbMarker{\awindowshorter{\rangeRel}{\rangeTup}(\rangeOf{t'})} +\winbotk{\rangeRel,\rangeTup}{\agga}(\rangeOf{t'}) \\
	\ubMarker{\restrwinishorter{\rangeRel}{\rangeTup}(\rangeOf{t'})} &=
	\lbMarker{\awindowshorter{\rangeRel}{\rangeTup}(\rangeOf{t'})} +
	\wintopk{\rangeRel,\rangeTup}{\agga}(\rangeOf{t'})\\
	\iftechreport{
	 \bgMarker{\restrwinishorter{\rangeRel}{\rangeTup}(\rangeOf{t'})} &= \bgMarker{\awindowshorter{\rangeRel}{\rangeTup}(\rangeOf{t'})}                                                     \\
	}
\winbotk{\rangeRel, \rangeTup}{\agga} & = \selection_{\ranka<\certaincount{\rangeRel,\rangeTup}}(\rank{\lbMarker{\agga}}{\ranka}(\selection_{\lbMarker{\agga}<0}(\lbMarker{\awindowshorter{\rangeRel}{\rangeTup}}))) \\
  \wintopk{\rangeRel, \rangeTup}{\agga} & = \selection_{\ranka<\certaincount{\rangeRel, \rangeTup}}(\rank{- \ubMarker{\agga}}{\ranka}(\selection_{\ubMarker{\agga}>0}(\ubMarker{\awindowshorter{\rangeRel}{\rangeTup}}))) \\
\end{align*}

\vspace{-15pt}

\mypar{Windowed Aggregation}
Using the filtered combined windows we are ready to define row-based windowed aggregation over \abbrAUDBs.
\BG{In the definition we use $\bgMarker{\rangeRel}$ to denote the select-guess world encoded by $\rangeRel$, i.e., replacing each uncertain attribute value $\uv{c_1}{c_2}{c_3}$ with $c_2$ and tuple annotation $(n_1,n_2,n_3)$ with $n_2$.}
To compute aggregation results, we utilize the operation $\mysmbNAU{f}$ defined in \cite{FH21} for aggregation function $f$ that combines the range-annotated aggregation attribute value of a tuple  with the tuple's multiplicity bounds. For instance, for \aggsum, $\mysmbNAU{\aggsum}$ is multiplication, e.g., if a tuple with $\agga$ value $\uv{10}{20}{30}$ has multiplicity $\ut{1}{2}{3}$ it contributes $\uv{10}{40}{90}$ to the sum. Here, $\bigoplus$ denotes the application of the aggregation function over a set of elements (e.g., $\sum$ for \aggsum). Note that, as explained above, the purpose of $\flattern{\rangeRel}$ is to split a tuple with $n$ possible duplicates  into $n$ tuples with a multiplicity of $1$. Furthermore, note that the bounds on the aggregation result may be the same for the $i^{th}$ and $j^{th}$ duplicate of a tuple. To deal with that we apply a final projection to merge such duplicate result tuples.

\begin{Definition}[Row-based Windowed Aggregation]\label{def:row-based-windowed-a}
    Let $\rangeRel$ be an \abbrAUDB relation. We define window operator $\awinrow$ as:
  {
  \revm{\begin{align*}
   \awinrow(\rangeRel)(\rangeTup) &= \projection_{\schemaOf(\rangeRel),X}(\wrowresultshort(\rangeRel))\\
    \wrowresultshort(\rangeRel)(\rangeTup \circ  \textendaggishorter{\rangeTup}) &= \flattern{\rangeRel}(\rangeTup) \\
    \textendaggishorter{\rangeTup} &= \bigoplus_{\rangeOf{t'}} \rangeOf{t'}.\agga \mysmbNAU{f} \restrwinishorter{\flattern{\rangeRel}}{\rangeTup}(\rangeOf{t'}) \\
    \flattern{\rangeRel} & = \projection_{\schemaOf(\rangeRel),\dummySortAttr}(\rank{\schemaOf(\rangeOf{R})}{\dummySortAttr}(\rangeRel)
                           )
  \end{align*}}}
\end{Definition}
\vspace{-2mm}
\begin{Example}[\abbrAUDB Windowed Aggregation]\label{ex:audbwindow}
Consider the \abbrAUDB relation $\rangeRel$ shown below and query $\winrow{sum(C)}{SumA}{A}{B}{-1}{0}(\rangeRel)$, i.e.,  
windowed aggregation partitioning by $A$, ordering on $B$, and computing $sum(C)$ over windows including $1$ preceding and the current row. For convenience we show an identifier for each tuple on the left. As mentioned above, we first expand each tuple with a possible multiplicity larger then one using sorting. Consider tuple $\rangeOf{t_3}$. Both $\rangeOf{t_1}$ and $\rangeOf{t_2}$ may belong to the same partition as $\rangeOf{t_3}$ as their $A$ value ranges overlap. There is no tuple that certainly belongs to the same partition as $\rangeOf{t_3}$. Thus, only tuple $\rangeOf{t_{3}}$ itself will certainly belong to the window. To compute the bounds on the aggregation result we first determine which tuples (in the expansion created through sorting) may belong to the window for $\rangeOf{t_{3}}$. These are the two tuples corresponding to the duplicates of $\rangeOf{t_1}$, because these tuples may belong to the partition for $\rangeOf{t_{3}}$ and their possible sort positions ($\uv{0}{0}{1}$ and $\uv{1}{1}{2}$) overlap with the sort positions possibly covered by the window for $\rangeOf{t_{3}}$ ($\uv{0}{1}{2}$). Since the size of the window is 2 tuples, the bounds on the sum are computed using the lower / upper bound on the $C$ value of $\rangeOf{t_{3}}$ ($\uv{2}{4}{5}$) and no additional tuple from the possible window (because the $C$ value of $\rangeOf{t_{1}}$ is positive) for the lower bound and the largest possible $C$ value of one copy (we can only fit one additional tuple into the window) of $\rangeOf{t_1}$ ($7$) for the upper bound. Thus, we get the aggregation result $\uv{2}{11}{12}$ as shown below.
\vspace{-3mm}

  \begin{minipage}{0.99\linewidth}
  \centering
  \scalebox{0.8}{
    \begin{tabular}{cc|c|cc}
      &\thead{A} & \thead{B} & \thead{C} & $\semN^3$ \\
      \cline{2-4}
      $\rangeOf{t_1}$&    1 & \uv{1}{1}{3} & 7 & (1,1,2)\\
      $\rangeOf{t_2}$&    \uv{2}{3}{3} & 15 & 4 & (0,1,1) \\
      $\rangeOf{t_3}$&      \uv{1}{1}{2} & 2 & \uv{2}{4}{5} & 1\\
    \end{tabular}
  }
  \end{minipage}

  \begin{minipage}{0.99\linewidth}
    \centering
    \scalebox{0.8}{
    \begin{tabular}{cc|c|c|cc}
      &\thead{A} & \thead{B} & \thead{C} & \thead{SumC} & $\semN^3$ \\
      \cline{2-5}
      $\rangeOf{r_1}$&      1 & \uv{1}{1}{3} & 7 & \uv{7}{7}{14}  & 1\\
      $\rangeOf{r_2}$&      1 & \uv{1}{1}{3} & 7 & \uv{7}{7}{14}  & (0,0,1)\\
      $\rangeOf{r_3}$&      \uv{1}{1}{2} & 2 & \uv{2}{4}{5} & \uv{2}{11}{12} & 1\\
      $\rangeOf{r_4}$&      \uv{2}{3}{3} & 15 & 4 & \uv{4}{4}{9} & (0,1,1)\\
     \end{tabular}
     }
  \end{minipage}
\end{Example}
\subsection{Bound Preservation}\label{sec:audb-windowed-agg-bound-preservation}

We now prove this semantics for group-based and row-based windowed aggregation over \abbrAUDBs to be bound preserving.

\begin{Theorem}[Bound Preservation for Windowed Aggregation]\label{theo:bound-preservation-f}
Consider an \abbrAUDB relation $\rangeRel$ and incomplete bag relation $\prel$ 
such that $\prel \dbbounds \rangeRel$, and $\sortattrs \subseteq \schemaOf(\prel)$. For any 
row-based windowed aggregation $\awinrow$, we have:
\begin{align*}
\awinrow(\prel) &\dbbounds \awinrow(\rangeRel)
\end{align*}\BG{If we nee d a shorter version, we can replace this with ``Group-based and row-based Windowed aggregation over \abbrAUDBs is bound preserving.}
\end{Theorem}
\detailedproof{ %
  As in the proof for sorting over \abbrAUDBs, we consider WLOG one of the possible worlds $\rela \in \prel$ and a tuple matching $\TM$ based on which $\rangeRel$ is bounding $\rela$. We then construct a tuple matching $\TM'$ for the output of windowed aggregation. In the proof, we utilize the fact that windowed aggregation produces one output tuple $\tup$ for each input tuple $\tup'$ such that $\tup$ extends the input tuple $\tup'$ with the aggregation result for $\tup'$'s window and  has the same multiplicity as the input tuple $\tup'$. Thus, we only need to show that the bounds on the aggregation function result bound the values in the result for the possible world $\rela$ and that tuples with multiplicity $n$ are split into $n$ output tuples with multiplicity $1$. 
}{ %
	For simplicity, we denote $\prel_o$ and $\rangeRel_o$ as the input incomplete bag relation and input \abbrAUDB relation s.t. $\prel_o  \dbbounds \rangeRel_o$. We use $\prel$ and $\rangeRel$ as the output of $\flattern{}$ s.t. $\prel \defas \flattern{prel_o}$ and $\rangeRel \defas \flattern{\rangeRel_o}$. By By \cite{FH21} with selection bounding proven and \Cref{theo:bound-preservation-o} with sort bounding proven, we have $\prel \dbbounds \rangeRel$. WLOG, since $\flattern{\rangeRel}$ return relation with multiplicity of 1s, we use $\rangeTup \in \rangeRel$ and $\tup \in \rela \in \prel$ s.t. $\TM_{\rela}(\rangeTup,\tup)=1$. By \cite{FH21}, we have $\shortpart{\rangeTup}{\rangeRel} \dbbounds \shortpart{\tup}{\rangeRel}$, so there exist valid tuple matching $\TM_{\mathcal{P}}$ bounds the output of $\shortpart{\tup}{\rangeRel}$ and $\shortpart{\rangeTup}{\rangeRel}$.\\
	For $\awindowshorter{\rangeRel}{\rangeTup}$ computing uncertain window for $\rangeTup$ and $\awindowshorter{\rela}{\tup}$ computing deterministic window for $\tup$, we first recast the definition of the \abbrAUDB window for $\rangeRel$ and the window for $\tup$ as a selection query. Define conditions,
	\begin{align*}
	\theta & \defas \tpos{\shortpart{\tup}{\rela}}{\sortattrs}{t'}{0} \subseteq \\ & [\tpos{\shortpart{\tup}{\rela}}{\sortattrs}{t}{0} +l, \tpos{\shortpart{\tup}{\rela}}{\sortattrs}{t}{0} + u] \\
	\ubMarker{\boldsymbol\theta} & \defas ([\lbMarker{\tpos{\shortpart{\rangeTup}{\rangeRel}}{\sortattrs}{\rangeOf{t'}}{0}}, \ubMarker{\tpos{\shortpart{\rangeTup}{\rangeRel}}{\sortattrs}{\rangeOf{t'}}{0}}]
             \\ & \cap [\lbMarker{\tpos{\shortpart{\rangeTup}{\rangeRel}}{\sortattrs}{\rangeTup}{0}} +l, \ubMarker{\tpos{\shortpart{\rangeTup}{\rangeRel}}{\sortattrs}{\rangeTup}{0}} + u]) \neq \emptyset \\
	\lbMarker{\boldsymbol\theta} & \defas [\lbMarker{\tpos{\shortpart{\rangeTup}{\rangeRel}}{\sortattrs}{\rangeOf{t'}}{0}}, \ubMarker{\tpos{\shortpart{\rangeTup}{\rangeRel}}{\sortattrs}{\rangeOf{t'}}{0}}] \subseteq \\ & \hspace{2mm} [\ubMarker{\tpos{\shortpart{\rangeTup}{\rangeRel}}{\sortattrs}{\rangeTup}{0}} +l, \lbMarker{\tpos{\shortpart{\rangeTup}{\rangeRel}}{\sortattrs}{\rangeTup}{0}} + u]
	\end{align*}
	By rule of uncertain scalar evaluation, $\theta \tmatch [\lbMarker{\boldsymbol\theta},\ubMarker{\boldsymbol\theta}]$, so deterministic window candidate condition $\theta$ bounded by $\lbMarker{\boldsymbol\theta}$ and $\ubMarker{\boldsymbol\theta}$.
	By \cite{FH21}, we have
	\begin{align*}
		\lbMarker{\awindowshorter{\rangeRel}{\rangeTup}(\rangeOf{t'})}= & \lbMarker{\shortpart{\rangeTup}{\rangeRel}(\rangeOf{t'})} \multK \lbMarker{\boldsymbol\theta} \\
		= & \selection_{\boldsymbol\theta}(\shortpart{\rangeTup}{\rangeRel})\lbMarker{(\rangeOf{t'})}
	\end{align*}
	So $\lbMarker{\awindowshorter{\rangeRel}{\rangeTup}}$ lower bounds $\awindowshorter{\rela}{\tup}$ and there exists valid tuple matching $\TM_{\mathcal{W}}$. \\
	analog to lower bound, 
	\begin{align*}
		\ubMarker{\awindowshorter{\rangeRel}{\rangeTup}(\rangeOf{t'})}+\ubMarker{\awindowshorter{\rangeRel}{\rangeTup}(\rangeOf{t'})}= & \ubMarker{\shortpart{\rangeTup}{\rangeRel}(\rangeOf{t'})} \multK \lbMarker{\boldsymbol\theta} \\
		= & \selection_{\boldsymbol\theta}(\shortpart{\rangeTup}{\rangeRel})\lbMarker{(\rangeOf{t'})}
	\end{align*} 
	So $\ubMarker{\awindowshorter{\rangeRel}{\rangeTup}(\rangeOf{t'})}+\ubMarker{\awindowshorter{\rangeRel}{\rangeTup}(\rangeOf{t'})}$ upper bounds $\awindowshorter{\rela}{\tup}$. Put differently, $\ubMarker{\awindowshorter{\rangeRel}{\rangeTup}}$ upper bounds $\awindowshorter{\rela}{\tup}-\lbMarker{\awindowshorter{\rangeRel}{\rangeTup}}$\\
	Define possible window matched tuples $\pwsymb_{\rela,\tup}(\tup')=\awindowshorter{\rela}{\tup}(t')-\sum_{t'}\TM_{\mathcal{W}}(\rangeTup',t')$, by definition of tuple matching, $\forall_{\rangeTup'}:\sum_{t'}\TM_{\mathcal{W}}(\rangeTup',t') \geq \lbMarker{\awindowshorter{\rangeRel}{\rangeTup}}(\rangeTup')$ we get $\awindowshorter{\rela}{\tup}-\lbMarker{\awindowshorter{\rangeRel}{\rangeTup}}$ upper bounds $\pwsymb_{\rela,\tup}$, so $\ubMarker{\awindowshorter{\rangeRel}{\rangeTup}}$ upper bounds $\pwsymb_{\rela,\tup}$\\
	For $\tup$, defined by \Cref{sec:windowed-aggregation}, we know that
	$$\textendaggishorter{\tup}=\bigoplus_{t'} t'.\agga \smpair_{f} \awindowshorter{\rela}{\tup}(t')$$
	For $\rangeTup$, substituting based on \Cref{def:row-based-windowed-a} we get
	$$\lbMarker{\textendaggishorter{\rangeTup}} = \lbMarker{\left(\bigoplus_{\rangeOf{t'}} \rangeOf{t'}.\agga \mysmbNAU{f} \restrwinishorter{\rangeRel}{\rangeTup}(\rangeOf{t'})\right)}$$
	Since $\bigoplus$ is point-wise for $\aggmin$, $\aggmax$ and $\aggsum$, 
	$$= \bigoplus_{\rangeOf{t'}}\lbMarker{(\rangeOf{t'}.\agga \mysmbNAU{f} \restrwinishorter{\rangeRel}{\rangeTup}(\rangeOf{t'}))}$$
	\begin{align*}
	= \bigoplus_{\rangeOf{t'}} min(\lbMarker{\rangeOf{t'}.\agga} \smpair_{f} \lbMarker{\restrwinishorter{\rangeRel}{\rangeTup}(\rangeOf{t'})}, & \ubMarker{\rangeOf{t'}.\agga} \smpair_{f} \lbMarker{\restrwinishorter{\rangeRel}{\rangeTup}(\rangeOf{t'})}, \\ & \lbMarker{\rangeOf{t'}.\agga} \smpair_{f} \ubMarker{\restrwinishorter{\rangeRel}{\rangeTup}(\rangeOf{t'})},\ubMarker{\rangeOf{t'}.\agga} \smpair_{f} \ubMarker{\restrwinishorter{\rangeRel}{\rangeTup}(\rangeOf{t'})})
	\end{align*}
	Since $\lbMarker{\restrwinishorter{\rangeRel}{\rangeTup}}$ is constructed to minimize $\rangeTup'.A$, 
	$$= \bigoplus_{\rangeOf{t'}} \lbMarker{\rangeOf{t'}.\agga} \smpair_{f} \lbMarker{\restrwinishorter{\rangeRel}{\rangeTup}(\rangeOf{t'})}$$
	By definition of $\lbMarker{\restrwinishorter{\rangeRel}{\rangeTup}(\rangeOf{t'})}$,
	$$ = \bigoplus_{\rangeOf{t'}} \lbMarker{\rangeOf{t'}.\agga} \smpair_{f} (\lbMarker{\awindowshorter{\rangeRel}{\rangeTup}(\rangeOf{t'})} +\winbotk{\rangeRel,\rangeTup}{\agga}(\rangeOf{t'})) $$
	By semimodule law \cite{FH21},
		$$ = \bigoplus_{\rangeOf{t'}} \lbMarker{\rangeOf{t'}.\agga} \smpair_{f}\lbMarker{\awindowshorter{\rangeRel}{\rangeTup}(\rangeOf{t'})} +_f \bigoplus_{\rangeOf{t'}} \lbMarker{\rangeOf{t'}.\agga} \smpair_{f} \winbotk{\rangeRel,\rangeTup}{\agga}(\rangeOf{t'})$$
	Since $t' \tmatch \rangeOf{t'}$ and $\lbMarker{\awindowshorter{\rangeRel}{\rangeTup}}$ lower bounds $\awindowshorter{\rela}{\tup}$, 
	$$	\leq \bigoplus_{t'} t'.\agga \smpair_{f} \sum_{t'}\TM_{\mathcal{W}}(\rangeOf{t'},t) +_f  \bigoplus_{\rangeOf{t'}} \lbMarker{\rangeOf{t'}.\agga} \smpair_{f} \winbotk{\rangeRel,\rangeTup}{\agga}(\rangeOf{t'})$$
	By using $k=\certaincount{\rangeRel,\rangeTup}$, we denote the deterministic top-k operation as \\ $\winbotk{\rangeRel,\rangeTup}{\agga}(\rangeTup') = mink_{\lbMarker{A}}(\ubMarker{\awindowshorter{\rangeRel}{\rangeTup}})(\rangeTup')$ and we get
	$$	= \bigoplus_{t'} t'.\agga \smpair_{f} \sum_{t'}\TM_{\mathcal{W}}(\rangeOf{t'},t) +_f  \bigoplus_{\rangeOf{t'}} \lbMarker{\rangeOf{t'}.\agga} \smpair_{f} mink_{\lbMarker{A}}(\ubMarker{\awindowshorter{\rangeRel}{\rangeTup}})(\rangeTup')$$
	Since $t' \tmatch \rangeTup$ and $\ubMarker{\awindowshorter{\rangeRel}{\rangeTup}}$ upper bounds $\pwsymb_{\rela,\tup}$, 
	$$	\leq \bigoplus_{t'} t'.\agga \smpair_{f} \sum_{t'}\TM_{\mathcal{W}}(\rangeOf{t'},t) +_f  \bigoplus_{t'} t'.\agga \smpair_{f} mink_{A}(\pwsymb_{\rela,\tup})(\tup)$$
	trivially,
	$$	\leq \bigoplus_{t'} t'.\agga \smpair_{f} \sum_{t'}\TM_{\mathcal{W}}(\rangeOf{t'},t) +_f  \bigoplus_{t'} t'.\agga \smpair_{f} \pwsymb_{\rela,\tup}(t')$$
	$$	= \bigoplus_{t'} t'.\agga \smpair_{f} \left(\sum_{t'}\TM_{\mathcal{W}}(\rangeOf{t'},t) + \awindowshorter{\rela}{\tup}(t')-\sum_{t'}\TM_{\mathcal{W}}(\rangeOf{t'},t)\right)$$
	$$ =  \bigoplus_{t'} t'.\agga \smpair_{f} \awindowshorter{\rela}{\tup}(t') = \textendaggishorter{\tup}$$
	So $\textendaggishorter{\rangeTup}$ lower bounds $\textendaggishorter{\tup}$. \\
	$\textendaggishorter{\rangeTup}$ upper bounds $\textendaggishorter{\tup}$ can be proven in analog way.\\
	Given that $\tup \tmatch \rangeTup$ and $\textendaggishorter{\tup} \tmatch \textendaggishorter{\rangeTup}$, we get
	$$\TM_{res}(\rangeTup \circ \textendaggishorter{\rangeTup}, \tup \circ \textendaggishorter{\tup})=\TM_{\rangeRel}(\rangeTup,\tup)$$
	As we made no assumptions about $\rangeTup$ and $\tup$ apart from $\TM_{\rangeRel}(\rangeTup,\tup) = 1$, this implies that $\TM_{res}$ is a tuple matching and, thus, this concludes the proof.
}
\section{SQL-based Implementation}
\label{sec:sql-based-operator}

We now present a rewrite-based implementation of the bound-preserving sorting and windowed aggregation semantics presented above as relational algebra expressions.
Specifically, we define a function \rewrUAA{\query} that translates a query  into an equivalent query that operates over a relational encoding of \abbrAUDBs.
Using the encoding of \abbrAUDBs from~\cite{FH21}, each uncertain attribute is encoded as a triple of selected guess and lower and upper bounds attributes. Furthermore, three additional attributes are used to store triples of row multiplicities $\uv{\lbMarker{\mult}}{\mult}{\ubMarker{\mult}}$. 

\newcommand{\sqlordisend}{\textsf{isend}}
\newcommand{\sqlordqlow}{\query_{lower}}
\newcommand{\sqlordqupper}{\query_{upper}}
\newcommand{\sqlordqsg}{\query_{sg}}
\newcommand{\sqlordqbounds}{\query_{bounds}}
\begin{figure}
\ifnottechreport{
{\small
	\begin{align*}
	  \rewrUAA{\rank{\sortattrs}{\ranka}(\query)} & \defas \gamma_{
                                                    \schemaOf(\rewrUAA{\query})
                                                    ,e_{pos}}(\sqlordqbounds)
	  \\
	  e_{pos} & \defas \aggsum(\lbMarker{pos}) \renameto \lbMarker{\ranka}, \aggsum(\ubMarker{pos}) \renameto \ubMarker{\ranka}
	  \\
	 \sqlordqbounds             & \defas  \omega^{[-\infty,0]}_{\lbMarker{e} \renameto \lbMarker{pos}, \ubMarker{e} \renameto \ubMarker{pos}; \emptyset; pt}(\sqlordqlow \union \sqlordqupper)
	 \\
	 \lbMarker{e}                     & \defas \ifte{\sqlordisend=0}{\aggsum(\sqlordisend \cdot \lbMarker{\mult})}{0}
	 \\
	 \ubMarker{e}                     & \defas \ifte{\sqlordisend=1}{\aggsum((1-\sqlordisend) \cdot \ubMarker{\mult})}{0}
	 \\
	 \sqlordqlow & \defas \projection_{\schemaOf(\rewrUAA{\query}), \lbMarker{\sortattrs} \renameto pt, 0 \renameto \sqlordisend}(\rewrUAA{Q})
	 \\
	 \sqlordqupper & \defas \projection_{\schemaOf(\rewrUAA{\query}), \ubMarker{\sortattrs} \renameto pt, 1 \renameto \sqlordisend}(\rewrUAA{Q})
	\end{align*}
}\\[-4mm]
} %
\iftechreport{
{\small
	\begin{align*}
	  \rewrUAA{\rank{\sortattrs}{\ranka}(\query)} & \defas \gamma_{
                                                    \schemaOf(\rewrUAA{\query})
                                                    ,e_{pos}}(\sqlordqbounds)
	  \\
	  e_{pos} & \defas \aggsum(\lbMarker{pos}) \renameto \lbMarker{\ranka}, \aggsum(\bgMarker{pos}) \renameto \bgMarker{\ranka}, \aggsum(\ubMarker{pos}) \renameto \ubMarker{\ranka}
	  \\
	 \sqlordqbounds             & \defas  \omega^{[-\infty,0]}_{\lbMarker{e} \renameto \lbMarker{pos}, \bgMarker{e} \renameto \bgMarker{pos}, \ubMarker{e} \renameto \ubMarker{pos}; \emptyset; pt}(\sqlordqlow \union \sqlordqupper \union \sqlordqsg)
	 \\
	 \lbMarker{e}                     & \defas \ifte{\sqlordisend=0}{\aggsum(\sqlordisend \cdot \lbMarker{\mult})}{0}
	 \\
	 \bgMarker{e}                     & \defas \ifte{\sqlordisend=-1}{\aggsum(\bgMarker{\mult})}{0}
	 \\
	 \ubMarker{e}                     & \defas \ifte{\sqlordisend=1}{\aggsum((1-\sqlordisend) \cdot \ubMarker{\mult})}{0}
	 \\
	 \sqlordqlow & \defas \projection_{\schemaOf(\rewrUAA{\query}), \lbMarker{\sortattrs} \renameto pt, 0 \renameto \sqlordisend}(\rewrUAA{Q})
     \sqlordqsg &\defas \projection_{\schemaOf(\rewrUAA{\query}), \bgMarker{\sortattrs} \renameto pt, -1 \renameto \sqlordisend}(\rewrUAA{Q})
	 \\
	 \sqlordqupper & \defas \projection_{\schemaOf(\rewrUAA{\query}), \ubMarker{\sortattrs} \renameto pt, 1 \renameto \sqlordisend}(\rewrUAA{Q})
	\end{align*}
}
}
\caption{Rewriting the sorting operator; The selected-guess position computation is omitted for clarity.}
\label{fig:orderByRewrite}
\end{figure}

\subsection{Rewriting Sorting Queries}\label{sec:sql-based-order}
The query rewrite rule for our ordering operator is shown in \Cref{fig:orderByRewrite}:
This approach creates two copies of the input relation: $\sqlordqlow$ containing the lower bounds of the sort attributes, and $\sqlordqupper$ with the upper bounds. 
These tuples represent endpoints of value ranges for $\sortattrs$, and so we refer to the tuples of the former as start tuples, and the latter as end tuples. 
The lower bound on the tuple's position wrt. $\afles$ is the number of tuples that certainly precede it: For a given start tuple, this is the total certain multiplicity ($\lbMarker{\mult}$) of end tuples that appear before it.
The upper bound is computed similarly from the total possible multiplicity ($\ubMarker{\mult}$) of start tuples that precede an end tuple.  
$\sqlordqbounds$ computes these values for each start and end tuple using windowed aggregation to find all end tuple / start tuples that precede a point and then sum up their certain / possible multiplicity. Note that we use windowed aggregation with more than one aggregation function here which can be expressed as two windowed aggregation operators in our formalism. The resulting start tuples store  lower bounds (resp.,  upper bounds for end tuples). The final rewrite is obtained by merging the start and end tuples back together using a group-by aggregate.
The selected-guess position, not shown above, is computed analogously using a second window specification as part of $\sqlordqbounds$ (see \cite{techreport} for the full version).

\subsection{Rewriting Windowed Aggregation Queries}
\label{subsec:sql-based-window}

\newcommand{\wsqlqlower}{\ensuremath{\query_{\lbMarker{pos}}}}
\newcommand{\wsqlqupper}{\ensuremath{\query_{\ubMarker{pos}}}}
\newcommand{\wsqlqbounds}{\ensuremath{\query_{bnds}}}
\newcommand{\wsqlqendpoints}{\ensuremath{\query_{endpoints}}}
\newcommand{\wsqlqpart}{\ensuremath{\query_{part}}}
\newcommand{\wsqlqpos}{\ensuremath{\query_{pos}}}
\newcommand{\wsqlqselfpos}{\ensuremath{\query_{withselfpos}}}
\newcommand{\wsqlqselffilter}{\ensuremath{\query_{selfpos}}}
\newcommand{\wsqlqposwin}{\ensuremath{\query_{winposs}}}
\newcommand{\wsqlqcertwin}{\ensuremath{\query_{markcert}}}
\newcommand{\wsqlqwinlower}{\ensuremath{\query_{agglower}}}
\newcommand{\wsqlqwinupper}{\ensuremath{\query_{aggupper}}}
\newcommand{\wsqlqwinfilter}{\ensuremath{\query_{inwindow}}}
\newcommand{\wsqlqresultbounds}{\ensuremath{\query_{aggbnds}}}

\newcommand{\wsqlselfpos}{\ensuremath{selfpos}}

\newcommand{\wsqlewposs}{\ensuremath{e_{isposs_w}}}
\newcommand{\wsqlewcert}{\ensuremath{e_{iscert_w}}}
\newcommand{\wsqlepartcert}{\ensuremath{e_{iscert_p}}}
\newcommand{\wsqlecert}{\ensuremath{e_{iscert}}}

\newcommand{\awinrowsum}{\winrow{\aggsum(A)}{X}{\gbAttrs}{\sortattrs}{l}{u}}

\begin{figure}
{\small
 	\begin{align*}
 \hspace*{-7mm}
 		\rewrUAA{\awinrowsum(Q)} \defas &\;
 			\projection_{\schemaOf(\wsqlqresultbounds),X}(\\ &\hspace{5mm}\wsqlqresultbounds \Join_{id_1=id} \awinrowsum(Q))
\end{align*}
\begin{align*} 		
		 \wsqlqresultbounds
      \defas &\; \gamma_{\schemaOf(\rewrUAA{Q});\,\aggsum(e_{lb}) \renameto \lbMarker{X},\aggsum(e_{ub}) \renameto \ubMarker{X}}(\wsqlqwinfilter)
		\\
		e_{lb} \defas &\; {\bf if}\,{
			\lbMarker{pos}<l+u \wedge (iscert \vee \lbMarker{\agga}<0)
		}\\&\;{\bf then}\,{
			(\ifte{\lbMarker{A} < 0}{ \lbMarker{\agga} \cdot \ubMarker{\mult_2}}{ \lbMarker{\agga} \cdot \lbMarker{\mult_2}})
		}\,{\bf else}\,{0}
		\\
		e_{ub} \defas &\; {\bf if}\,{
			\ubMarker{pos}<l+u \wedge (iscert \vee \ubMarker{\agga}>0)
		}\\&\;{\bf then}\,{
  (\ifte{\ubMarker{A} > 0}{ \ubMarker{\agga} \cdot \ubMarker{\mult_2}}{ \ubMarker{\agga} \cdot \lbMarker{\mult_2}})
		}\,{\bf else}\,{0}
		\\  \hline
		\wsqlqwinfilter \defas &\; \selection_{\lbMarker{pos}<l+u \;\lor\; \ubMarker{pos}<l+u}(\wsqlqwinupper)
		\\
      \wsqlqwinupper \defas & \winrow{count(*)}{\ubMarker{pos}}{id_1}{iscert,\ubMarker{A}}{-\infty}{0}(\wsqlqwinlower) \\
      \wsqlqwinlower \defas & \winrow{count(*)}{\lbMarker{pos}}{id_1}{iscert,\lbMarker{A}}{-\infty}{0}(\wsqlqcertwin)\\
 \hline
 \end{align*}
\begin{align*}
		\wsqlqcertwin \defas &\; \projection_{\schemaOf(\wsqlqposwin),\wsqlecert \renameto iscert}(\wsqlqposwin)
		\\
		\wsqlecert \defas &\; \wsqlepartcert \wedge \wsqlewcert
		\\
		\wsqlepartcert \defas &\; \lbMarker{\query_1.\gbAttrs}=\ubMarker{\query_1.\gbAttrs}=\lbMarker{\query_2.\gbAttrs}=\ubMarker{\query_2.\gbAttrs}
		\\
		\wsqlewcert \defas  &\; \lbMarker{\ranka} >= \ubMarker{\wsqlselfpos}-l \wedge \ubMarker{\ranka} <= \lbMarker{\wsqlselfpos}+u
		\\
		\wsqlqposwin \defas &\; \selection_{\wsqlewposs}(\wsqlqselfpos)
		\\
		\wsqlewposs \defas &\; \lbMarker{\wsqlselfpos}-l \leq \ubMarker{\ranka} \wedge \ubMarker{\wsqlselfpos}+u \geq \lbMarker{\ranka}
		\\
		\wsqlqselfpos \defas &\; \wsqlqpos \Join_{id_1=id} \wsqlqselffilter
		\\
		\wsqlqselffilter \defas &\; \projection_{id_1 \renameto id,\lbMarker{\ranka} \renameto \lbMarker{\wsqlselfpos},\ubMarker{\ranka} \renameto \ubMarker{\wsqlselfpos}}(\selection_{id_1=id_2}(\wsqlqpos)) \\ 
\hline 
\end{align*}
\begin{align*}
		\wsqlqpos \defas &\; \aggregation{\schemaOf(\wsqlqbounds)}{e_{pos}}(\wsqlqbounds)
		\\
		e_{pos} \defas &\; \aggsum(\lbMarker{pos}) \renameto \lbMarker{\ranka},\aggsum(\ubMarker{pos}) \renameto \ubMarker{\ranka}\\
		\wsqlqbounds \defas &\; \omega_{e_{c} \renameto \lbMarker{pos}, e_{p} \renameto \ubMarker{pos};id_1;pt}(\wsqlqendpoints)
		\\
	 	e_{c} \defas &\; \ifte{\sqlordisend=1}{\aggsum(\sqlordisend \cdot \lbMarker{\mult_2})}{0}
	 	\\
	 	e_{p} \defas &\; \ifte{\sqlordisend=1}{\aggsum((1-\sqlordisend) \cdot \ubMarker{\mult_2})}{0}
	 	\\
		\wsqlqendpoints \defas &\; \wsqlqlower \union \wsqlqupper
		\\
		\wsqlqlower \defas &\; \projection_{\schemaOf(\wsqlqpart), \lbMarker{\sortattrs} \renameto pt, 0 \renameto \sqlordisend}(\wsqlqpart)
		\\
		\wsqlqupper \defas &\; \projection_{\schemaOf(\wsqlqpart), \ubMarker{\sortattrs} \renameto pt, 1 \renameto \sqlordisend}(\wsqlqpart)
		\\
		\wsqlqpart \defas &\; \rho_{\{B_1 \gets B \mid B \in \schemaOf(\rewrUAA{\query}) \}} (\rewrUAA{\query}) \\  &\hspace{3mm}\Join_{\theta_{join}} \rho_{\{B_2 \gets B \mid B \in \schemaOf(\rewrUAA{\query}) \} }(\rewrUAA{\query})
		\\        
		\theta_{join} \defas &\; \lbMarker{\query_1.\gbAttrs} \leq \ubMarker{\query_2.\gbAttrs} \wedge \ubMarker{\query_1.\gbAttrs} \geq \lbMarker{\query_2.\gbAttrs}
	\end{align*}
} 
 	\caption{Rewrite rule for windowed aggregation operator}
 	\label{fig:windowAggregateRewrite}
 \end{figure}

The ranged window aggregation query rewriting rule is shown in \Cref{fig:windowAggregateRewrite}.
The method uses a range overlap self-join on partition-by attributes to link partition-definition tuple from $\query_1$ with potential members of the group from $\query_2$; The result is denoted $\query_{join}$. 
The relation $\query_{rank}$ is defined in a manner analogous to \Cref{fig:orderByRewrite}, assigning each tuple to a position within its partition.
$\query_{window}$ builds the window, first filtering out all tuples that certainly do not belong to the window, and then $\query_{cert}$ labels tupls with whether they definitely belong to the window.
Next $\query_{out}$ computes the window aggregate, and then the final rewritten query includes a join with a computation of the selected guess result.


\section{Native Algorithms}
\label{sec:native-algorithms-au}

\newcommand\mycommfont[1]{\footnotesize\ttfamily\textcolor{blue}{#1}}
\SetCommentSty{mycommfont}
\SetKwComment{tcc}{// }{}
\SetAlFnt{\small}
\newcommand{\calcaggbounds}{\texttt{compBounds}\xspace}

We now introduce optimized algorithms for ranking and windowed aggregation over UA-DBs that are more efficient than their SQL counterparts presented in~\cite{techreport}.
Through a \emph{connected heap} data structure, these algorithms leverage the fact that the lower and upper position bounds are typically close approximations of one another to avoid performing multiple passes over the data.
\revm{We assume a physical encoding of an \abbrAUDB relation $\rangeRel$ as a classical relation~\cite{FH21} where each range-annotated value of an attribute $A$ is stored as three attributes $\lbMarker{A}$, $\bgMarker{A}$, and $\ubMarker{A}$. In this encoding, attributes $\lbMarker{\rangeTup.\multa}$, $\bgMarker{\rangeTup.\multa}$, and $\ubMarker{\rangeTup.\multa}$ store the tuple's multiplicity bounds.}


\subsection{Non-deterministic Sort, Top-k}\label{sec:nondet-sort-topk}
\Cref{alg:topkOnepass} sorts an input \abbrAUDB $\rangeRel$. 
The algorithm assigns to each tuple its position $\ranka$
given as lower and upper bounds: $\texttt{t}.\lbMarker{\ranka}, \texttt{t}.\ubMarker{\ranka}$, respectively\footnote{
  The selected guess position $\bgMarker{\ranka}$ is trivially obtained using an additional linked heap, and omitted here for clarity.
}.
Given a parameter $\texttt{k}$, the algorithm can also be used to find the top-\texttt{k} elements; otherwise we set $\texttt{k} = \revm{\card{R}}$ (the size of the input relation).
\revm{\Cref{alg:topkOnepass} \iftechreport{(\Cref{fig:sortOnePassLifecycle})} takes as input the relational encoding of an \abbrAUDB relation $\rangeRel$ sorted on $\lbMarker{\sortattrs}$, the lower-bound of the sort order attributes.
Recall from \Cref{eq:upos-lb} that to determine a lower bound on the sort position of a tuple $\rangeTup$ we have to sum up the smallest multiplicity of tuples $\rangeOf{s}$ that are certainly sorted before $\rangeTup$, i.e., where $\ubMarker{\rangeOf{s}.\sortattrs} \afles \lbMarker{\rangeTup.\sortattrs}$. Since $\lbMarker{\rangeOf{s}.\sortattrs} \afles \ubMarker{\rangeOf{s}.\sortattrs}$ holds for any tuple, we know that these tuples are visited by \Cref{alg:topkOnepass} before $\rangeTup$. We store tuples in a min-heap \texttt{todo} sorted on $\ubMarker{\sortattrs}$ and maintain a variable $\lbMarker{\texttt{rank}}$ to store the current lower bound. For every incoming tuple $\rangeTup$, we first determine all tuples $\rangeOf{s}$ from \texttt{todo} certainly preceding $\rangeTup$  ($\ubMarker{\rangeOf{s}.\sortattrs} < \lbMarker{\rangeTup.\sortattrs}$) and update $\lbMarker{\texttt{rank}}$ with their multiplicity. Since $\rangeTup$ is the first tuple certainly ranked after any such tuple $\rangeOf{s}$ and all tuples following $\rangeTup$ will also certainly ranked after $\rangeOf{s}$, we can now determine the upper bound on $\rangeOf{s}$'s position. Based on \Cref{eq:upos-ub} this is the sum of the maximal multiplicity of all tuples that may precede $\rangeOf{s}$. These are all tuples $\rangeOf{u}$ such that $\ubMarker{\rangeOf{s}.\sortattrs} \geq \lbMarker{\rangeOf{u}.\sortattrs}$, i.e., all tuples we have processed so far. We store the sum of the maximal multiplicity of these tuples in a variable $\ubMarker{\texttt{rank}}$ which is updated for every incoming tuple. We use a function \texttt{emit} to compute $\rangeOf{s}$'s upper bound sort position, adapt $\lbMarker{\rangeOf{s}.\multa}$ (for a top-k query, $\rangeOf{s}$ may not exist in the result if its position may be larger than $k$), add $\rangeOf{s}$ to the result, and adapt $\lbMarker{\texttt{rank}}$ (all tuples processed in the following are certainly ranked higher than $\rangeOf{s}$). Function \texttt{split} splits a tuple with $\mult > 1$ into multiple tuple as required by \Cref{def:au-db-sorting-operat}.
  If we are only interested in the top-k results, then we can stop processing the input once $\lbMarker{\texttt{rank}}$ is larger than $k$, because all following tuples will be certainly not in the top-k. Once all inputs have been processed, the heap may still contain tuples whose relative sort position wrt. to each other is uncertain. We flush these tuples at the end. 
}

\setlength{\textfloatsep}{3pt}
\setlength{\floatsep}{0pt}
\begin{algorithm}[t]
  \SetKwProg{Fn}{def}{}{}
  \KwIn{$\rangeRel$ (sorted by $\lbMarker{\sortattrs}$), $\texttt{k} \in \domN$ (or $\texttt{k} = \textbar R\textbar$)}
  $\texttt{todo} \gets \texttt{minheap}(\ubMarker{\sortattrs})$ ;\ $\lbMarker{\texttt{rank}} \gets 0$ ;\ $\ubMarker{\texttt{rank}} \gets 0$ ;\ $\texttt{res} \gets \emptyset$ \\
  \For{$\texttt{\rangeTup} \in \rangeRel$}{
    \While(\tcc*[f]{emit tuples}){$\texttt{todo}.|peek|().\ubMarker{\sortattrs} < \rangeTup.\lbMarker{\sortattrs}$}{
      $|emit|(\texttt{todo}.|pop|())$\\
      \If(\tcc*[f]{tuples certainly out of top-k?}){$\lbMarker{\texttt{rank}}>\texttt{k}$}{
        \Return{\texttt{res}}
      }
	}
	$\rangeOf{t}.\lbMarker{\ranka} \gets \lbMarker{\texttt{rank}}$ \tcc*[f]{set position lower bound}\\
	$\texttt{todo}.|insert|(\rangeTup)$ \tcc*[f]{insert into todo heap}\\
	$\ubMarker{\texttt{rank}}\ \texttt{+=}\;\ubMarker{\mult}$ \tcc*[f]{update position upper bound}
  }
  \While(\tcc*[f]{flush remaining tuples}){\textbf{\upshape not} $\texttt{todo}.|isEmpty|()$}{
    $|emit|(\texttt{todo}.|pop|())$\\
  }
  \Return{\texttt{res}}\\
  \BlankLine
  \Fn{\texttt{emit}$(\rangeOf{s})$}{
      $\rangeOf{s}.\ubMarker{\ranka} \gets \texttt{min}(\texttt{k},\;\ubMarker{\texttt{rank}})$ \tcc*[f]{position upper bound capped at $k$}\\
      \If(\tcc*[f]{$\rangeOf{s}$ may not be in result if $\rangeOf{s}.\ubMarker{\ranka} > k$}){$\ubMarker{\texttt{rank}} >\texttt{k}$}{
        $\rangeOf{s}.\lbMarker{\multa} \gets 0$
      }
      $\texttt{res} \gets \texttt{res} \cup |split|(\{\rangeOf{s}\})$ \\
      $\lbMarker{\texttt{rank}}\ \texttt{+=}\ \lbMarker{\rangeOf{s}.\multa}$ \tcc*[f]{update position lower bound}
    }
  \caption{Non-deterministic sort on $\sortattrs$ (top-k)}
  \label{alg:topkOnepass}
\end{algorithm}

\iftechreport{
\begin{algorithm}[h]
\SetKwProg{Fn}{Function}{:}{end}
\Fn{|split|(t)}{
\For{$i \in [1,\ubMarker{\texttt{t}.\mult}]$}{
	$\texttt{t}_i = |copy|(t)$ \;
	$\bgMarker{\texttt{t}_i.\mult} = \bgMarker{\texttt{t}.\mult}<i:1?0$ \;
	$\ubMarker{\texttt{t}_i.\mult} = 1$ \;
	$\lbMarker{\texttt{t}_i.\mult} = \lbMarker{\texttt{t}.\mult}<i:1?0$ \;
	$\ubMarker{\texttt{t}_i.\ranka} += i$ \;
	$\lbMarker{\texttt{t}_i.\ranka} += i$ \;
	\Return{$\texttt{t}_i$} \;}}
\caption{\revm{Split bag tuple}}
\label{alg:topkSplit}
\end{algorithm}
}

\iftechreport{
\begin{figure}
  \includegraphics[width=0.6\columnwidth]{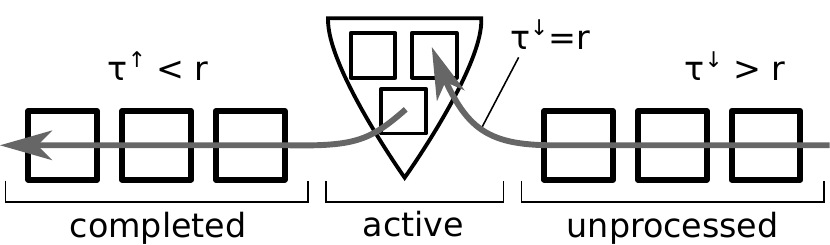}
  \vspace{-3mm}
  \setlength{\belowcaptionskip}{-2pt}
  \caption{The lifecycle of tuples in \Cref{alg:topkOnepass}}
  \label{fig:sortOnePassLifecycle}
\end{figure}
}

\iftechreport{
\revm{
\Cref{alg:topkSplit} defines function $split(\rangeTup)$ split multiplicities of range tuple $\rangeTup$ into multiplicity of ones using the semantics in \Cref{ fig:uadbSortSemantics}.
}
}

\ifnottechreport{
   \mypar{Complexity Analysis}
        \revm{Let $n = \card{\rangeRel}$. The algorithm requires $O(n \cdot \log n)$ to sort the input. Since the heap size is bound by $n$ and \texttt{pop} and \texttt{insert} are logarithmic in the heap size, the overall worst-case runtime is $O(n \cdot \log n)$. \Cref{alg:topkOnepass}'s worst-case memory requirement is $O(n)$.}
  }
\iftechreport{
\mypar{Complexity Analysis}
    \revm{Let $n = \card{\rangeRel}$. The algorithm requires $O(n \cdot \log n)$ to sort the input. It then processes the data in one-pass. For each tuple, we compare $\rangeTup.\ubMarker{\sortattrs}$ in $O(1)$ with the root of the heap and insert the tuple into the heap in $O(\log \card{heap})$. Tuples are removed from the heap just once in $O(\log \card{heap})$. In the worst-case, if the sort positions of all tuples may be less than $k$, then the heap will contain all $n$ tuples at the end before flushing. Thus, $\card{heap}$ is bound $n$ and we get $O(n \cdot \log n)$ as the worst-case runtime complexity for our algorithm requiring $O(n)$ memory. However, in practice, heap sizes are typically much smaller.}
}

\subsection{Connected Heaps}\label{sec:connected-heaps}
\revm{In our algorithm for windowed aggregation that we will present in \cref{sec:ranged-window-agg}, we need to maintain the tuples possibly in a window ordered increasingly on $\ubMarker{\ranka}$ (for fast eviction), sorted on $\lbMarker{\agga}$ to compute $\winbotk{\rangeRel}{\agga}$, and sorted decreasingly on $\ubMarker{\agga}$ to compute $\wintopk{\rangeRel}{\agga}$.
  We could use separate heaps to access the smallest element(s) wrt. to any of these orders efficiently. 
  However, if a tuple needs to be deleted, the tuple will likely not be the root element in all heaps which means we have to remove non-root elements from some heaps which is inefficient (linear in the heap size). Of course it would be possible to utilize other data structures that maintain order such as balanced binary trees. However, such data structures do not achieve the $O(1)$ lookup performance for the smallest element that heaps provide.  \iftechreport{Furthermore, trees are typically are not as efficient in practice as heaps which can be implemented as arrays.}
  Instead, 
  we introduce a simple, yet effective, data structure we refer to as a \emph{connected heap}. 
}
A \textit{connected heap} is comprised of $H$ heaps which store pointers to a shared set of records. Each heap has its own sort order. A record stored in a connected heap consists of a tuple (the payload) and $H$ backwards pointers that point to the nodes of the individual heaps storing this tuple. These backward pointers enable efficient deletion \revm{($O(H \cdot \log n)$)} of a tuple from all heaps when it is popped as the root of one of the component heaps. \ifnottechreport{\revm{In \cite{techreport} we explain how the standard sift-up and sift-down heap operations are used to restore the heap property in $O(\log n)$ when removing a non-root element  from a component heaps.}}
When a tuple is inserted into a connected heap, it is inserted into each component heap in \revm{$O(\log n)$} in the usual way with the exception that the backwards pointers 
are populated.
\ifnottechreport{
	\revb{
      In~\cite{techreport}, we experimentally comparing heaps with connected heaps. Even for small databases (10k tuples) and a small fraction of tuples with uncertain order-by values (1\%), connected heaps outperform heaps by a factor of $\sim 2$. Larger databases / more uncertain data result in larger heaps and, thus, even better performance.
		}
}
\iftechreport{
\mypar{Preliminary experiments}
To measure the impact of the backpointers in connected heaps on performance, we did a preliminary experimental comparison with using a set of independent heaps.
  Without the backlinks, removing an non-root element from a heap is linear in the size of the heap in the worst-case, because it may require a search over the whole heap   to find the position of such an element. Afterwards, the element can be deleted and the heap property can be restored in $O(\log n)$. Using the backlinks, finding the positions of an element in other heaps is $O(1)$ and so popping the root element of one heap and removing it from all other heaps is in $O((\log n) \cdot m)$ where $n$ is the size of the largest heap and $m$ is the number of heaps. The table below shows the execution times in milliseconds using connected heaps (back pointers) versus classical unconnected heaps ( linear search).  
This experiment was run on a database with 50k tuples and 1\%-5\% uncertainty (amount of tuples that are uncertain) varying the size of the ranges for the attribute we are aggregating over. The main factor distinguish linear search performance from back pointers is the heap size which for our windowed aggregation algorithm is affected by attribute range size, percentage of tuples which have uncertain order-by and data size. Even though in this experiment the amount of uncertain data and database size are quite low, we already see 25\% up to a factor of $\sim 10$ improvement. For larger databases or larger percentage of uncertain data, the sizes of heaps will increase and, thus, we will see even more significant performance improvements.

{\centering
\begin{tabular}{ l|l|c|c }
	\textbf{Uncert} & \textbf{Range} & \textbf{Connected heaps} & \textbf{Unconnected heaps} \\
	& & \textbf{(Back pointers)(ms)} & \textbf{(Linear search)(ms)} \\
	\hline
	1\% & 2000 & 1979.272 & 3479.042\\
	1\% & 15000 & 2045.162 & 6676.732\\
	1\% & 30000 & 2103.974 & 9646.330\\
	5\% & 2000 & 1976.651 & 4078.487\\
	5\% & 15000 & 2149.990 & 15186.657\\
	5\% & 30000 & 2191.823 & 22866.713\\
\end{tabular}
}
}

\iftechreport{
      \mypar{Deletion from a connected heap} When a node is popped from one of the component heaps 
the nodes of the other heaps storing the tuple are identified in $O(H)$ using the backwards pointers. Like in standard deletion of nodes from a heap, a deleted node is replaced with the right-most node at the leaf level. Standard sift-down and sift-up are then used to restore the heap property in $O(\log n)$.
      Recall that the heap property for a min-heap requires that for each node in the heap its value is larger than the value of its parent. Insertion of a new node $v$ into a heap places the new element at the next free position at the leaf level. This may violate the heap property. The heap property can be restored in $O(\log n)$ using sift-up (repeatedly replacing a node with its parent). Similarly, to delete the root of a heap, we replace the root with the right-most child. This again may violate the heap property if the new root is larger than one of its children. The heap property can be restored in $O(\log n)$ steps using sift-down, i.e., replacing a node that is larger than a child with the smaller of its children.
For a connected heap, deletion may cause a node to be deleted that is currently not the root of the heap. Like in standard heaps, we replace the node $v$ to be deleted with the right-most node $v_l$ from the leaf level. This may violate the heap property (every child is larger than its parent) in two possible ways: either $v_l$ is smaller than the parent of $v$ or $v_l$ is larger than one of the children of $v$. Note that it is not possible for both cases to occur at the same time, because the heap was valid before and, thus, if $v_l$ is larger than a child of $v$, then it has to be larger than the parent of $v$.
If $v_l$ is smaller than the parent of $v$, then it has to be smaller than all other nodes in the subtree rooted at $v$. We can restore the heap property by sifting up $v_l$.
    Now consider the case where $v_l$ is larger than one of the children of $v$ and let $v_c$ denote that child (or the smaller child if $v_l$ is larger than both children). Note that the subtree rooted at $v$ was a valid heap. Thus, replacing $v$ with $v_l$ is replacing the root element of this subheap and the heap property for the subheap can be restored using sift-down. Since $v_l$ is larger than the parent of $v$ this restores the heap property for the whole heap.}


\tikzset{
  treenode/.style = {align=center, inner sep=0pt, text centered,
    font=\sffamily, circle, black, draw=black,
    text width=1.5em, very thick},
  tuple/.style = {align=center, inner sep=3pt, text centered,
    font=\sffamily, rectangle, rounded corners, black, draw=black, thick, rectangle split, rectangle split horizontal,
                      rectangle split parts=2},
  link/.style = {thick, dashed,<->},
}

\begin{Example}[Connected heap]\label{ex:connected-heap}
  Consider the connected heap shown below on the left storing tuples $t_1 = (1,3)$, $t_2 = (2,6)$, $t_3 = (3,2)$, and $t_4=(4,1)$. Heap $h_1$ ($h_2$) is sorted on the first (second) attribute.
  Calling \texttt{pop()} on $h_1$ removes $t_1$ from $h_1$. Using the backwards pointer from $t_1$ to the corresponding node in $h_2$ (shown in red), we also remove $t_1$ from $h_2$. The node pointing to $t_1$ from $h_2$ is replaced with the right most leaf node of $h_2$ (pointing to $t_2$). In this case the heap property is not violated and, thus, no sift-down / up is required.

\begin{minipage}{1\linewidth}
	\centering
    \begin{minipage}{0.5\linewidth}
      \centering
      \scalebox{0.6}{
		\begin{tikzpicture}[level/.style = {sibling distance = 1.2cm,
            level distance = 1cm}]]
          \node [treenode, label=left:$\bf h_1$] (l1) at (0,1.5) {1}
          child { node [treenode] (l2) {2} edge from parent[very thick]
            child { node [treenode] (l4) {4} edge from parent[very thick] }
          }
          child { node [treenode] (l3) {3} edge from parent[very thick]};
          \node [tuple, label=left:t1] (t1) at (2.1,1.5) {
            \nodepart{one} 1
            \nodepart{two} 3
          };
          \node [tuple, label=left:t2] (t2) at (2.1,1) {
            \nodepart{one} 2
            \nodepart{two} 6
          };
          \node [tuple, label=left:t3] (t3) at (2.1,0.5) {
            \nodepart{one} 3
            \nodepart{two} 2
          };
          \node [tuple, label=left:t4] (t4) at (2.1,0) {
            \nodepart{one} 4
            \nodepart{two} 1
          };
          \node [treenode, label=left:$\bf h_2$] (r1) at (4,1.5) {1}
          child { node [treenode] (r2) {2} edge from parent[very thick]
            child {node [treenode] (r6) {6} edge from parent[very thick] }
          }
          child { node [treenode] (r3) {3} edge from parent[very thick]};
          \draw [link, red] (l1.north) to[out=50, in=125] (t1.one north);
          \draw [link, red] (t1.two north) to[out=50, in=90, distance=1.4cm] (r3.north);
          \draw [link, black] (t2.north west) to[out=160, in=30] (l2.east);
          \draw [link, black] (t2.east) to[out=10, in=180] (r6.west);
          \draw [link, black] (t3.south west) to[out=-170, in=-20] (l3.south east);
          \draw [link, black] (t3.south east) to[out=-30, in=-120] (r2.west);
          \draw [link, black] (t4.one south) to[out=-100, in=0] (l4.east);
          \draw [link, black] (t4.two south) to[out=-45, in=-90, distance=2.2cm] (r1.south);
        \end{tikzpicture}
      }
      \end{minipage}
      \begin{minipage}{0.49\linewidth}
        \centering
        \textbf{Result of \texttt{h1.pop()}}
      \scalebox{0.6}{
		\begin{tikzpicture}[level/.style = {sibling distance = 1.2cm,
            level distance = 1cm}]]
          \node [treenode, label=left:$\bf h_1$] (l2) at (0,1.5) {2}
          child { node [treenode] (l4) {4} edge from parent[very thick]}
          child { node [treenode] (l3) {3} edge from parent[very thick]};
          \node [tuple, label=left:t2] (t2) at (2.2,1.5) {
            \nodepart{one} 2
            \nodepart{two} 6
          };
          \node [tuple, label=left:t3] (t3) at (2.2,1) {
            \nodepart{one} 3
            \nodepart{two} 2
          };
          \node [tuple, label=left:t4] (t4) at (2.2,0.5) {
            \nodepart{one} 4
            \nodepart{two} 1
          };
          \node [treenode, label=left:$\bf h_2$] (r1) at (4,1.5) {1}
          child { node [treenode] (r2) {2} edge from parent[very thick]}
          child { node [treenode] (r6) {6} edge from parent[very thick]};
          \draw [link, black] (t2.north west) to[out=160, in=30] (l2.north east);
          \draw [link, black] (t2.north east) to[out=35, in=90, looseness = 1.3] (r6.north);
          \draw [link, black] (t3.south west) to[out=-150, in=0] (l3.east);
          \draw [link, black] (t3.south east) to[out=-30, in=-180, looseness = 1.4] (r2.west);
          \draw [link, black] (t4.south west) to[out=-140, in=-20] (l4.south east);
          \draw [link, black] (t4.south east) to[out=-20, in=-90, distance=1.4cm] (r1.south);
        \end{tikzpicture}
      }
    \end{minipage}
  \end{minipage}
 \\[-5mm]
\end{Example}
\subsection{Ranged Windowed Aggregation}\label{sec:ranged-window-agg}
\newcommand{\agglb}{\lbMarker{\texttt{pagg}}\xspace}
\newcommand{\aggub}{\ubMarker{\texttt{pagg}}\xspace}
\newcommand{\cheap}{\texttt{cert}\xspace}
\newcommand{\pheap}{\texttt{poss}\xspace}
\newcommand{\certlb}{\lbMarker{\texttt{c-rank}}\xspace}
\newcommand{\resultlb}{\lbMarker{X}}
\newcommand{\resultub}{\ubMarker{X}}
\newcommand{\ubpheap}{\texttt{ub\_\pheap}}
\newcommand{\lbpheap}{\texttt{lb\_\pheap}}
\newcommand{\openw}{\texttt{openw}\xspace}

Without loss of generality, we focus on window specifications with only a \lstinline|ROWS PRECEDING| clause; a \lstinline|FOLLOWING| clause can be mimicked by offsetting the window, i.e., \revm{a window bound of $[-N,0]$.}
\revm{\Cref{alg:window_precOnePass} uses a function \calcaggbounds to compute the bounds on the aggregation function result based on the certain and possible content of a window.}
\ifnottechreport{
We present the definition of this function for several aggregation functions in~\cite{techreport}.}
\iftechreport{
\revm{We discuss the code for these functions below for aggregation functions \aggmin, \aggmax, and \aggsum (\aggcount uses the same algorithm as \aggsum using $\uv{1}{1}{1}$ instead of the values of an attribute $\agga$).}
}
\revm{\Cref{alg:window_precOnePass} follows a sweeping pattern similar to \Cref{alg:topkOnepass} to compute the windowed aggregate in a single pass over the data which has been preprocessed by applying $\rank{\sortattrs}{\ranka}(\rangeRel)$ and then has been sorted on $\lbMarker{\ranka}$. This algorithm uses a minheap \openw which is sorted on $\ubMarker{\ranka}$ to store tuples for which have not seen yet all tuples that could belong to their window. Additionally, the algorithm maintains the following data structures:
\cheap is a map from a sort position $i$ to a tree storing tuples $\rangeTup$ that certainly exist and for which $\rangeTup.\lbMarker{\ranka} = i$ sorted on $\ubMarker{\ranka}$. This data structure is used to determine which tuples certainly belong to the window of a tuple; $(\pheap,\agglb,\aggub)$ is a connected minheap with $\pheap$, \agglb, and \aggub  are sorted on $\ubMarker{\ranka}$, $\ubMarker{\agga}$, $\ubMarker{\agga}$, respectively. This connected heap stores tuples possibly in a window. The different sort orders are needed to compute bounds on the aggregation function result for a window efficiently (we will expand on this later). Finally, we maintain a watermark \certlb for the lower bound of the certain part of windows.
}

\begin{figure}[t]
  \centering
  \begin{tikzpicture}
    [
    edg/.style={->,line width=0.4mm},
    ]
    \def\xscaler{0.6}
    \def\yscaler{0.4}
    \def\axispos{-5.5 * \yscaler}

    \draw[|->, thick] (-0,\axispos) -- (10 * \xscaler,\axispos);

    \foreach \x in {0,...,9}
        \draw[thick] (\x * \xscaler,0.3 * \yscaler + \axispos) -- (\x * \xscaler,-0.3 * \yscaler + \axispos) node[below] {\x};

		\draw[edg,blue,|-|]   (5 * \xscaler,-1 * \yscaler * 1)  node[left,black]{} -- node[above,black]{$\rangeOf{s}$} (7 * \xscaler,-1 * \yscaler * 1);


    \draw[edg,darkgreen,|-|] (0  * \xscaler, -1 * \yscaler * 1.7) -- node[above,black]{\textcolor{darkgreen}{\texttt{poss}}} (7  * \xscaler, -1 * \yscaler * 1.7);

    \draw[edg,red,|-|] (2  * \xscaler, -1 * \yscaler * 1.7) -- node[below,black]{\textcolor{red}{cert}} (5  * \xscaler, -1 * \yscaler * 1.7);

    \fill[darkgreen, very nearly transparent] (0,-1 * \yscaler * 1.7) rectangle (7 * \xscaler,\axispos);

    \fill[red, very nearly transparent] (2 * \xscaler,-1 * \yscaler * 1.7) rectangle (5 * \xscaler,\axispos);

    \draw[thick,blue,|-|]   (0 * \xscaler,-1 * \yscaler * 2.7)  node[left,black]{} -- node[above,black]{$\rangeOf{t_1}$} (3 * \xscaler,-1 * \yscaler * 2.7);

    \draw[thick,blue,|-|]   (2 * \xscaler,-1 * \yscaler * 3.7)  node[left,black]{} -- node[above,black]{$\rangeOf{t_2}$} (3 * \xscaler,-1 * \yscaler * 3.7);

    \draw[black,<-]   (3.1 * \xscaler,-1 * \yscaler * 3.7)  node[left,black]{} -- node[above right,black,xshift=3.5mm]{\footnotesize\texttt{cert}[2]} (4.5 * \xscaler,-1 * \yscaler * 2.7);

    \draw[thick,blue,|-|]   (3 * \xscaler,-1 * \yscaler * 4.7)  node[left,black]{} -- node[above,black]{$\rangeOf{t_3}$} (4 * \xscaler,-1 * \yscaler * 4.7);

    \draw[black,<-]   (4.1 * \xscaler,-1 * \yscaler * 4.7)  node[left,black]{} -- node[above right,black,xshift=3.5mm]{\footnotesize\texttt{cert}[3]} (5.5 * \xscaler,-1 * \yscaler * 3.7);

    \draw[thick,blue,|-|]   (8 * \xscaler,-1 * \yscaler * 4)  node[left,black]{} -- node[above,black]{$\rangeOf{t_4}$} (9 * \xscaler,-1 * \yscaler * 4);

    \draw[black,->]   (1 * \xscaler, \axispos+\xscaler)  node[left,black]{} -- node[above left,black,xshift=-1.5mm,yshift=1.5mm]{\footnotesize\texttt{{c-rank}\lbMarker{}}} (1.9 * \xscaler, \axispos+0.1);

  \end{tikzpicture}
  \vspace{-10pt}
  \caption{\revb{State example for \Cref{alg:window_precOnePass}, N=5, $\lbMarker{\texttt{c-rank}}$=2.}}\label{fig:wind-alg-state-example}
\end{figure}

\begin{algorithm}[t]
  \SetKwProg{Fn}{def}{}{}
  \SetSideCommentLeft{}{}{}
  \KwIn{$f$, $X$,$\sortattrs$, $N$, $\agga$, $\rank{\sortattrs}{\ranka}(\rangeRel)$ \textbf{sorted on} $\lbMarker{\ranka}$}
  $\openw \gets |minheap|(\ubMarker{\ranka})$ \tcc*[f]{tuples with open windows}\\
  $\cheap \gets |Map|(|int|, |Tree|(\ubMarker{\ranka}))$ \tcc*[f]{certain window members by pos.} \\
  $(\pheap,\agglb,\aggub) \gets \texttt{connected-minheap}(\ubMarker{\ranka}, \lbMarker{\agga}, \ubMarker{\agga})$ \\
  $\certlb \gets 0$ \tcc*[f]{watermark for certain window}\\
  $|res| \gets \emptyset$\\
  \For{$\rangeTup \in \rangeRel$}{
    $|\openw|.|insert|(\rangeTup)$ \label{algl:nwin-insert-openw}\\
    \If(\tcc*[f]{insert into potential certain window}){$\lbMarker{\rangeOf{t}.\multa} > 0$}{
      $\cheap[\lbMarker{\rangeTup.\ranka}].|insert|(\rangeTup) $ \\
    }
    \While(\tcc*[f]{close windows}){$|\openw|.|peek|().\ubMarker{\ranka} < \rangeTup.\lbMarker{\ranka}$}{
      $\rangeOf{s} \gets |\openw|.|pop|()$ \label{algl:nwin-emit}\\
      \While(\tcc*[f]{evict certain win.}){$\certlb < \rangeOf{s}.\ubMarker{\ranka} - N$}{ \label{algl:nwin-cert-evict}
        $\cheap[\certlb] = \NULL$ \\
        $\certlb++$\\
      }
      $\rangeOf{s}.|X| \gets |\calcaggbounds|(f,\rangeOf{s},|\cheap|, |\pheap|)$ \tcc*[f]{compute agg.} \label{algl:nwin-calc-agg}\\
      \While(\tcc*[f]{evict poss. win.}){$|\pheap|.|peek|.\ubMarker{\ranka} < \rangeOf{s}.\lbMarker{\ranka} -N$}{
        $|\pheap|.|pop|()$\\
      }
      $|res| \gets |res| \cup \{\rangeOf{s}\}$\\
    }
    $|\pheap|.|insert|(\rangeTup)$ \tcc*[f]{insert into poss. win.}\\
  }

  \caption{Aggregate $f(A) \to X$, sort on $\sortattrs$, $N$ preceding}
  \label{alg:window_precOnePass}
\end{algorithm}
\iftechreport{

    \begin{algorithm}[t]
      \SetKwProg{Fn}{def}{}{}
      \SetSideCommentLeft{}{}{}
      \Fn(\tcc*[f]{compute bounds on $\aggsum(\agga)$}){$\texttt{\calcaggbounds}(f,\rangeOf{t},\cheap, \pheap)$}{  \label{algl:nwin-calc-agg}
        \If{$f = \aggsum$}{
          \Return |computeSumBounds|($\rangeOf{t}$,\cheap, \pheap) \\
        }
        \If{$f = \aggmin$}{
          \Return |computeMinBounds|($\rangeOf{t}$,\cheap, \pheap) \\
        }
        \If{$f = \aggmax$}{
          \Return |computeMaxBounds|($\rangeOf{t}$,\cheap, \pheap) \\
        }
        \If{$f = \aggcount$}{
          \Return |computeCountBounds|($\rangeOf{t}$,\cheap, \pheap) \\
        }
        \If{$f = \aggavg$}{
          \Return
          |computeMinBounds|($\rangeOf{t}$,\cheap, \pheap) \\
        }
      }

      \caption{Computing bounds for $f(A) \to X$ for tuple $\rangeTup$}
      \label{alg:sum-bounds-algo}
    \end{algorithm}

    \begin{algorithm}[t]
      \SetKwProg{Fn}{def}{}{}
      \SetSideCommentLeft{}{}{}
      \Fn(\tcc*[f]{compute bounds on $\aggsum(\agga)$}){$\texttt{ComputeSumBounds}(\rangeOf{t},|\cheap|, |\pheap|)$}{  \label{algl:nwin-calc-agg}
        $n \gets  N - 1; \resultlb \gets \rangeOf{t}.\ubMarker{\agga}; \resultub \gets \rangeOf{t}.\ubMarker{\agga}$ \tcc*[f]{pos. and bounds}\\
        \For{$x \in [\ubMarker{\rangeOf{t}.\ranka} - N, \lbMarker{\rangeOf{t}.\ranka}]$}{
          \For{$\rangeOf{s} \in |\cheap|[x]$}{
            \If(\tcc*[f]{belongs to cert. window of $\rangeOf{s}$}){$\ubMarker{\rangeOf{s}.\ranka} \leq \lbMarker{\rangeOf{t}.\ranka}$}{
              $\resultub += \ubMarker{\rangeOf{s}.\agga}; \resultlb += \lbMarker{\rangeOf{s}.\agga}$ \\
              $n--$\\
            }
            \lElse{
              \textbf{break}
            }
          }
        }
        $\lbpheap \gets |copy|(\agglb); \ubpheap \gets |copy|(\aggub)$\\
        $n_{lb} \gets n; n_{ub} \gets n$ \tcc*[f]{max. num. of tuples possibly in win. } \\
        \While(\tcc*[f]{compute $\lbMarker{X}$}){$n_{lb}>0 \land \neg |\lbpheap|.|isEmpty|()$}{
          $\rangeOf{s} \gets |\lbpheap|.|pop|()$\\
          \If(\tcc*[f]{only values $<0$ contribute to $\lbMarker{X}$}){$\lbMarker{\rangeOf{s}.\agga} < 0$}{
            $\resultlb += \lbMarker{\rangeOf{s}.\agga}$ \\
            $n_{lb}--$\\
          }
          \lElse
          {
            \textbf{break}
          }
        }
        \While(\tcc*[f]{compute $\ubMarker{X}$}){$n_{ub}>0 \land \neg |\ubpheap|.|isEmpty|()$}{
          $\rangeOf{s} \gets |\ubpheap|.|pop|()$\\
          \If(\tcc*[f]{only values $>0$ contribute to $\ubMarker{X}$}){$\ubMarker{\rangeOf{s}.\agga} > 0$}{
            $\resultub += \ubMarker{\rangeOf{s}.\agga}$ \\
            $n_{ub}--$\\
          }
          \lElse{
            \textbf{break}
          }
        }
        \Return $[\resultlb,\resultub]$ \\
      }

      \caption{Computing bounds for $\aggsum(A) \to X$ for tuple $\rangeTup$}
      \label{alg:sum-bounds-algo}
    \end{algorithm}

    \begin{algorithm}[t]
      \SetKwProg{Fn}{def}{}{}
      \SetSideCommentLeft{}{}{}
      \Fn(\tcc*[f]{compute bounds on $\aggmin(\agga)$}){$\texttt{computeMinBounds}(\rangeOf{t},|\cheap|, |\pheap|)$}{  \label{algl:nwin-calc-agg}
        $n \gets  N - 1; \resultlb \gets \rangeOf{t}.\ubMarker{\agga}; \resultub \gets \rangeOf{t}.\ubMarker{\agga}$ \tcc*[f]{pos. and bounds}\\
        \For{$x \in [\ubMarker{\rangeOf{t}.\ranka} - N, \lbMarker{\rangeOf{t}.\ranka}]$}{
          \If(\tcc*[f]{min of certain lower-bound}){$|\cheap|[x]$}{
           		$\resultub \gets x$ \\
           		\textbf{break} \\
          }
        }
        $\resultlb = \agglb.|peek|().\lbMarker{\agga}$ \tcc*[f]{min of possible lower-bound}\\
        \Return $[\resultlb,\resultub]$ \\
      }

      \caption{Computing bounds for $\aggmin(A) \to X$ for tuple $\rangeTup$}
      \label{alg:min-bounds-algo}
    \end{algorithm}
  }



\revm{
\Cref{alg:window_precOnePass} first inserts each incoming tuple into \openw
(\Cref{algl:nwin-insert-openw}). If the tuple certainly exists, it is inserted
into the tree of certain tuples whose lower bound position is
$\rangeTup.\lbMarker{\ranka}$. Note that each of these trees is sorted on
$\ubMarker{\ranka}$ which will be relevant later. Next the algorithm determines
for which tuples from \openw, their windows have been fully observed. These are
all tuples $\rangeOf{s}$ which are certainly ordered before the tuple
$\rangeTup$ we are processing in this iteration ($\rangeOf{s}.\ubMarker{\ranka}
< \rangeTup.\lbMarker{\ranka}$). To see why this is the case first observe that
(i) we are processing input tuples in increasing order of $\lbMarker{\ranka}$
and (ii) tuples are ``finalized'' by computing the aggregation bounds in
monotonically increasing order of $\ubMarker{\ranka}$. Given that we are using a
window bound $[-N,0]$, all tuples $\rangeOf{s}$ that could possibly belong to
the window of a tuple $\rangeTup$ have to have $\rangeOf{s}.\lbMarker{\ranka}
\leq \rangeTup.\ubMarker{\ranka}$. Based on these observations, once we
processed a tuple $\rangeTup$ with $\rangeTup.\lbMarker{\ranka} >
\rangeOf{s}.\ubMarker{\ranka}$ for a tuple $\rangeOf{s}$ in \openw, we know that
no tuples that we will process in the future can belong to the window for
$\rangeOf{s}$. In \Cref{algl:nwin-emit} we iteratively pop such tuples from
\openw. For each such tuple $\rangeOf{s}$ we evict tuples from \cheap and update
the high watermark \certlb (\Cref{algl:nwin-cert-evict}). Recall that for a
tuple $\rangeOf{u}$ to certainly belong to the window for $\rangeOf{s}$ we have
to have $\rangeOf{s}.\ubMarker{\ranka} -N \geq \rangeTup.\lbMarker{\ranka}$.
Thus, we update \certlb to $\rangeOf{s}.\ubMarker{\ranka} -N$ and evict from
\cheap all trees storing tuples for sort positions smaller than
$\rangeOf{s}.\ubMarker{\ranka} -N$. Afterwards, we compute the bounds on the
aggregation result for $\rangeOf{s}$ using \cheap and \pheap (we will describe
this step in more detail in the following). Finally, evict tuples from \pheap
(and, thus, also \agglb and \aggub) which cannot belong to any windows we will
close in the future. These are tuples which are certainly ordered before the
lowest possible position in the window of $\rangeOf{s}$, i.e., tuples
$\rangeOf{u}$ with $\rangeOf{u}.\ubMarker{\ranka} < \rangeOf{s}.\lbMarker{s} -
N$ (see \Cref{fig:possible-and-certain-wind}). Evicting tuples from \pheap based
on the tuple for which we are currently computing the aggregation result bounds
is safe because we are emitting tuples in increasing order of
$\ubMarker{\ranka}$, i.e., for all tuples $\rangeOf{u}$ emitted after
$\rangeOf{s}$ we have $\rangeOf{u}.\ubMarker{\ranka} >
\rangeOf{s}.\ubMarker{\ranka}$.
\revb{\Cref{fig:wind-alg-state-example} shows an example state for the algorithm when tuple $\rangeOf{s}$ is about to be emitted.  Tuples fully included in the red region ($\rangeOf{t_2}$ and $\rangeOf{t_3}$) are currently in $\cheap[i]$ for sort positions certainly  in the window for $\rangeOf{s}$. Tuples with sort position ranges overlapping with green region are in the possible window (these tuples are stored in $\pheap$). Tuples like $\rangeOf{t_4}$ with upper-bound position higher than $\rangeOf{s}$ will be popped and processed after $\rangeOf{s}$.} Once all input tuples have been processed, we
have to close the windows for all tuples remaining in \openw. This process is the same as emitting tuples before we have processed all inputs and, thus, is omitted form \Cref{alg:window_precOnePass}. 
}

\ifnottechreport{
\revm{
  \Cref{alg:window_precOnePass} uses function \calcaggbounds to compute the bounds on the aggregation function result for a tuple $\rangeTup$ using \cheap, \agglb and \aggub following the definition from \cref{sec:audb-win-agg-semantics}. First, we fetch all tuples that are certainly in the window from \cheap based on the sort positions that certainly belong to the window for $\rangeTup$ ([\rangeTup.\ubMarker{\ranka} - N, \rangeTup.\lbMarker{\ranka}]) and aggregate their $\agga$ bounds. Afterwards, we use \agglb and \aggub to efficiently fetch up $\certaincount{\rangeRel, \rangeTup}$ tuples possibly in the window for $\rangeTup$ to calculate the final bounds based on \wintopkname and \winbotkname. As mentioned before the detailed algorithm and further explanations are presented in~\cite{techreport}.
  }
}
\iftechreport{
\Cref{alg:window_precOnePass} uses function \calcaggbounds to compute the bounds on the aggregation function result for a tuple $\rangeTup$ using \cheap, \agglb and \aggub following the definition from \cref{sec:audb-win-agg-semantics}.
}

\BG{openw is minheap on $\tau^{\uparrow}$ cert is maxheap on $\tau^{\downarrow}$ and poss is minheap on $\tau^{uparrow}$}

\mypar{Complexity Analysis}
\revm{
\Cref{alg:window_precOnePass} first sorts the input in $O(n \log n)$ time using \Cref{alg:topkOnepass} followed by a deterministic sort on $\lbMarker{\ranka}$. Each tuple is inserted into \openw, \pheap, and \cheap at most once and poped from \openw exactly once. The size of the heaps the algorithm maintains is certainly less than $n$ at all times. To compute the aggregation function bounds, we have to look at the certain tuples in $\cheap[i]$ at most $size([N,0]) = N+1$ sort positions $i$ and at most $N+1$ tuples from $\pheap$ that can be accessed using the connected heaps in $O(N \cdot \log n)$. Thus, the overall runtime of the algorithm is $O(N \cdot n \cdot \log n)$.
  }

\section{Experiments}
\label{sec:experiment}

\newcommand{\competitorMCDBten}{\textit{MCDB10}\xspace}
\newcommand{\competitorMCDBtwenty}{\textit{MCDB20}\xspace}
\newcommand{\competitorMCDB}{\textit{MCDB}\xspace}
\newcommand{\competitorDet}{\textit{Det}\xspace}
\newcommand{\competitorRewr}{\textit{Rewr}\xspace}
\newcommand{\competitorRewrIndex}{\textit{Rewr(Index)}\xspace}
\newcommand{\competitorImp}{\textit{Imp}\xspace}
\newcommand{\competitorSymb}{\textit{Symb}\xspace}
\newcommand{\competitorPTk}{\textit{PT-k}\xspace}
\newcommand{\competitorRankAUDB}{\competitorImp/\competitorRewr\xspace}

We evaluate the efficiency of our rewrite-based approach and the native implementation of the algorithms presented in \Cref{sec:native-algorithms-au} \revm{in Postgres} and the accuracy the of approximations they produce.  

\mypar{Compared Algorithms}
We compare against several baselines: \competitorDet evaluates queries deterministically ignoring uncertainty in the data. We present these results to show the overhead of the different incomplete query evaluation semantics wrt. deterministic query evaluation;
\textit{MCDB}~\cite{jampani2008mcdb} evaluates queries over a given number of possible worlds sampled from the input incomplete database using deterministic query evaluation. \competitorMCDBten and \competitorMCDBtwenty are MCDB with 10 and 20 sampled worlds, respectively. \revb{For tests, we treat the highest and lowest possible value for all samples as the upper and lower bounds and compare against the tight bounds produced by the compared algorithms (since computing optimal bounds is often intractable). Given a tightest bound $[c,d]$, we define the recall of a bound $[a,b]$ as $\frac{min(b,d)-max(a,c)}{d-c}$ and the accuracy of  $[a,b]$ as $\frac{max(b,d)-min(a,c)}{min(b,d)-max(a,c)}$. The recall/accuracy for a relation is then the average recall/accuracy of all tuples.} %
\iftechreport{
\competitorPTk~\cite{10.1145/1376616.1376685} only supports sorting and returns all answers with a probability larger than a user-provided threshold of being among the top-k answers. By setting the threshold to 1 (0) we can use this approach to compute all certain (possible) answers.
\competitorSymb represents aggregation results, rank of tuples, and window membership as symbolic expressions which compactly encode the incomplete database produced by possible world semantics using the model from \cite{amsterdamer2011provenance} for representing aggregation results and a representation similar to \cite{AB14} to encode uncertainty in the rank of tuples. We use an SMT solver (Z3~\cite{moura-08-z}) to compute tight bounds on the possible ranks / aggregation results for tuples.
\competitorRewr is a rewrite-based approach we implemented uses self-unions for sorting queries and self-joins for windowed aggregation queries.
\competitorImp is the native implementation of our algorithms in Postgres.
All experiments are run on a 2$\times$6 core 3300MHz 8MB cache AMD Opteron 4238 CPUs, 128GB RAM, 4$\times$1TB 7.2K HDs (RAID 5) with the exception of \competitorPTk which was provided by the authors as a binary for Windows only. We run \competitorPTk on a separate Windows machine with a 8 core 3800MHz 32MB cache AMD Ryzen 5800x CPU, 64G RAM, 2TB HD. Because the \competitorPTk implementation is single-threaded and in-memory, we consider our comparisons are in favor of \competitorPTk. We implement our algorithms as an extension for Postgres 13.3 and evaluate all algorithms on Postgres. We report the average of 10 runs.
} %
\ifnottechreport{ %
For \competitorPTk~\cite{10.1145/1376616.1376685}, we set its threshold to 1 (0) to compute all certain (possible) answers.
\competitorSymb represents ranking and aggregation results as symbolic expressions similar to ~\cite{amsterdamer2011provenance,AB14}. We use an SMT solver (Z3~\cite{moura-08-z}) to compute tight bounds on the possible sort positions / aggregation results for tuples.
\competitorRewr is our rewrite-based approach~\cite{techreport} that to process the input relation twice for sorting and uses self-joins to determine the content of windows.
\competitorImp is implemented as a native extension for Postgres 13.3.
All experiments are run on a 2$\times$6 core 3300MHz 8MB cache AMD Opteron 4238 CPUs, 128GB RAM, 4$\times$1TB 7.2K HDs (RAID 5) with the exception of \competitorPTk which was provided by the authors as a Windows binary. We run \competitorPTk on a separate Windows machine with an 8-core 3800MHz 32MB cache AMD Ryzen 5800x CPU, 64G RAM, and 2TB HD. \competitorPTk is single-threaded and in-memory. Since we deactivated intra-query parallelism in Postgres, but still have to go to disk, the comparison is in favor of \competitorPTk. We report the average of 10 runs.
}

\subsection{Microbenchmarks on Synthetic Data}
To evaluate how specific characteristics of the data affect our system's performance and accuracy, we generated synthetic data consisting of a single table with 2 attributes for sorting and 3 attributes for windowed aggregation.
Attribute values are uniform randomly distributed.
\OK{Should be randomly distributed \textbf{values in the range ...} to provide context for the range-scaling graphs.}
Except where noted, we default to 50k rows and 5\% uncertainty with maximum 1k attribute range on uncertain values. 


\subsubsection{Sorting and Top-k Queries}

\vspace{-2mm}
\mypar{Scaling Data Size}
\Cref{fig:rankaltsize} shows the runtime of sorting,
varying the dataset size.
Since \competitorSymb and \competitorPTk perform significantly worse, we only include these methods for smaller datasets (\Cref{fig:rank-altsize-low}).
MCDB and our techniques significantly outperform \competitorSymb and \competitorPTk ($\sim$2+ OOM). \competitorRewr is roughly on par with \competitorMCDBtwenty while \competitorImp outperforms \competitorMCDBten.
Given their poor performance and their lack of support for windowed aggregation, we exclude \competitorSymb and \competitorPTk from the remaining microbenchmarks.

\mypar{Varying k, Ranges, and Rate}
\Cref{fig:rankmicro} shows runtime of top-k (k is specified) and sorting queries (k is not specified) when varying (i) the number of tuples returned $k$, (ii) the size of the ranges of uncertain order-by attributes (\emph{range}), and (iii) the fraction of tuples with uncertain order-by attributes.
\competitorImp is the fastest method, with an overhead of deterministic query processing between 3.5 (top-k) and 10 (full sorting).
\competitorRewr has higher overhead over \competitorDet than  \competitorMCDB.
Notably, the performance of \competitorMCDB and \competitorRewr is independent of all three varied parameters. Uncertainty and \emph{range} have small impact on the performance of \competitorImp while computing top-k results is significantly faster than full sorting when $k$ is small.  

\vspace{-5pt}
	\begin{figure}[h]
  \centering
  \begin{minipage}{1.0\linewidth}
  \centering
\resizebox{1\linewidth}{!}{
{
  \begin{tabular}{c|r|r|r|r|r}
 \thead{Configurations}                                                                     & \multicolumn{1}{|c}{\thead{\competitorDet}} & \multicolumn{1}{|c}{\thead{\competitorImp}} & \multicolumn{1}{|c}{\thead{\competitorRewr}} & \multicolumn{1}{|c}{\thead{\competitorMCDBten}} & \multicolumn{1}{|c}{\thead{\competitorMCDBtwenty}} \\ \hline
	r=1k,u=5\% 	& 31.5ms 	& 233.1ms 	& 786.7ms 	& 310.1ms	& 639.3ms 	\\
    \hline
	r=10k,u=5\%  & 30.9ms 	& 286.1ms 	& 792.6ms 	&  314.3ms	& 621.2ms	\\
	\hline
	r=1k,u=20\%  & 31.8ms 	& 266.3ms 	& 794.9ms 	& 325.8ms	& 651.2ms 	\\
	\hline
	r=1k,u=5\%,k=2  & 13.4ms 	& 48.3ms 	& 750.4ms 	& 149.1ms	& 295.2ms 	\\
	\hline
	r=1k,u=5\%,k=10  & 13.4ms 	& 48.2ms 	& 751.1ms 	& 150.4ms	& 296.1ms 	\\
	\hline
  \end{tabular}
  }}
  \resizebox{.5\textwidth}{!}{
  \textbf{Range(r),Uncertainty(u),k or full sorting}
  }
  \\[-4mm]
\setlength{\belowcaptionskip}{-13pt}
\bfcaption{\footnotesize Sorting and Top-K Microbenchmarks - Performance}
\label{fig:rankmicro}
\end{minipage}
\end{figure}

\vspace{-20pt}
\begin{figure}[h]
  \centering
  \begin{minipage}{1.0\linewidth}
\begin{subfigure}[b]{.5\linewidth}
	\centering
  \includegraphics[width=0.99\textwidth, trim=0cm 3cm 0cm 2cm, clip]{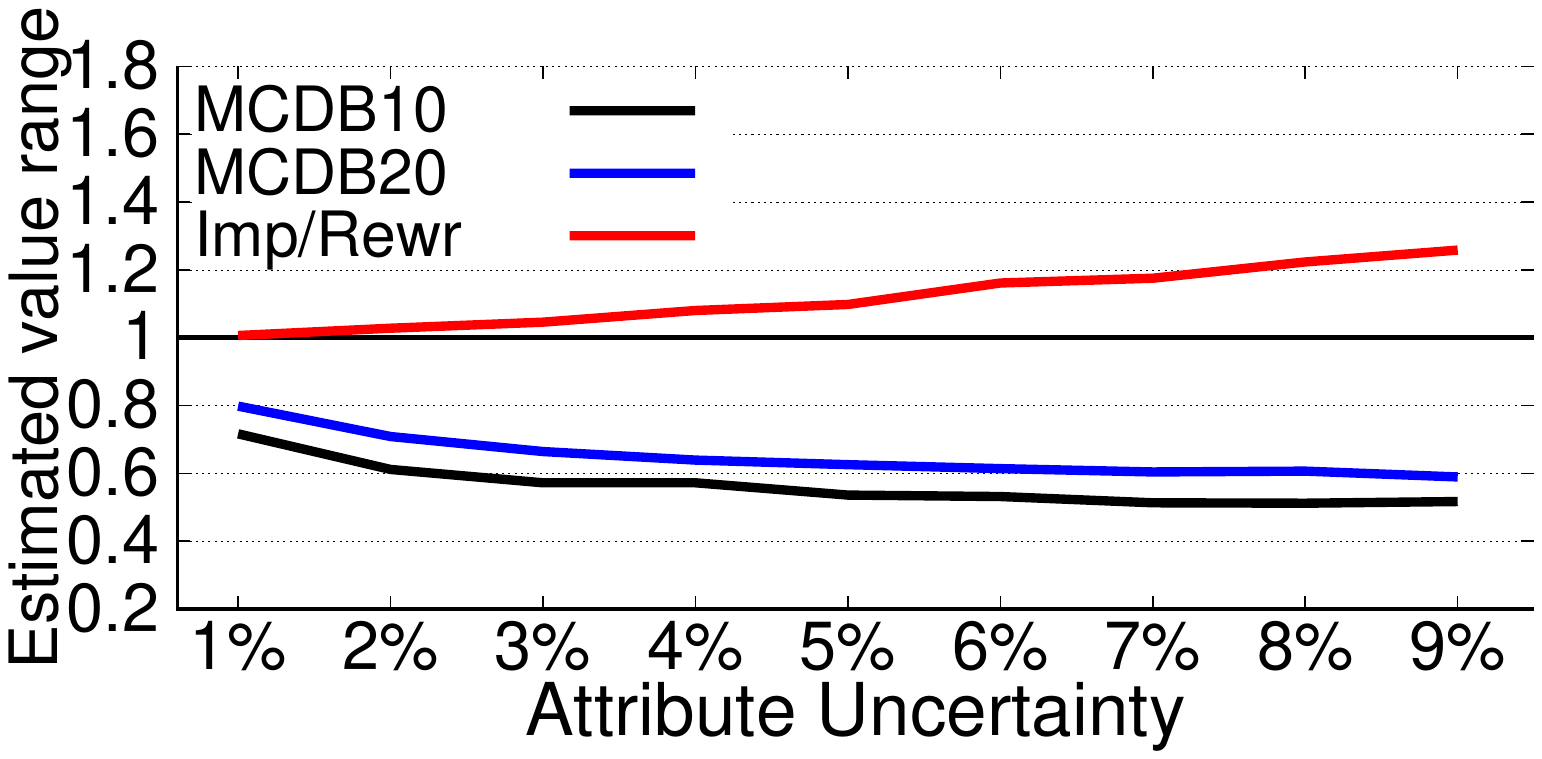}
  \vspace*{-7mm}
  \bfcaption{Varying uncertainty}
  \label{fig:microrankacc1}
\end{subfigure}
\begin{subfigure}[b]{.5\linewidth}
	\centering
  \includegraphics[width=0.99\textwidth, trim=0cm 3cm 0cm 2cm, clip]{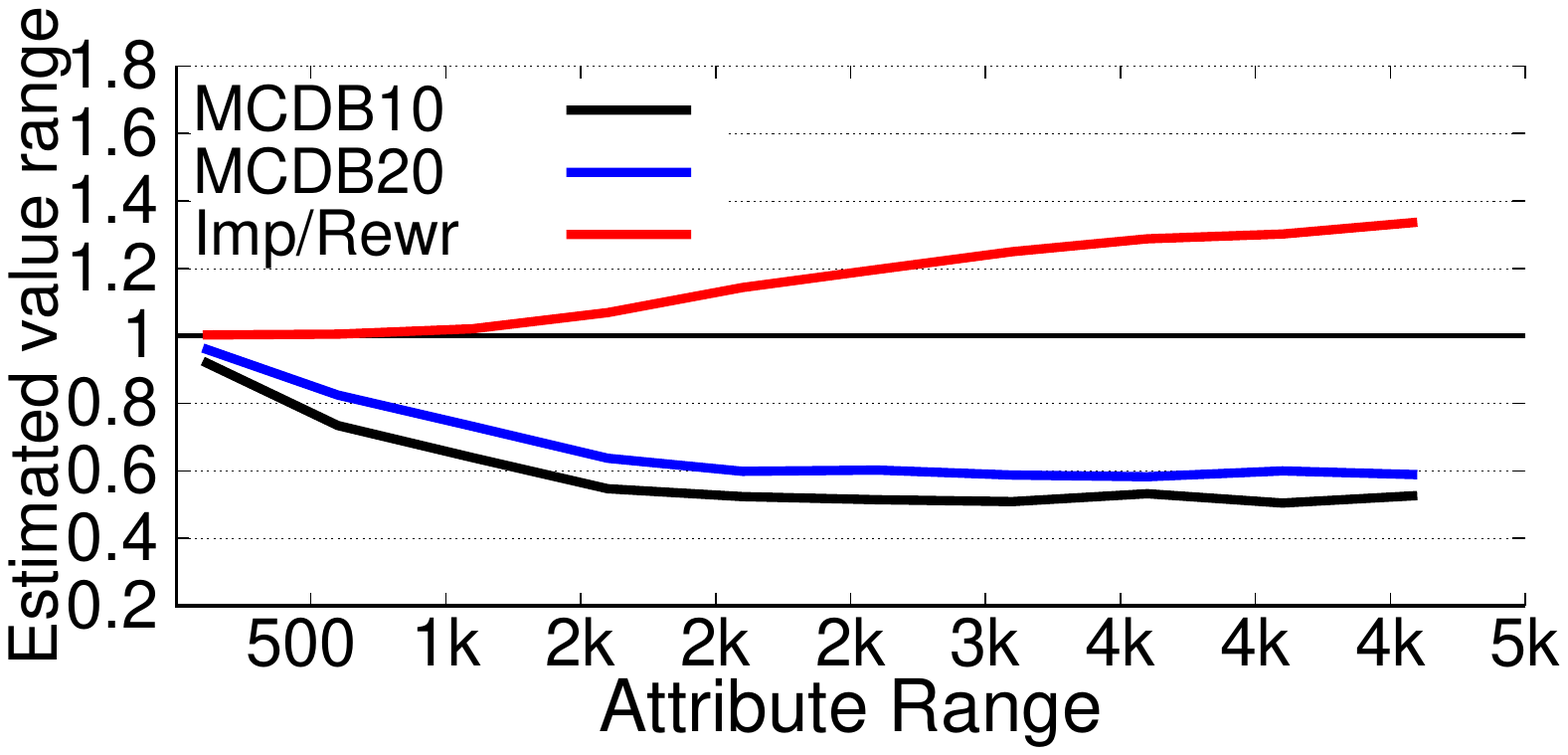}
  \vspace*{-7mm}
  \bfcaption{Varying range}
  \label{fig:microrankacc2}
\end{subfigure}\\[-8mm]
\setlength{\belowcaptionskip}{-10pt}
\bfcaption{\footnotesize Sorting microbenchmarks - approximation quality}
\label{fig:microrankacc}
\end{minipage}
\end{figure}

\begin{figure}[t]
  \centering
  \begin{minipage}{1.0\linewidth}
\begin{subfigure}[b]{.5\linewidth}
	\centering
  \includegraphics[width=0.99\textwidth, trim=0cm 3cm 0cm 2cm, clip]{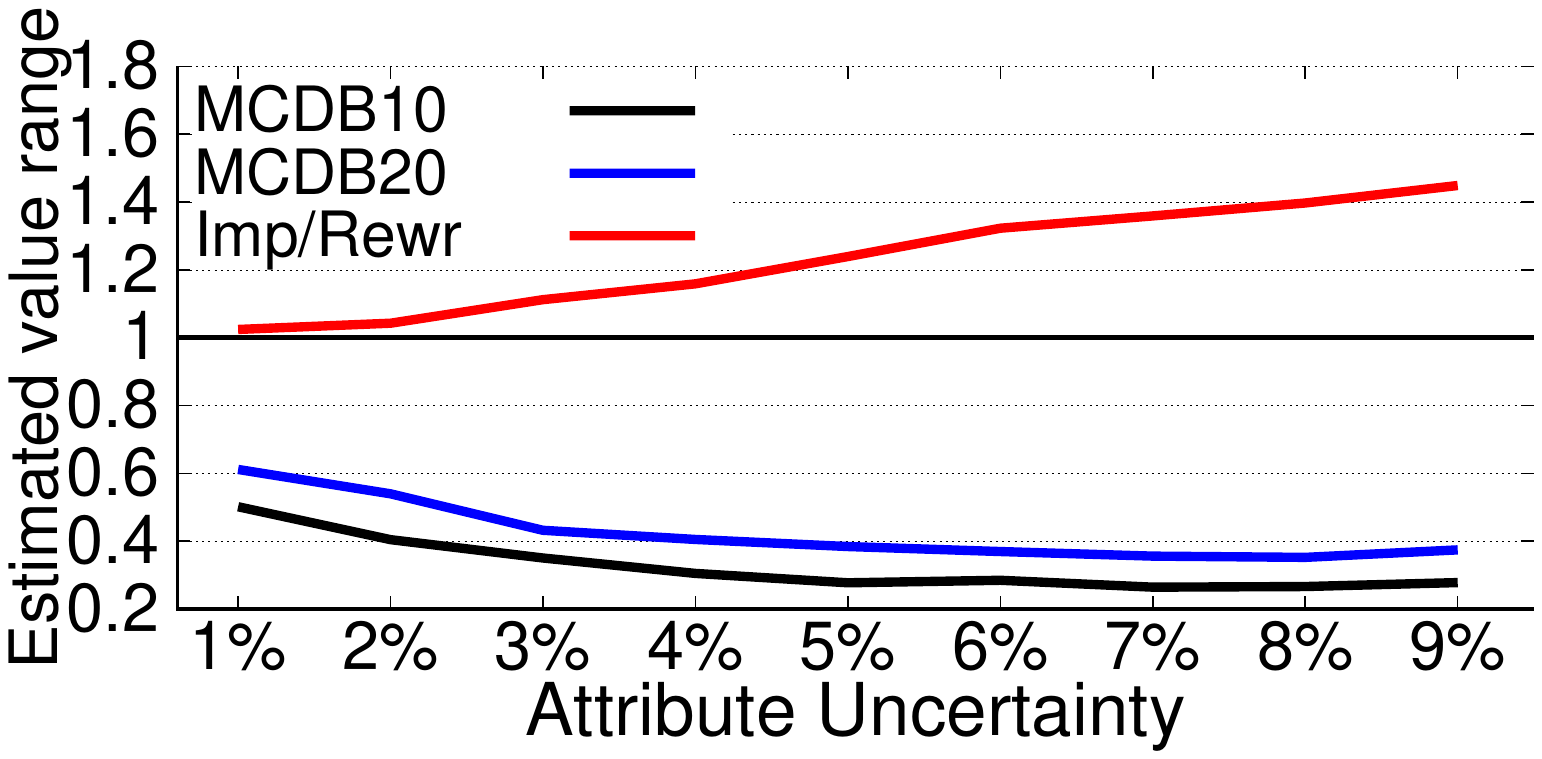}
  \vspace*{-7mm}
  \bfcaption{Varying uncertainty}
  \label{fig:microwindowacc1}
\end{subfigure}
\begin{subfigure}[b]{.5\linewidth}
	\centering
  \includegraphics[width=0.99\textwidth, trim=0cm 3cm 0cm 2cm, clip]{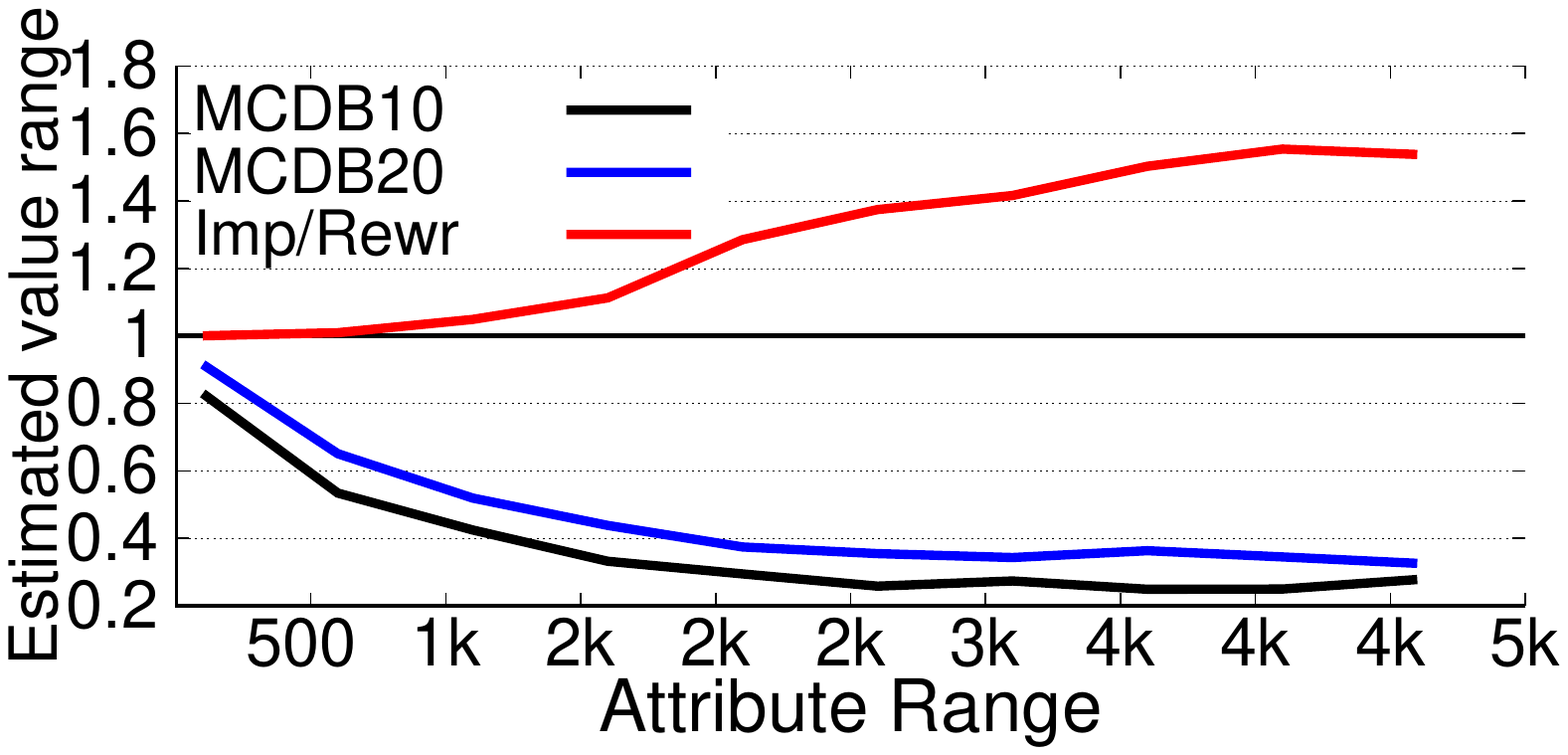}
  \vspace*{-7mm}
  \bfcaption{Varying range}
  \label{fig:microwindowacc2}
\end{subfigure}\\[-8mm]
\setlength{\belowcaptionskip}{-10pt}
\bfcaption{\footnotesize Window microbenchmarks - approximation quality}
\label{fig:microwindowacc}
\end{minipage}
\end{figure}

\begin{figure}[h]
	\begin{minipage}{1\linewidth}
  	\begin{subfigure}[b]{.5\linewidth}
	\centering
  \includegraphics[width=.99\textwidth, trim=0cm 1.5cm 0cm 0cm, clip]{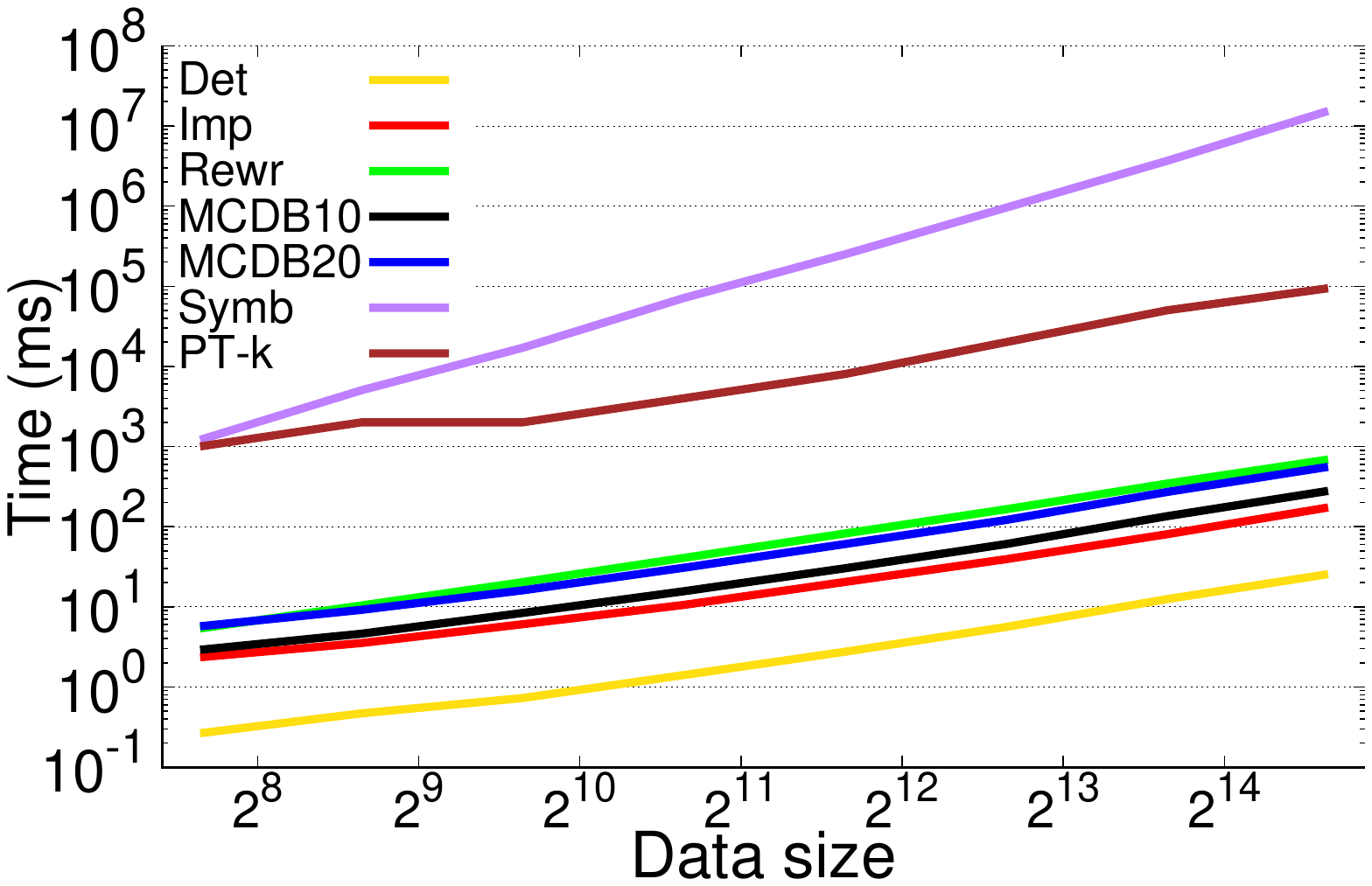}
  \vspace*{-7mm}
  \bfcaption{Smaller datasets}
  \label{fig:rank-altsize-low}
\end{subfigure}
\begin{subfigure}[b]{.5\linewidth}
	\centering
  \includegraphics[width=.99\textwidth, trim=0cm 1.5cm 0cm 0cm, clip]{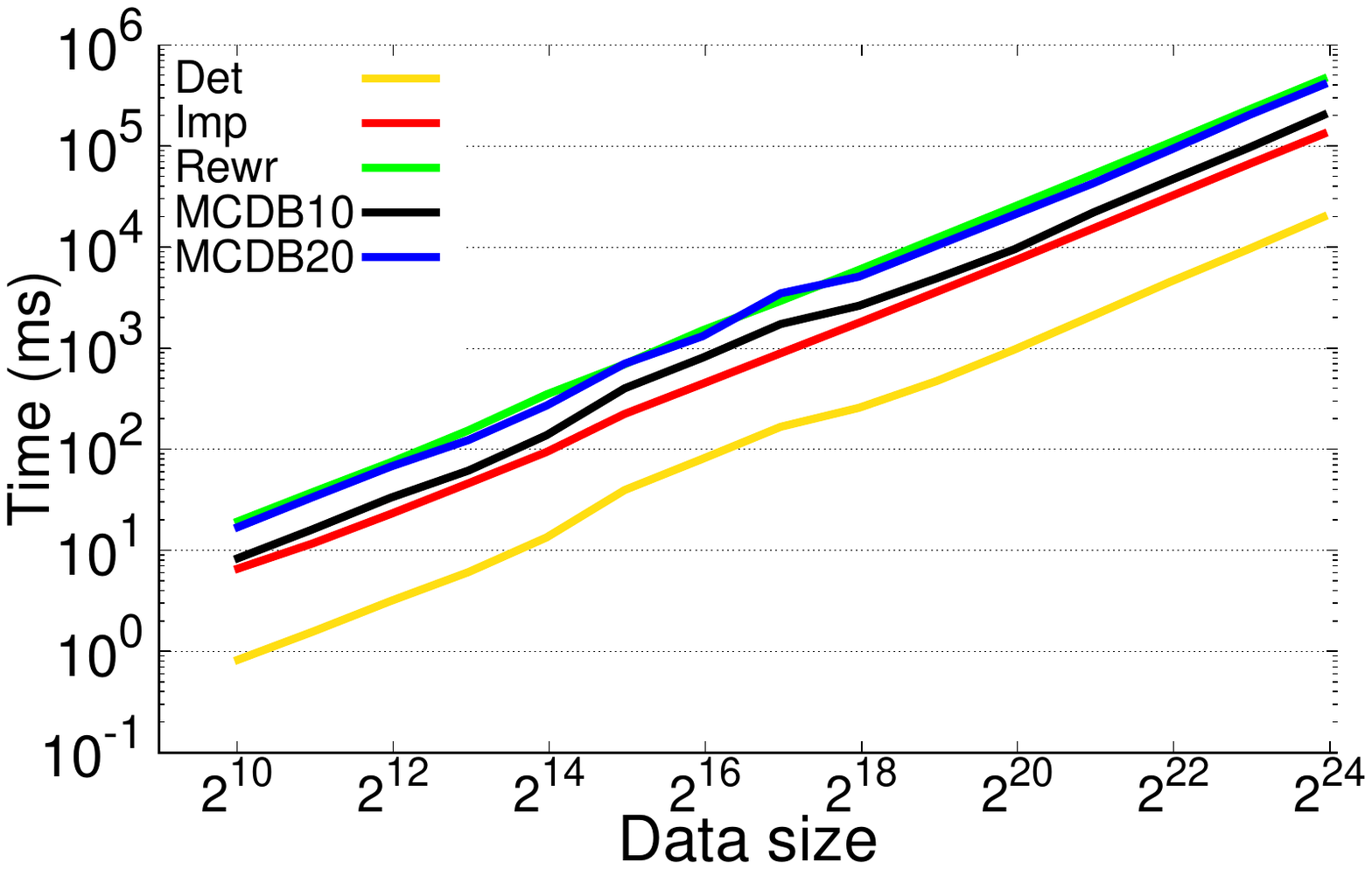}
  \vspace*{-7mm}
  \bfcaption{Larger datasets}
  \label{fig:rank-altsize-high}
\end{subfigure}\\[-8mm]
\BGI{Perhaps add size $\backslash$verb!$\backslash$n! (relative)}
\setlength{\belowcaptionskip}{-5pt}
  	\caption{\footnotesize Sorting performance varying dataset size \label{fig:rankaltsize}}
	\end{minipage}
\end{figure}

\begin{figure}[h]
	\begin{minipage}{1\linewidth}
  	\begin{subfigure}[b]{.5\linewidth}
	\centering
  \includegraphics[width=.99\textwidth, trim=0cm 1.5cm 0cm 0cm, clip]{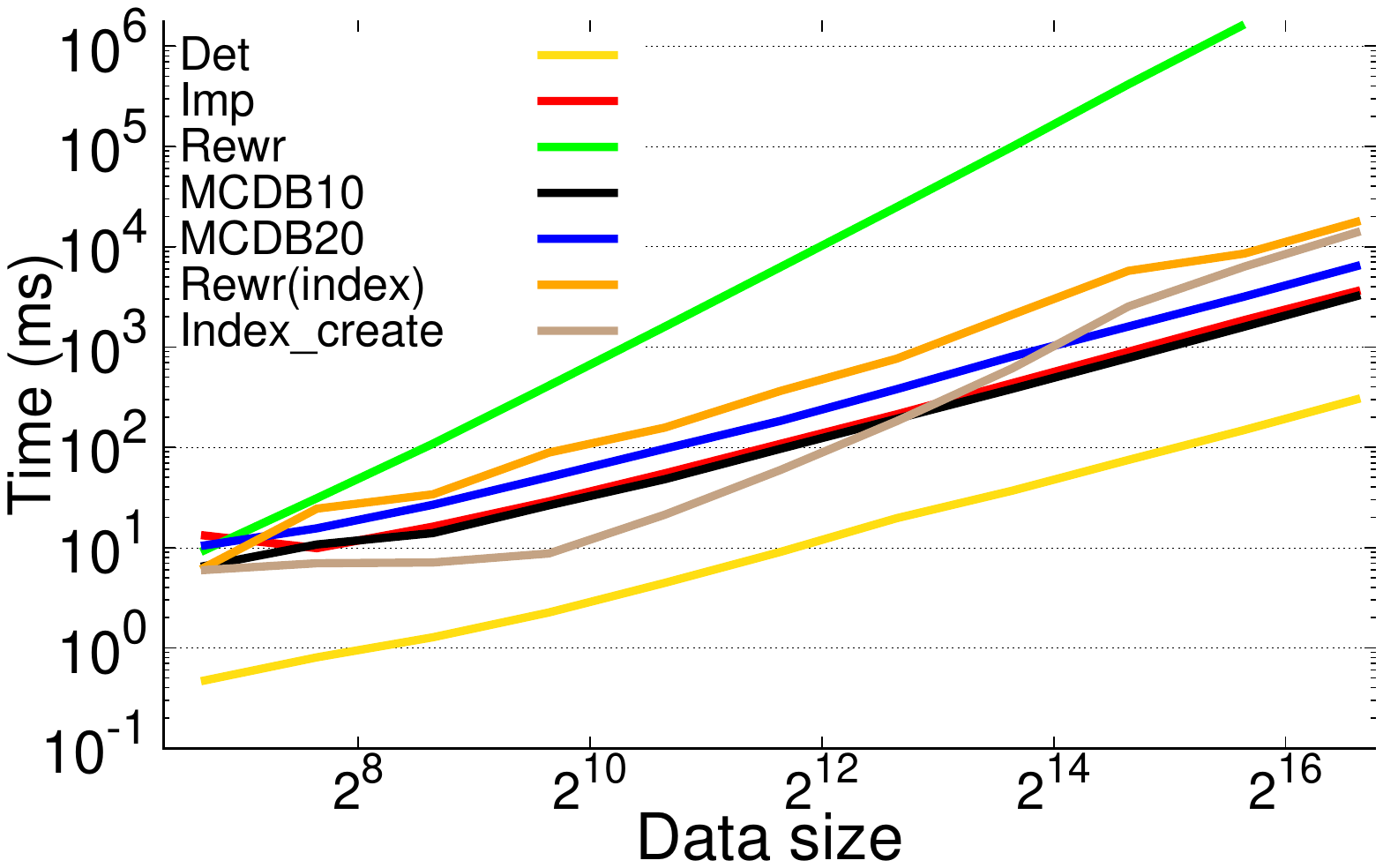}
  \vspace*{-7mm}
  \bfcaption{\revc{Smaller datasets}}
  \label{fig:window-altsize-low}
\end{subfigure}
\begin{subfigure}[b]{.5\linewidth}
	\centering
  \includegraphics[width=.99\textwidth, trim=0cm 1.5cm 0cm 0cm, clip]{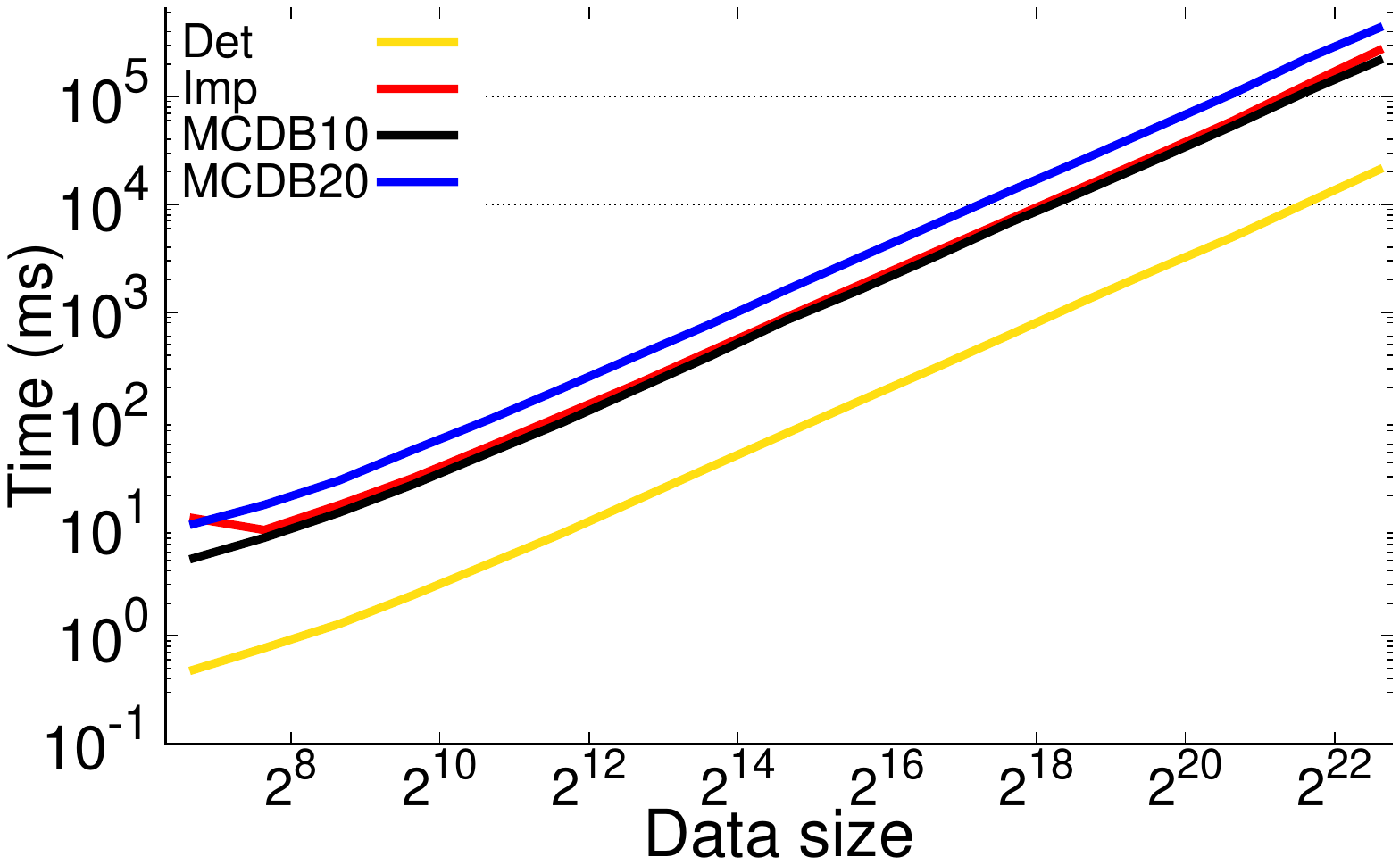}
  \vspace*{-7mm}
  \bfcaption{Larger datasets}
  \label{fig:window-altsize-high}
	\end{subfigure}\\[-8mm]
	\setlength{\belowcaptionskip}{-2pt}
  	\caption{\footnotesize Windowed aggregation performance varying dataset size \label{fig:windowaltsize}}
	\end{minipage}
\end{figure}

\begin{figure}[h]
  \centering
  \begin{subfigure}{\linewidth}
  \centering
  \resizebox{0.95\linewidth}{!}{
{\footnotesize
  \begin{tabular}{cc|r|r|r|r}
 \multicolumn{2}{c|}{\thead{Configurations}}                                                                     & \multicolumn{1}{|c}{\thead{\competitorDet}} & \multicolumn{1}{|c}{\thead{\competitorImp}} & \multicolumn{1}{|c}{\thead{\competitorMCDBten}} & \multicolumn{1}{|c}{\thead{\competitorMCDBtwenty}} \\ \hline
    \multirow{4}{*}{\rotatebox[origin=c]{0}{\parbox[c]{2cm}{\centering Order-by\\+ Window size                                                                                                                                                                 }}}
	& w=3,r=1k,u=5\% 	& 85.3ms 	& 895.3ms 	& 948.6ms 	& 1850.4ms 	\\
    \cline{2-6}
	& w=3,r=10k,u=5\%  & 87.1ms 	& 899.7ms 	& 931.3ms 	&  1877.5ms	\\
	\cline{2-6}
	& w=3,r=1k,u=20\%  & 88.7ms 	& 903.2ms 	& 944.7ms 	& 1869.7ms 	\\
	\cline{2-6}
	& w=6,r=1k,u=5\%  & 86.2ms 	& 1008.3ms 	& 953.1ms 	& 1885.1ms 	\\
	\hline
  \end{tabular}
  }}\\[-2mm]
  \bfcaption{Order-by, Window size (w), Range (r), Uncertainty (u)}
  \label{fig:microw-orderby}
  \end{subfigure}
  \begin{subfigure}{\linewidth}
  \centering
  \resizebox{0.95\linewidth}{!}{
{\footnotesize
  \begin{tabular}{cc|r|r|r|r}
 \multicolumn{2}{c|}{\thead{Configurations}}                                                                     & \multicolumn{1}{|c}{\thead{\competitorDet}} & \multicolumn{1}{|c}{\thead{\competitorRewr}} & \multicolumn{1}{|c}{\thead{\competitorMCDBten}} & \multicolumn{1}{|c}{\thead{\competitorMCDBtwenty}} \\ \hline
    \multirow{3}{*}{\rotatebox[origin=c]{0}{\parbox[c]{2cm}{\centering Order-by\\+ Partiton-by\\+ Window size                                                                                                                                                                 }}}
	& w=3,r=1k,u=5\% 	& 105.1ms 	& 73.5s 	& 1209.4ms 	& 2127.1ms 	\\
    \cline{2-6}
	& w=3,r=10k,u=5\%  & 101.7ms 	& 75.2s 	&  1231.3ms	&  2142.9ms	\\
	\cline{2-6}
	& w=3,r=1k,u=20\%  & 104.2ms 	& 81.1s 	&  1201.1ms & 2102.3ms	\\
	\hline
  \end{tabular}
  }}\\[-2mm]
  \bfcaption{Order-by + partition-by, Window size (w), Range (r), Uncertainty (u)}
  \label{fig:microw-partitionby}
  \end{subfigure}\\[-4mm]
  \setlength{\belowcaptionskip}{2pt}
  \bfcaption{\footnotesize Windowed aggregation microbenchmarks - Performance}
\label{fig:microw}
\end{figure}


\mypar{Accuracy}
\Cref{fig:microrankacc} shows  the error of the bounds generated by \competitorImp (\competitorRewr produces identical outputs), and \competitorMCDB.
Recall that \competitorImp is guaranteed to over-approximate the correct bounds, while \competitorMCDB is guaranteed to under-approximate the bounds, \revm{because it does not compute all possible results}. We measure the size of the bounds related to the size of the correct bound (as computed by \competitorSymb and \competitorPTk), and then take the average over all normalized bound sizes.
In all cases our approach produces bounds that are closer to the exact bounds than \competitorMCDB (~30\% over-approximation versus ~70\% under-approximation in the worst case).
We further note that an over-approximation \revm{of possible answers is often} preferable to an under-approximation because no possible results will be missed.

\subsubsection{Windowed Aggregation}
\mypar{Scaling Data Size}
\Cref{fig:windowaltsize} shows the runtime of windowed aggregation when varying dataset size.  
We compare two variants of our rewrite-based approach which uses a range overlap join to determine which tuples could possibly belong to a window.
\competitorRewrIndex uses a range index supported by Postgres.  
\revm{We show index creation time and query time separately}.
We exclude \competitorSymb, because for more than 1k tuples,  Z3 exceeds the maximal allowable call stack depth and crashes.
The performance of \competitorImp is roughly on par with \competitorMCDBten.
\competitorRewrIndex is almost as fast as \competitorMCDBtwenty, but is 5 $\times$ slower than \competitorImp.

\mypar{Varying window spec, Ranges, and Rate}
\Cref{fig:microw} shows the runtime of windowed aggregation varying  attribute uncertain value ranges (on all columns), percentage of uncertain tuples, and window size. For \competitorImp
(\Cref{fig:microw-orderby})
we use a query without partition-by. We also compare runtime of our rewriting based approach
(\Cref{fig:microw-partitionby})
 using 
 both partition-by and order-by on 8k rows.
\competitorImp exhibits similar runtime to \competitorMCDBten and outperforms \competitorMCDBtwenty.
 \iftechreport{Doubling the window size have only a slight impact (about 10\%) on our implementation performance.}
 \competitorRewr is slower than \competitorMCDB by several magnitudes due to the range-overlap join. 
\iftechreport{Our techniques are not significantly affected by the range and uncertainty rate.}


\begin{figure}[t]
  \centering
\resizebox{1\linewidth}{!}{
{
  \begin{tabular}{cc|r|r|r|r|r|r}
    \multicolumn{2}{c|}{\thead{Datasets}}                                                                    & \multicolumn{1}{|c}{\thead{\competitorImp}}  & \multicolumn{1}{|c}{\thead{\competitorDet}} & \multicolumn{1}{|c}{\thead{\competitorMCDBtwenty}} & \multicolumn{1}{|c}{\thead{\competitorRewr}} & \multicolumn{1}{|c}{\thead{\competitorSymb}} & \multicolumn{1}{|c}{\thead{\competitorPTk}}                                     \\
 \multicolumn{2}{c|}{\thead{\& Queries}}                                                                     & \multicolumn{1}{|c}{\thead{(time)}} & \multicolumn{1}{|c}{\thead{(time)}}  & \multicolumn{1}{|c}{\thead{(time)}} & \multicolumn{1}{|c}{\thead{(time)}} & \multicolumn{1}{|c}{\thead{(time)}} & \multicolumn{1}{|c}{\thead{(time)}}       \\ \hline
    \multirow{2}{*}{\rotatebox[origin=c]{0}{\parbox[c]{2cm}{\centering \textbf{Iceberg}~\cite{icebergdata}                                                                                                                                                               \\(1.1\%, 167K)}}}
	& Rank 	& 0.816msms 	& 0.123ms 	& 2.337ms 	& 1.269ms 	& 278ms   & 1s   \\
    \cline{2-8}
	& Window & 2.964ms 	& 0.363ms 	& 7.582ms 	& 1.046ms 	& 589ms   & N.A.   \\
	\hline
    \multirow{2}{*}{\rotatebox[origin=c]{0}{\parbox[c]{2cm}{\centering \textbf{Crimes}~\cite{crimesdata}                                                                                                                                                               \\(0.1\%, 1.45M)}}}
	& Rank 	&  1043.505ms 	&  94.306ms 	& 2001.12ms 	& 14787.723ms 	& >10min   & >10min   \\
    \cline{2-8}
	& Window & 3.050ms 	& 0.416ms  	& 8.337ms 	& 2.226ms 	& >10min 	& N.A.   \\
	\hline
    \multirow{2}{*}{\rotatebox[origin=c]{0}{\parbox[c]{2cm}{\centering \textbf{Healthcare}~\cite{healthcaredata}                                                                                                                                                               \\(1.0\%, 171K)}}}
    & Rank 	& 287.515ms & 72.289ms 	& 1451.232ms 	& 4226.260ms 	&   15s & 8s   \\
    \cline{2-8}
	& Window & 130.496ms 	& 15.212ms  	& 323.911ms 	& 13713.218ms 	& >10min & N.A.   \\
	\hline
  \end{tabular}
  }}\\[-4mm]
	\setlength{\belowcaptionskip}{2pt}
  \bfcaption{\footnotesize Real world data - performance}
  \label{fig:realq_perf}
\end{figure}

\begin{figure}[t]
  \centering
\resizebox{0.75\linewidth}{!}{
{
  \begin{tabular}{cc|l|l|l}
   \multicolumn{2}{c|}{\thead{Datasets \& Measures}}                                                                    & \multicolumn{1}{|c}{\thead{\competitorRankAUDB}} & \multicolumn{1}{|c}{\thead{\competitorMCDBtwenty}}  & \multicolumn{1}{|c}{\thead{\competitorPTk/\competitorSymb}}                                       \\ \hline
    \multirow{2}{*}{\rotatebox[origin=c]{0}{\parbox[c]{1.5cm}{\centering \textbf{Iceberg}\\~\cite{icebergdata}                                                                                                                                                                 }}}
	& bound accuracy 	&   0.891 	& 1 & 1	\\
    \cline{2-5}
	& bound recall  & 1 	&  	0.765 & 1\\
	\hline
    \multirow{2}{*}{\rotatebox[origin=c]{0}{\parbox[c]{1.5cm}{\centering \textbf{Crimes}\\~\cite{crimesdata}                                                                                                                                                                }}}
	& bound accuracy 	& 0.996 	& 1 & 1	\\
    \cline{2-5}
	& bound recall  & 1 	& 0.919  & 1	\\
	\hline
    \multirow{2}{*}{\rotatebox[origin=c]{0}{\parbox[c]{2cm}{\centering \textbf{Healthcare}\\~\cite{healthcaredata}                                                                                                                                                                }}}
	& bound accuracy 	& 0.990 	& 1  & 1	\\
    \cline{2-5}
	& bound recall  & 1 	& 0.767  & 1	\\
	\hline
  \end{tabular}
  }}\\[-4mm]
  \setlength{\belowcaptionskip}{2pt}
  \bfcaption{\footnotesize Real world data - sort position accuracy and recall}
  \label{fig:realq-rankacc}
\end{figure}

\begin{figure}[t]
  \centering
\resizebox{1\linewidth}{!}{
{
  \begin{tabular}{cc|r|r|r|r}
   \multicolumn{2}{c|}{\thead{Datasets}}                                                                    & \multicolumn{1}{|c}{\thead{Grouping/Order}}  & \multicolumn{1}{|c}{\thead{Grouping/Order}} & \multicolumn{1}{|c}{\thead{Aggregation}} & \multicolumn{1}{|c}{\thead{Aggregation}}                                      \\
 \multicolumn{2}{c|}{\thead{\& Methods}}                                                                     & \multicolumn{1}{|c}{\thead{accuracy}} & \multicolumn{1}{|c}{\thead{recall}}  & \multicolumn{1}{|c}{\thead{accuracy}} & \multicolumn{1}{|c}{\thead{recall}} \\ \hline
    \multirow{3}{*}{\rotatebox[origin=c]{0}{\parbox[c]{1.5cm}{\centering \textbf{Iceberg}\\~\cite{icebergdata}                                                                                                                                                                 }}}
	& \competitorRankAUDB 	& 0.977 	& 1 	& 0.925 	& 1 	\\
    \cline{2-6}
	& \competitorMCDBtwenty  & 1 	& 0.745 	& 1 	&  0.604	\\
	\cline{2-6}
	& \competitorSymb  & 1 	& 1 	& 1 	& 1 	\\
	\hline
    \multirow{3}{*}{\rotatebox[origin=c]{0}{\parbox[c]{1.5cm}{\centering \textbf{Crimes}\\~\cite{crimesdata}                                                                                                                                                               }}}
	& \competitorRankAUDB 	& 0.995 	& 1 	& 0.989 	& 1 	\\
    \cline{2-6}
	& \competitorMCDBtwenty  & 1 	& 0.916 	& 1 	& 0.825 	\\
	\cline{2-6}
	& \competitorSymb  & 1 	& 1 	& 1 	& 1 	\\
	\hline
    \multirow{3}{*}{\rotatebox[origin=c]{0}{\parbox[c]{2cm}{\centering \textbf{Healthcare}\\~\cite{healthcaredata}                                                                                                                                                                }}}
	& \competitorRankAUDB 	& 0.998 	& 1 	& 0.998 	& 1 	\\
    \cline{2-6}
	& \competitorMCDBtwenty  & 1 	& 0.967 	& 1 	&  0.967	\\
	\cline{2-6}
	& \competitorSymb  & 1 	& 1 	& 1 	& 1 	\\
	\hline
  \end{tabular}
  }}
  \setlength{\belowcaptionskip}{-2pt}
  \bfcaption{\footnotesize Real world data - windowed aggregation accuracy and recall}
  \label{fig:realq-windacc}
\end{figure}
\subsection{Real World Datasets}
We evaluate our approach on real datasets (Iceberg~\cite{icebergdata}, Chicago crime data~\cite{crimesdata}, and Medicare provide data~\cite{healthcaredata}) using realistic sorting and windowed aggregation queries~\cite{aval}.\BG{Where do we get these queries from?}\BG{Show basic stats for these datasets} To prepare the datasets, we perform data cleaning methods (entity resolution and missing value imputation) that output a AU-DB encoding of the space of possible repairs. \Cref{fig:realq_perf} shows the performance of real queries on these datasets reporting basic statistics (uncertainty and \#rows).

\iftechreport{
We use the following queries. \\
\mypar{iceberg} \\
Find top 3 sizes of ice-bergs mostly observed.
\begin{lstlisting}
	SELECT size,count(*) AS ct FROM iceberg
	GROUP BY size
	ORDER BY ct DESC LIMIT 3;
\end{lstlisting}
Window: For each day, find rolling sum of number of icebergs observed on that day and following 3 days.
\begin{lstlisting}
	SELECT date, sum(number) OVER (ORDER BY date
		BETWEEN CURRENT ROW AND 3 FOLLOWING) AS r_sum
	FROM iceberg;
\end{lstlisting}
\mypar{Crimes} \\
Rank: Find top three days with most incidents of crimes.
\begin{lstlisting}
	SELECT date, count(*) AS ct
		FROM crimes GROUP BY date
		ORDER BY ct DESC LIMIT 3;
\end{lstlisting}
Window: For each crime in 2016, find the earliest year among the crime itself and nearest crime at north and south of it.
\begin{lstlisting}
	SELECT rid, min(year) OVER
		(ORDER BY latitude BETWEEN
			1 PRECEDING AND 1 FOLLOWING) AS min_year
	FROM crimes WHERE year='2016';
\end{lstlisting}
\mypar{healthcare} \\
Rank: Find top 5 facility with highest score on MRSA Bacteremia.
\begin{lstlisting}
	SELECT facility_id,facility_name,score FROM healthcare
	WHERE measure_name = 'MRSA Bacteremia'
	ORDER BY score LIMIT 5;
\end{lstlisting}
Window: get in-line rank of facility on MRSA Bacteremia scores.
\begin{lstlisting}
	SELECT facility_id,facility_name,count(*) OVER
	(ORDER BY score DESC) AS rank
	FROM healthcare
	WHERE measure_name = 'MRSA Bacteremia';
\end{lstlisting}


}

For sorting and top-k queries \revb{that contain aggregation which commonly seen in real use-cases, we only measure the performance of the sorting/top-k part over pre-aggregated data} (see~\cite{FH21} for an evaluation of the performance of aggregation over AU-DBs) 
In general, our approach (\competitorImp) is faster than \competitorMCDBtwenty. \competitorSymb and \competitorPTk are significantly more expensive. \Cref{fig:realq-rankacc} shows the approximation quality for our approach and \competitorMCDB. Our approach has precision close to 100\% except for sorting on the Iceberg dataset  which has a larger fraction of uncertain tuples and wider ranges of uncertain attribute values due to the pre-aggregation. \competitorMCDB has lower recall on Iceberg and Healthcare sorting queries since these two datasets have more uncertain tuples (10 times more than the Crimes dataset). \Cref{fig:realq-windacc} shows the approximation quality of our approach and \competitorMCDB for windowed aggregation queries.
We measured both the approximation quality of grouping of tuples to windows and for the aggregation result values.
For Crimes and Iceberg, the aggregation accuracy is affected by the partition-by/order-by attribute accuracy and the uncertainty of the aggregation attribute itself. The healthcare query computes a count, i.e., there is no uncertainty in the aggregation attribute and approximation quality is similar to the one for sorting. Overall, we provide good approximation quality at a significantly lower cost than the two exact competitors.


\section{Conclusions and Future Work}
\label{sec:concl-future-work}

In this work, we present an efficient approach for under-ap\-prox\-i\-mat\-ing certain answers and over-ap\-prox\-i\-mat\-ing possible answers for top-k, sorting, and windowed aggregation queries over incomplete databases. 
Our approach based on AU-DBs~\cite{FH21} is unique in that it supports windowed aggregation, is also closed under under full relational algebra with aggregation, and is implemented as efficient one-pass algorithms in Postgres.
Our approach significantly outperforms existing algorithms for ranking uncertain data while being applicable to more expressive queries and bounding all certain and possible query answers.
\iftechreport{Thus, our approach enables the efficient evaluational of complex queries involving sorting over incomplete databases. We present a SQL-based implementation as well as the aforementioned one-pass algorithms. Using an implementation of these algorithms in Postgres, we demonstrate that our approach significantly outperforms the SQL-based implementation and for windowed aggregations we have performance close to sampling based approach with 10 samples while 10 sample produces low recall comparing with the accuracy our approach have. Furthermore, our approach significantly outperforms an existing algorithm for ranking uncertain data while being applicable to more expressive queries and bounding all certain and possible query answers.}
In future work, we plan to extend our approach to deal more expressive classes of queries, e.g., recursive queries, and will investigate index structures for AU-DBs to further improve performance.


\bibliographystyle{abbrv}
\bibliography{main.bib}

\end{document}